2021

# The Struggle with Inequality

FROM ROUSSEAU TO PIKKETY

SHIN-ICHIRO INABA (MEIJIGAKUIN UNIVERSITY)

# Contents

*This is an English translation of Shin-ichiro Inaba, *Fubyoudou tono Tatakai* (in Japanese, Tokyo, Bungei Shunjuu, 2016) by the author himself, with assistance of www.DeepL.com/Translator (free version).

<div style="text-align: right;">
(!st draft, April 3, 2021.)

(2nd draft, April 16, 2021.)
</div>

# Prologue

When did the theory of economic inequality begin?

By the time French economist Thomas Piketty's magnum opus, *Capital in the 21st Century*, was published in French in 2013, he should have already established a reputation among economists interested in inequality, both in France and abroad. What we did not expect was that a 1,000-page book with extensive notes, however plainly written for a general audience, would quickly become a bestseller in English translation and in Japanese and other foreign editions, even though the French edition initially showed only moderate sales, and the aftermath must have flowed back to France, creating a worldwide boom.

How did it become such a big boom? It is hard to say, because a boom is a matter of chance, but as I mentioned earlier, prior knowledge of the work of Piketty, who is active internationally and writes in English, was already shared by economists interested in the subject in English-speaking countries - or rather, around the world. Among them were Joseph Stiglitz and Paul Krugman, who won the Nobel Prize in Economics, and were pushing hard before the English edition of *Capital in the 21st Century* was published.

Yes, the groundwork had already been laid. For a long time, economists, as well as sociologists and political scientists, had not paid much attention to economic inequality, especially domestic inequality in developed countries that experienced rapid growth after World War II. But, at least in the social sciences, and to some extent in journalism and what Krugman calls "policy promoters" (people who propose and sell policies, whether academics, intellectuals, journalists, or bureaucrats), there has begun a clear "inequality renaissance" in the transition from the 20th to the 21st century. The boom of the "inequality society" in Japan can also be said to have been part of this trend, although of course it probably reflects Japan's own special circumstances. Piketty also grew up as a scholar in the midst of this trend.

However, the significance of Piketty's work does not lie solely in making the world public aware of the meaning of these thick undercurrents that already existed. Piketty's statistical empirical research, which he began by digging up basic data after returning to France from the Massachusetts Institute of Technology in the United States, despite the high salary he was earning, and which he presents to the public rather than to academia in *Capital in the Twenty-First Century*, has been a major factor in the so-called "inequality renaissance" since the 1990s. It was clearly a new, if not an anti-mainstream, movement in the "inequality renaissance" of the 1990s.

What is it? Naturally, I would have to talk about this "inequality renaissance" in economics since the 1990s, but since I have been given the space of a whole volume, I would like to go back further and start from the beginning. However, it is a little too much to talk about the "beginning of inequality" as a real problem, so I will focus on inequality in the Piketty sense, that is, inequality in the capitalist market economy, and the disparities in income and wealth/assets that arise among the people who participate in the market and engage in economic activities. In addition, I will focus on the history of economics, in particular, the disciplines used to understand the inequality rather than the reality of such market inequality itself.

So let's go back about two and a half centuries to the 18th century in Western Europe, the era of the so-called Enlightenment. In other words, the "controversy" between Jean-Jacques Rousseau, the navel-gazer from Geneva who is said to have "ideologically prepared the French Revolution," and Adam Smith, the Scotsman who was never directly in contact with Rousseau and was later called the "father of economics" - or more precisely, Smith, who was impressed by Rousseau but only wanted to surpass him, is the benchmark.

In his "Letter to the Editor's Companion," Smith wrote to the Scottish Review of Edinburgh in 1756, urging them to pay attention to the so-called Enlightenment, the latest science and thought represented by *L'Encyclopédie*, in which Rousseau's name appears. According to Smith, the tradition of natural law theory and governance theory developed by Thomas Hobbes in *Leviathan* and John Locke in *Two Treatises of Governnent* was inherited more in France than in England, and Rousseau is among them.

In the seventeenth century, the era of the scientific revolution and the starting point of early modern philosophy independent of Christian theology, Hobbes, Hugo Grotius and others added a new meaning to the term "natural law. In the past, "natural law," as "the law established by God and imprinted on all creatures, including human beings, prior to the laws made by human beings," had been thought of as a model for man-made laws, institutions, and nations, and that man-made laws and the institutions of actual human society were imperfect imitations. However, since the time of Hobbes, the dominant view has been that the natural law commanded by God only defines the basic nature of nature and human beings, and that the specific law and order of the human world is to be created by human beings. The so-called "social contract theory" of Hobbes and Locke provides an exemplar of a theory of law, state and governance based on such an idea.

Smith, while criticizing Rousseau's arguments as "rhetorical excesses and useless for analysis," introduces them with great enthusiasm, and quotes at length from Rousseau's seminal work, *Discours sur l'origine et les fondements de l'inégalité parmi les hommes,* shortly after its publication. In fact, it is quite possible to interpret Smith's *The Wealth of Nations* as a refutation of Rousseau's *l'inégalité parmi les hommes,* in view of the theory of "government" in *Lectures on Jurisprudence*, which became the prototype of *The Wealth of Nations*, and the existing early drafts. In other words, while Rousseau ascribes the cause of social inequality and the misery of the poor to private ownership, the division of labor based on it, and the market economy, Smith admits that private ownership and the market economy are the cause of inequality, but argues that they do not necessarily produce miserable results, but rather the opposite.

Rousseau's Theory of the Origin of Human Inequality
First, let's take a look at *l'inégalité parmi les hommes*.
It is clear that this work aims to critique Hobbes's and Locke's theories of governance and the contractual state, but it rather accepts their arguments as an interpretation of reality, reverses the moral evaluation of them, and goes beyond them by exposing the further premises of their arguments. I aim to do this by exposing the further premises of their arguments. The following statement from the beginning of the second part of *l'inégalité parmi les hommes*. is a perfect example of this.

> The true founders of political society were those who enclosed a piece of land and found the first people simple enough to think and believe that it was theirs.   How many crimes, how many wars, how many murders, how many miseries, how many horrors, could he have spared mankind?

According to Rousseau, the establishment of the system of private property rights and the development of the division of labor resulting from it are the basic causes of the development of inequality between the rich and the poor, between the powerful and the powerless common people in human society. Rousseau accepts the basic framework of Hobbes' argument in its entirety, saying that in order to establish the order of property rights, the establishment of a centralized governing power, the state, is indispensable, and that the establishment of such a state can only be achieved through the agreement and cooperation of people who aim to establish property rights, in other words, through a "social contract. Rousseau accepts the basic framework of Hobbes' argument almost in its entirety. The dividing line is that, unlike Hobbes and Locke, Rousseau finds it horrifying.

Rousseau's argument was based on the idea that "the 'state of nature' of Hobbes and Locke is by no means a primordial, primordial state, and that there is an earlier, more primordial 'state of nature. Hobbes and Locke's "state of nature" is nothing more than our current civilized society without the power of the state, which Hobbes calls "a state of war" because it is already densely populated and has accumulated enough wealth to be fought over. Rousseau, on the other hand, considers a true "state of nature" to be a situation where the population is small and people are isolated and scattered in nature, living a life of near self-sufficiency. The "state of nature" described by Hobbes and Locke is, in Rousseau's view, a "social state.

Rousseau admits that there are sufficient grounds for Hobbes' and Locke's arguments as to why people want to establish property rights and therefore establish a state. It is only natural that those who seek survival and self-preservation should seek the security of property as the basis of their livelihood. Therefore, Rousseau did not deny the social contract, deny the state, and deny law and order, but sought the possibility of a better state and a better social contract, and discussed the state system and the constitution in his Theory of the Social Contract, and also discussed public administration in his Theory of Political Economy.

Smith, The Wealth of Nations

But in any case, it is clear that Rousseau seeks the fundamental cause of inequality in human society, especially in civilized society, in the system of property rights backed by state power. Smith, too, has no great objection to Rousseau in this respect. He states in The Wealth of Nations that

> Civil government, so far as it is instituted for the security of property, is in reality instituted for the defence of the rich against the poor, or of those who have some property against those who have none at all. (*The Wealth of Nations,* Book V, Chapter 1, Section 2, Paragraph 12.)

But on the other hand, Smith also argues as follows.

> [I]t may be true, perhaps, that the accommodation of an European prince does not always so much exceed that of an industrious and frugal peasant, as the accommodation of the latter exceeds that of many an African king, the absolute master of the lives and liberties of ten thousand naked savages. (*The Wealth of Nations,* Book 1, Chapter 1, Paragraph 11.)

As for the effects of the establishment of property rights, Rousseau first looks at the inequalities it perpetuates, but Smith, like Hobbes and Locke, focuses more on the fact that the establishment of property rights makes people feel secure in the use of their property and encourages them to work and invest, which in turn raises their productivity and thus their standard of living. In other words, it raises productivity and thus living standards. More than that, it emphasizes the tremendous effect of including transactions with others in the utilization of such property. Law and order will ensure not only the security of one's property, but also the security of one's dealings with others. The development of trade not only facilitates the effective use of property in society as a whole, which would not be possible if people were isolated, such as exchanging each other's unwanted items. Through exchange, people will be able to specialize and devote themselves to the work they are good at, which in turn will increase the productivity of the society as a whole.

David Ricardo's theory of "comparative advantage"-that all participants in a transaction can benefit by ceasing to be self-sufficient, specializing in tasks they are relatively good at, and obtaining through trade what they cannot provide themselves-is known as an argument in favor of free trade, but it is not limited to international trade. It is actually an argument for the division of labor and the benefits of trade in general.

In civilized societies, where the state power has established law and order and commercial transactions have developed as a result of the increased productivity of society as a whole, inequality has increased, as Rousseau said, but the lives of the people at the bottom of that unequal society have absolutely improved - that is, the richest people in societies where commerce has not developed. But the lives of those at the bottom of the inequality are absolutely improving - they are better off than the richest people in less commercial societies. Smith raises the question in this way.

The Origins of the "Growth or Inequality" Debate
Let me get this straight. Rousseau sought to locate the origin of inequality in the establishment of the property rights system (and the state power that supported it). Smith did not directly deny the argument, but he pointed out that the development of commercial transactions and market economy could further expand the inequality established by the property rights system. But on the other hand, Smith asked, "But isn't that inequality less of a problem, since it occurs in parallel with growth and the improvement of people's lives, including those at the bottom? He said it is true that the property rights system and the commercial and market economy based on it have brought about the inequality that Rousseau criticized, but they have also raised the overall level of the economy. This provides a prototype for today's "growth or inequality" debate.
If we read Smith's challenge to Rousseau a little differently, we can ask the following question,

although he does not ask it directly. Namely.

> What if the current inequalities could be alleviated, but only through a system of property rights and regulation of the market economy, which in turn caused a decline in the productive forces? In particular, what if it were to reduce the standard of living of the poor as well as the rich? What if the solution was to eliminate inequality in the form of "everyone getting poorer together?

And. This kind of argument is called the "leveling-down objection" in modern philosophy, and we will look at it in more detail at the end of this book. Although Smith does not throw the "leveling-down objection" itself at Rousseau, I think it is safe to regard him as its de facto forerunner.

Furthermore, we can raise an objection to Rousseau from another direction. To put it plainly, Rousseau's concept of the "state of nature" allows for the possibility of inequalities that fall into its blind spot, and I suspect that a Hobbesian perspective would be more meaningful in dealing with them.
Rousseau regarded the "state of nature" described by Hobbes and Locke as a "social state" at best, and tried to envision a more fundamental level, a situation where people are equal as isolated beings. However, is there nothing between this Hobbesian and Lockean situation and the Rousseauian situation? In the first place, isn't the Hobbesian "state of nature" = Rousseau's "state of society" also different from the Rousseauian "state of nature" and rather unequal? Doesn't Rousseau's toolkit, however, fail to capture the problem at this level? Isn't there a difference in the nature of inequality between the Hobbesian "state of war," the Lockean "state of nature," where law and order are loosely maintained by the self-help and voluntary mutual aid of the productive, and a state with a clear sovereignty?
Of course, if we compare this to the "state of nature" in Rousseau's sense, civil society with an established state would be in an unequal situation, but what happens if we compare this to the Hobbesian "state of nature" = "state of war" or the Lockean "state of nature" = "civil society without a state"? This is what I mean. If we follow the way of Hobbes, Locke, and Smith, such a "state of nature" must be overcome before we can question equality/inequality, because the productive forces and living standards are low to begin with. However, if we depart from Rousseau's trick (though it is a harsh word), we may find it interesting from the perspective of "equality/inequality.

If you read Hobbes more deeply, you will find a different point of view.

Hobbes argues that the state (or Commonwealth, as he calls it) that is established by overcoming the state of nature as a state of war is not necessarily established by the equal consent of people, by contract. In fact, Hobbes goes so far as to say that more often than not, states are formed as a result of conquest by powerful individuals.

This may seem surprising for Hobbes, who is an equality theorist (I won't use the word "egalitarian" here), who believes that people are equal in nature (they are equal in intellect and physical strength) and in reality, before "should" and idealism such as "people should be respected equally. It is not unusual, but rather normal, for a state to be formed as a clique of many henchmen under one master, not through an equal agreement or contract, but through violent struggle for power, perhaps even controlled by luck. It is normal. Therefore, Hobbes is not only comparing the "state of nature" as an anarchic "state of war" without a state in the first place, but also the civil society where a state is established. According to him, there are two types of states: the contract type and the conquest type (in Hobbes's own words, "commonwealth by institution" and "commonwealth by acquisition"), and it is meaningful to compare the two.

It is natural in the case of Locke, who explicitly recognizes the right of resistance and revolution, but even in the case of Hobbes, who emphasizes the absoluteness of sovereignty, the goal of sovereignty is only to uphold the law, and the arbitrary exercise of power is denied. In a conquering state, however, the arbitrary exercise of governing power, unconstrained by law, is quite possible. How would inequality fare under such a so-called "tyrannical" state? (We'll come back to this in Chapter 10.

The Inequality Debate in Capitalism

Let me summarize again here. When we discuss economic inequality within Japan, at the level of a country, we are basically assuming that the mechanism that runs the Japanese economy is a free market economy based on the foundation of a system of private property rights. On top of that, in order for the various resources (material resources, labor, knowledge, etc.) to be used efficiently in the economy and society as a whole, and for economic growth and improvement in the average standard of living to be achieved, it is desirable that free competition be sustained while observing the rules of ownership and the market.

However, free competition in the market creates not only inequality as a result of those who work hard and achieve commensurate results (i.e., obtain revenues appropriate to the costs invested), but also inequality due to differences in outcomes caused by pure bad luck. In other words, free competition in the market, if left unchecked, will increase inequality in society.

Therefore, if we want to correct the inequality, we have to intervene or restrict the free competition in the market. However, if excessive intervention in the market weakens competition, slows down the efficient use of resources, technological innovation, and the rise of productive forces, or even causes an absolute decline in them, then the goal (or at least one of the most important goals) aimed at by correcting inequality and reducing inequality, the improvement of the living standards of the people at the bottom, will fail. Therefore, the difficult question is how to strike a balance in this trade-off situation of "if you put up one side, you can't put down the other." ― The inequality debate in capitalist society, including the inequality debate in Japan's "lost 20 years," often takes this configuration. Even in the case of Thomas Piketty's blockbuster, it is safe to say that this is true to some extent.

The prototype for this kind of debate - or, to put it crudely, the debate between growth and inequality, or the trade-off between growth and inequality reduction - was established, roughly speaking, in the hypothetical birch-yard debate between Rousseau and Smith (?), as described here. The original form of the "trade-off between growth and easing of inequality" argument was, roughly speaking, made in the virtual debate between Rousseau and Smith described here. However, if we look at it a little more broadly in the context of the time, we can see that there are problems of law and order as a precondition for a market economy, and related problems of inequality and economic growth of a different kind than those created by competitive markets.

And in fact, these problems are as important as ever in today's world. In other words, in order to think about the global inequality between nations and the disparity between developed and developing countries, which Piketty dares to leave out of his subject matter, it is necessary to connect with such issues.

Structure of this book

The following is a brief sketch of the overall structure of this book.

The first and second chapters that follow focus on Smith and Karl Marx, in the broad classical tradition of economics, where it is believed that there is an inseparable relationship between production and distribution, economic growth and inequality. Of course, Smith affirmed the capitalist market economy and Marx criticized it, but they shared the recognition that the distribution of income and wealth affects production and economic growth.

However, in neoclassical economics, which originated around the end of the 19th century and has become the mainstream today, the issues of production and distribution are discussed separately. The benchmark is the recognition that "no matter what the distribution of income and wealth, if the market is efficient, the resources in society will be used efficiently and maximum production will be achieved". This is despite the fact that the economists considered to be the founders of the neoclassical school had shown an active interest in social issues, especially worker poverty. Solving this apparent puzzle is the task of Chapters 3 and 4.

Toward the end of the 20th century, however, there was a renewed awareness within economics of the problem of the relationship between production and distribution. In particular, when focusing on investment in technological innovation and human capital, there was a growing recognition that production and distribution are not necessarily unrelated, and that the distribution of income and wealth can affect short-term production and long-term economic growth. The young Piketty's beginnings as an economist are set against this backdrop - what this book calls the "inequality renaissance. Chapters 5 to 8 will explain the circumstances of the restoration of classical concerns within the neoclassical framework (in a sense).

Then, in chapters 9 and 10, I will discuss the fact that Piketty's work is a new development in this "inequality renaissance," in the context of "Capital in the 21st Century," and try to gain a perspective on future trends in the debate.

## Chapter 1: Adam Smith and the Classical Political Economy

The newness of Adam Smith - the image of "capitalism

In the previous chapter, I said, "The benchmarks are Rousseau and Smith!" So, let's look at Smith a little more.

There is no end to what was revolutionary about Smith's *Wealth of Nations,* but the idea of the "invisible hand" and the analysis of the market mechanism that balances supply and demand through changes in prices were of course much more sophisticated than before, but they were not started by Smith. It had already been argued before Smith that crops and industrial products are subject to commerce and that the competitive price mechanism works there. What is noteworthy here is that Smith clearly identified what he later called the " market of factor of production".

Smith's clear novelty is that the resources that go into producing such products, the fundamental sources of productive power that include human "labor(labour)," "land," which is a nature that cannot be produced by humans (but is nevertheless subject to human ownership), and "capital" (which Smith never seemed to grasp its essential nature). He clearly presented that they are traded in the market.

To be more precise, it is doubtful that Smith's focus on labor, capital, and land as factors of production, as distinct from ordinary products, is original. Jacques Turgot, a late Bourbon fiscal official who is often put in the box of Physiocracy in the history of economic theory and thought, clearly stands tall as a forerunner of Smith on this point. Nevertheless, if Smith dared to go further than Turgot, it was firstly because he interpreted the market for these factors of production as a price mechanism, and secondly because he did not see the economy in terms of specific and discrete orders of goods, industries, and technologies, such as grain, iron, agriculture, and industry. The second point is that we have come to think of the economy not in terms of grain, iron, agriculture, industry, or other concrete and individual goods, industries, and technologies, but in terms of more abstract and general orders.

This second point may be a little difficult to understand, so I will explain. First of all, when it comes to "labor," Smith is not talking about specific tasks such as farming, livestock care, or supervision in the field of agriculture, or the manual work of skilled craftsmen, supervision, or simple menial work in the field of manufacturing. All these kinds of work, no matter how diverse, are all done by the same person in the end. In terms of manual labor, simple tasks such as carrying a load or weeding a field may be quite different in terms of specific physical exercise, but if we measure the degree of fatigue per hour, they are almost the same - something that can be compensated for by similar rest and nutritional supplementation. If the work is more complex and requires prior knowledge or training, we may not be able to simply equate them, but if the cost of training and learning can be measured in the same way, in terms of money and time, we can still reduce them to the same kind of "labor.

The "labor(labour)" that Smith considers to be traded at the price of wages, and for which supply and demand are adjusted in the market, is ultimately this kind of abstract, general "labor" that any human being (with proper training) could probably do, not this or that kind of concrete "work. The above applies to "land" and "capital" in exactly the same way. If we look at the contents of "capital," we can see that it is a collection of miscellaneous and concrete things, such as production equipment (fixed capital), inventories of raw materials, intermediate goods, work in progress, etc., or cash (current capital), but they are all of the "same" quality insofar as they are valued in monetary terms and are invested in the market to create monetary value. And in this sense, they are the same.

In this sense, labor, capital, and land are traded in competitive markets, and their supply and demand are balanced according to the price mechanism, according to Smith. The price of "labor" is the wage, the price of "capital" is the interest, and the price of "land" is the land rent.

What is immediately noticeable here is that, at least for "capital" and "land," the issue here is actually not the exchange of the whole thing, the buying and selling, but the leasing. This means that "labor" is also not a whole commodity to be bought and sold, but rather an object to be rented and borrowed, but Smith does not go too deeply into this area. However, Smith did not delve too deeply into this area, or rather, the entire history of economics has largely overlooked this issue. But let's postpone this issue for now.

### Inequality under Capitalism

The image of the modern market economy, or "capitalism" in Marxist parlance, which Smith provided the prototype for and which was later carried on by David Ricardo, John Stuart Mill, and even Karl Marx in the 19th century, is largely complete here. It is not just a market economy. It is not just commodities that people create and then use up. Assets that are used to create such commodities but are not necessarily consumed themselves, but remain durable - capital and land - are also commodities that are traded through the market. (Whether labor is such a commodity or not, I will reserve that for now.) And the society in which all people live in this market is structured according to the different types of assets each person owns: landowners, who own the land; capitalists, who own the capital; and workers, who own the labor (or do they own nothing?). In other words, society is divided into three major classes: landowners who own land, capitalists who own capital, and workers who own labor (or nothing?).

Let's assume that the total amount of "land" is given by nature and cannot be produced by humans, so it is unchanging. Then, the only things that can change in quantity through trade - that can be produced or consumed by humans - are capital and labor. It is easy to see that "capital" is the obvious one here, but it may be difficult to understand that "labor" is also the obvious one. However, in *The Wealth of Nations,* Smith describes how the supply of labor changes in response to changes in the price of wages, not only through "inventory adjustment" of existing labor, i.e., increases and decreases in working hours and workload, but also through "production adjustment," i.e., increases and decreases in the working population itself.

> The liberal reward of labour, by enabling them to provide better for their children, and consequently to bring up a greater number, naturally tends to widen and extend those limits.(..) If this demand is continually increasing, the reward of labour must necessarily encourage in such a manner the marriage and multiplication of labourers, as may enable them to supply that continually increasing demand by a continually increasing population. If the reward should at any time be less than what was requisite for this purpose, the deficiency of hands would soon raise it; and if it should at any time be more, their excessive multiplication would soon lower it to this necessary rate. The market would be so much under–stocked with labour in the one case, and so much over–stocked in the other, as would soon force back its price to that proper rate which the circumstances of the society required. It is in this manner that the demand for men, like that for any other commodity, necessarily regulates the production of men; quickens it when it goes on too slowly, and stops it when it advances too fast. (*The Wealth of Nations*, Book 1, Chapter 8, Paragraph 40)

In layman's terms, when wages rise, marriages speed up and increase, more children are born, the population grows, and the labor supply increases, while when wages stagnate or fall, procreation is curtailed and the labor supply declines. This may sound too far-fetched, but later historical research has shown that this was the reality in early modern Britain. It seems that there was a relationship between wage trends and marriage trends.
However, some people may think that this is too slow and that the time lag is too large to be called a "mechanism for adjusting labor supply and demand (and thus the population itself). That's true. But that is not what I want to emphasize here. What Smith discovered here, and later supported by quantitative historical research, is the mechanism by which rising wages and thus rising incomes of workers lead to higher fertility and population growth rather than to higher living standards. This mechanism may remind you of the name of Thomas Malthus in his "Population Theory", but it is a recognition that has been consistently inherited throughout the so-called "classical political economy", from Smith to Malthus to Ricardian.

Both Smith and Ricardo seem to have had the cool-headed understanding that while it is good for workers to have higher wages and a higher standard of living, there is a limit to this, and in reality wages will only rise and fall around their "natural level" at best. Like Malthus, Ricardo took Malthus's theory of population very seriously and criticized the subsidies for the poor under the Poor Law of the time as distorting the population control mechanism of the labor market and creating overpopulation, and thus low wages and poverty. Ricardo was quite prescient in considering the improvement of workers' living standards as desirable from the perspective of contributing to population control, but he did not seem to be able to reach the idea of human capital (as it was later called) from there.

What was inherited from Smith to Malthus, Ricardo, and their critic Marx was the recognition that there is a limit to the increase in wages and the income of workers, and that it cannot go beyond the level of maintaining a standard of living - somehow making a living and having offspring. From the other side of the coin, this also means that the capitalist is the one who is responsible for investment and capital accumulation. In other words, They thought that the possibility of workers saving from their wage income to create capital and become capitalists could be ignored as a mass phenomenon, if not as an individual example. The idea of "human capital," as it was later called, was also still very young. Unlike "capital," that is, things, capital goods that are tangible assets, labor is consumed after use, and wages may compensate for that consumption, but they do not add anything more - they do not enhance the worker's ability.

Capital, on the other hand, may be depleted a little in the short term, but it is never completely exhausted. This is why capitalists are able to save and invest their profits instead of consuming them for their own living. The same is true for landowners in that they can afford to save from their income, but in the classical economic sense, "land" is "nature" that cannot be produced or consumed by humans, and thus cannot be the object of investment. Investment in land improvement is not the improvement of "land" itself, but the addition of "capital" to "land", in which case the landowner as the subject of investment is considered to be acting as a capitalist.

Inequality was formulated as that between the haves and have-nots under the system of property rights in the Rousseauian scheme described above, but in classical economics after Smith, it is further formulated as whether or not one has capital as accumulated wealth under capitalism, and whether or not one is a subject of capital accumulation. Of course, we have to carefully consider whether this is a different way of looking at the same phenomenon of inequality, or whether they are different kinds of inequality. However, since then, in the 19th and 20th centuries, the phenomenon of inequality in capitalist economies has basically been understood and discussed around this post-Smith framework.

However, there is still a critical component missing: sustained productivity improvement and technological innovation.

In the next chapter, we will look at Karl Marx, who inherited the economic theories of Smith and Ricardo but tried to apply them to a critique of the capitalist market economy and a society dominated by the capitalist class. The importance of Marx lies in the fact that he was the first serious theorist of this "technological innovation".

Of course, Marx was a revolutionary and a radical critic of capitalism and market economy. Normally, when inequality and inequality are discussed, Marx has been privileged as the original thinker to be returned to rather than Rousseau. So, in the context of this book, what kind of evaluation can be given to him?

## Chapter 2: Marx and the commodity labor power

### Marx's Pioneering Nature - The Discovery of "Labor Power

If, as we have seen, the framework for the theory of inequality under capitalism was outlined by Smith, then the question naturally arises: What is the difference between Smith - and more specifically, between classical economists and Marx?　The question naturally arises. In fact, some bad-mouthing economists say, "Marx is just Ricardian with a few extras." Of course, both Smith and Ricardo were great, and it is debatable whether Marx is greater than them in terms of the history of economics.

If we dare to say that Marx was "great," the first point that comes to mind is that he was working on a comprehensive system of social science and philosophy of history that was not limited to economics, and in this respect, Smith, or G. W. F. Hegel, who is closely related to Marx but cannot be called an economist, was also great. In this respect, Smith had a grand systematic concept. In general, being comprehensive and systematic is not necessarily evidence of greatness, since this was before the establishment of specialized social sciences.

If we take Marx's "greatness" as a problem in this context, apart from the issue of "technological innovation" that I mentioned at the end of the previous chapter, it is the perspective of "alienation" that he introduced after learning from Hegel's philosophy. The word "alienation" itself is mainly found in unpublished manuscripts and has almost disappeared from the works published before his death, but the word "labor power" in "Capital" clearly projects this idea.

The idea of "labor power" and the concept of "commodity labor power commodity" were introduced precisely to explain the inequality between capitalists and workers. Since transactions in a free market are usually based on the voluntary agreement of both parties, the things exchanged there should be equivalent insofar as they are equal in value. The same should be true when a capitalist hires a worker. Nevertheless, one side accumulates more and more profit, while the other side remains poor. Why is this?　Marx sought the secret in the peculiar nature of the "labor-power commodity.

The point is that labor power is assumed to be the ability to work, the source from which labor comes, and is distinguished from the activity of labor itself. I have already pointed out that the economic concept of "labor" after Smith is something more abstract and general, distinct from this and that concrete work or task. This distinction was continued in Marx as the distinction between "concrete useful labor" and "abstract human labor," but Marx also made a distinction between "labor" and "the capacity to labor" or "the source of labor. He argued that it is actually not "labor" that is bought and sold, but rather "labor power" that is the object of trade. How can this be so?

Viewing "exploitation" in terms of "labor value

First, there is the issue of the quantitative level. According to Marx, what is bought and sold in employment transactions is not labor but a labor power as commodity, and wages, despite their appearance (often paid for "labor" rather than "labor power," such as time or performance), are not compensation for labor but for labor power. So how is that compensation determined?

According to Marx, the price of a commodity is determined in the short run by supply and demand in the market, but in the long run by the cost of producing it. (This idea was inherited and developed by Smith and Ricardo.) The breakdown of the cost would be the cost of the materials used in production, the rent of the capital equipment and land used in production, the wages paid to the workers, and so on. If we ignore land for the time being, we can consider the breakdown of the cost of materials and capital equipment, which are also produced by humans, and the cost of the materials, capital equipment, and labor invested in them. If we repeat this process endlessly, we can theoretically consider the total amount of labor (or more precisely, labor power) directly and indirectly invested in the production of a certain commodity. Or we can think of something called "cost converted on a labor basis".

This "cost converted on a labor basis" can be called "labor value". This is a purely theoretical abstraction that does not appear in the real economy, and it is usually not the same as the price of a commodity when it is actually traded in the market, but there is a proper regular relationship.

The reason why we have to assume such an incomprehensible abstract concept as "labor value" is to capture the occurrence of inequality through equal trade, or "exploitation" in Marx's sense. In short, for any commodity produced by human beings, apart from its price, we can calculate its total labor input, its "embodied labor value". Now, what is the "embodied labor value" of a labor power as commodity? Marx, following the tradition since Smith, considers it to be the cost of subsistence.

If a father's labor is enough to support his wife and children, it would be the cost of living for several people. (It could be lower if the wages of the wife and children are taken into account, though.) If the wages of ordinary workers were to fall far below the level sufficient to support their livelihood, society would not be able to survive. Thus, the "natural level" of wages tends to correspond roughly to the cost of living for a standard family. This is what Marx meant when he said, "Wages are the price of labor power commodities." Furthermore, if we think about the reality of the cost of living, we can say that it is the money spent on food, clothing, shelter, education, and entertainment, and in other words, it is a group of commodities such as food, clothing, and entertainment services that are purchased. In other words, Marx's idea of the "embodied labor value of labor-power commodities" is the "embodied labor value of the total commodities purchased and consumed by workers (families) in the course of their ordinary lives."

What Marx is trying to say is that the "embodied labor value of labor-power commodities" and the "total labor value produced by labor-power commodities" are completely different quantities. Under normal circumstances, the latter should exceed the former, otherwise capitalists would not be able to make a profit by hiring workers, and society would not be able to survive, macroscopically speaking.

Very simply, if we measure the amount of labor in terms of time, it looks like this. Let's say that the value of labor produced by one worker is 8 hours worth per day. Then, what is the labor value of that worker's labor-power commodity, i.e., the total labor value of the commodity that is sufficient to sustain his livelihood? According to Marx, it should be no more than eight hours, no less than five hours, no less than six hours, no less than eight hours. Otherwise, the capitalist will never be able to make a profit. In other words, the difference between the two, the difference between the "invested labor value of the labor commodity" and the "total labor value produced by the labor commodity," is what Marx calls "surplus value," and is the source of the capitalist's profit.

Marx himself was not good at mathematics, and the state of development of mathematics at this time was not at a stage that would allow for the mathematical modeling of the economy that Marx envisioned. In the second half of the 20th century, after about 100 years, a mathematical model of Marx's economic system was developed.

Exploitation of Labor and Profit

What we came to understand was that, to a large extent, Marx was making sense, but he was missing the point. First of all, considering the standard case, it is certain that a positive surplus value is a necessary and sufficient condition for a positive profit - the former is true if and only if the latter is true - that is, if there is no exploitation of labor, there is no profit, and if there is profit, there is also exploitation of labor. In other words, if there is no exploitation of labor, there is no profit. (This is called the "Fundamental Theorem of Exploitation," "Marx's Fundamental Theorem," etc.) However, what became clearer was that this "exploitation" was not limited to labor in any way.

Marx singled out labor because most of what we give economic value to is created with the involvement of human labor; that is, unlike other things, labor is put into almost every commodity. That is why Marx thought that labor, unlike all other things, is a special thing that can give economic value to things, a source of value, and since it does, it is also a source of surplus value. But this was a kind of illusion. If we try to formulate it theoretically, especially if we consider not only direct but also indirect inputs, it is common to find that in the production of most commodities, a large number of commodities - including not only other commodities but also the commodities themselves - are input.

In other words, just as we can consider the "embodied labor value" of a commodity that is not directly produced by labor (but in which labor is invested in the raw materials and capital used in its production), we can also consider the "embodied petroleum value" of a commodity in which petroleum is not directly used in the process, either as fuel, raw material, catalyst, etc. (This is also true for oil itself, of course.)

In theory, we can calculate the "value of goods embodied" for almost any commodity - for example, the " embodied value of wheat ", "embodied value of rice", "embodied value of gasoline" for wheat, ...... for rice, etc. - for almost any commodity. And in order for ordinary capitalist business to be profitable, for profits to be positive, for the economy as a whole to be viable - i.e., for an economy at least as large as today's to reproduce and survive tomorrow and beyond - there must be not only the exploitation of surplus value, measured on a labor basis, but also the exploitation of all other commodities. It has been argued that not only is the exploitation of surplus value measured on a labor basis a necessary and sufficient condition for the reproduction and survival of an economy of at least the same size and quality as the economy of tomorrow, but also the exploitation of surplus value measured on a commodity basis for all other commodities.

In other words, the "embodied X value" of any commodity X in the entire economy, including labor, must be less than one - that is, the amount of X directly and indirectly required to produce one unit of X must be less than one unit - the "embodied Prius value" of one automobile, say a Prius, must also be less than one. This was a necessary and sufficient condition for capitalists to make a profit, and for the economy as a whole to survive and grow. (This is called the "generalized commodity exploitation theorem.")

If this is the case, the Marxian doctrine that "labor is the source of economic value" will not only degenerate into a mere illusion and delusion, but it will also lead to the degeneration of criticism of capitalism based on the theory of exploitation that "the workers who are responsible for the labor are unfairly exploited by the capitalists who do not work, and the apparent sale and purchase of labor power as an equivalent exchange masks this." The critique of capitalism based on the theory of exploitation also loses much of its persuasive power. Nevertheless, "exploitation of labor" is still valid if it means that "the capitalist gets more labor value in his income than he has labored for (a commodity in which he has invested), while the worker gets less labor value in his income than he has labored for (a commodity in which he has invested)." If we assume that the meaning of labor is not "that it is the source of the economic value of all commodities" but that it is important to human welfare and social justice in some other way (for example, by establishing a moral principle such as "those who do not work should not eat"), then the theory of exploitation has not lost its meaning. However, the original Marxist theory of labor value no longer has any basis.

Why did Marx insist on the existence of "labor power"?
Now that the "theory of value" has fallen into disuse, what is more important is the second, qualitative aspect. In other words, labor and labor power are completely different, and it is really labor power that is being bought and sold, but people do not realize this. Let's dig a little deeper into this area.

In the first place, the problem had already arisen when we looked at Smith in the previous chapter. In other words, the transaction price of capital and land was not a whole sale price, but rather a clear image of interest and ground rent as rental fees in the lease, while the situation was not clear for labor and wages. In the first place, whether labor was to be bought and sold as a whole or leased like land and capital was not properly considered.

In the world of law (private law, property law) and jurisprudence, buying and selling, renting and hiring are distinct types of transactions, but in economics, the distinction is actually vague. To put it very roughly, in economics after Smith, the basic form of transaction is understood as an exchange based on the voluntary agreement of both parties, and in the large-scale market economy, which is the central subject of economics, buying and selling is the main form of exchange using the medium of money. (This analogy breaks down, however, in the case of what the Civil Code calls lending and borrowing, which typically involves lending and borrowing money at interest, but we'll leave that for now.) Labor transactions such as hiring and contracting were also discussed in the image of such anomalous "buying and selling" (buying and selling of goods in the form of services by workers), but the analogy was not really complete.

In other words, if the reward for the use of land and capital is land rent and interest, the reward for labor is wages, but what corresponds to the price of the land itself (land price) or the price of the entire capital (total stock price or corporate value in the case of a stock company)? In the first place, what are the objects to which such whole prices are applied? Neither Smith nor most other economists since then have explored this question.

In a sense, Marx seems to have been trying to make a thorough analogy. But the trouble is that Marx says that labor power is a "commodity" rather than "capital. In other words, Marx believed that labor power, unlike land or capital, is consumed when it is sold. This is not clear from just reading *Capital*, but in *Grundrisse*, a draft of *Capital* that was not published before his death, he says it more clearly.

If "labor power" also wears out and disappears just like "labor," is it worth the trouble to distinguish it from "labor"? Of course, Marx had a theory of exploitation, which we saw earlier, so he must have thought that there was a clear "Yes!" but now that we no longer appreciate it, we have to be more skeptical. Nevertheless, I still think that Marx's insistence on the existence of "labor power" is meaningful. What does this mean?

This is expressed in a very simple way by Yoshihiko Uchida, a Japanese historian of economic thoughts.

> Workers have the right to dispose of their labor power, but they do not have the right to dispose of their labor at all. If you think this is a lie, try disposing of your labor at work at your leisure. It will be you who will be disposed of. It is capital that disposes of labor freely. The more modern the factory, the more clearly this is the case. (Yoshihiko Uchida, *Shihonron no Sekai* (*The World of **Das Kapital***), p. 78)

The worker in employment is, after all, on the continuum from the slave and the indentured servant, a debt slave with a fixed term, whose person is controlled by the employer under certain restrictions. Of course, an indentured servant, who voluntarily enters into a free contract and is paid a wage, is a legitimate worker in the eyes of economics, but he is subject to the same comprehensive control of his employer as a slave, albeit within the limits of the contract. The freedom of the hired worker is primarily exercised when entering and exiting the relationship itself, and is considerably constrained while in the relationship.

To put it simply, the relationship between employer and employee is a status relationship as an asymmetrical relationship of dominance and subordination. However, the status relationship is different from what we assume about slavery and serfdom in the so-called status-based societies of antiquity and the Middle Ages, where people were originally divided into free people and slaves, and slaves were owned as property of free people and subject to sale and lease. Slaves were owned as property of free people, to be bought, sold, or borrowed. Even in such societies, people moved (or were forced to move) across the barriers of status, falling from free people to slaves when they became prisoners of war or were unable to pay their debts, and then becoming free people when they were recruited by family or friends, paid their debts on their own, or were favored by their masters. Modern employment is a mechanism that allows people to move back and forth between freedom and obedience (subordination to the rule of others) in a different way.

In modern civil society, including Marx's contemporaries, employment relations are based on the assumption that "citizens" of the same status enter into status relations within certain limits through voluntary contracts. In this way, through the form of a voluntary contract, instead of asking an independent artisan to do a specific job (which can be specified in advance in the contract), they can do a variety of jobs flexibly and according to the will of their employer (as domestic slaves or servants). In other words, they can be asked to do various jobs flexibly and according to the will of the employer (i.e., without being specifically specified in the contract in advance).

The reason why Marx had to insist that "what is bought and sold in employment is not labor but the commodity of labor power" was to answer, in part, the question, "If the capitalist and the worker trade their labor in an equivalent exchange, where does the capitalist's profit come from?" But that is not the end of the story. It was also to answer the question, "If labor is traded on the basis of the free will of the worker, why is it controlled by the will of the employer and not the worker?" As we have already seen, the former question has already been dismantled, but the latter is quite persistent. However, we have to wait until the latter half of the 20th century to see the emergence of the problem of "alienation" among "affluent workers" who have long since been freed from the fear of starvation brought about by the rapid economic growth of developed countries.

Will technological innovation continue to create unemployment?

Once again we come back to the quantitative question, so to speak. We have already mentioned that Marx made a distinction between labor and labor power, but regarded labor power as a "commodity" (or "flow" in the modern economic sense) and not as "capital" (or "stock"). In other words, Marx, like Smith, Malthus, and Ricardo, believed that there was a definite limit to the rise of wages. However, Marx looked for the reason in other places.

According to Smith, Malthus, and Ricardo, the main reason why wages do not rise much above the edge of the workers' survival level is the working class behavior that higher wages and incomes encourage people to marry earlier and have more children, which eventually increases the labor supply ....... The main reason for this is the behavior of the working class. Marx, however, rejected such ideas. But Marx rejected such an idea, of course, for moral and political reasons rather than as an objective factual judgment. However, Marx did not just emotionally reject the classical theory of the labor market, but he also prepared an alternative theory. This is the so-called "industrial reserve army theory" or "relative overpopulation theory. The industrial reserve army is, in essence, the unemployed. The fact that it is a "relative overpopulation" means that the productive capacity of society as a whole is sufficient to support people and keep them alive, i.e., there is no "absolute excess" of population, and therefore, the unemployed are merely in excess of the demand for labor, i.e., there is a "relative excess. In other words, it's a criticism of Malthusianism. Marx's theory is that in a capitalist economy, there is always a certain number of unemployed people, and their existence is a dead weight and downward pressure on wages. So, why do the unemployed always exist in capitalism?  It is ultimately because of technological innovation, as he later put it.

In a capitalist economy, fierce competition puts constant pressure on capitalists to rationalize their operations. And the capitalists try to cut costs as much as possible. So, of course, there is always the option of simply strengthening labor - extending working hours, cutting wages, or increasing labor density (Marx called this the "production of absolute surplus value") - but this is met with severe opposition from the workers. That is why technological innovation is used to increase productivity. In post-industrial capitalism, investment and capital accumulation is not just for the purpose of expanding the scale of business, but also for the purpose of increasing productivity through mechanization and other means. (This is the "production of relative surplus value.")

However, one of the possible consequences of increasing productivity through technological innovation is the choice to reduce manpower, since it is now possible to produce the same thing with less manpower. This is quite possible if sales remain unchanged. Needless to say, the mildest way to reduce manpower is to shorten working hours, and if it becomes more difficult, of course, people will be laid off. Thus, in a capitalist economy, fierce competition forces companies to constantly strive for technological innovation, which in turn constantly creates unemployment - this is roughly what Marx argued. The combination of this and the business cycle, or boom-bust cycles, means that in a capitalist economy, there is always a significant number of unemployed people in search of work, and their wages are constantly reduced to the point of subsistence.

Marx would like to argue that the number of unemployed is never zero not only during recessions but also during booms due to constant rationalization and labor saving, but he has not been able to prove this. In other words, I have not been able to convincingly argue that the increase in labor demand during a boom must be offset, or even surpassed, by the decrease in labor demand due to rationalization.

Capital accumulation is the cause of inequality

Now, let me briefly summarize the flow from Smith to Marx. The basic form of the argument has already been established by Smith, but in essence, in a capitalist economy, a society where not only commodities in the ordinary sense but also labor, capital and land are controlled by the market mechanism, the central mechanism that generates economic inequality is ultimately capital accumulation and economic growth. In the so-called classical economics after Smith, as well as in Marx, who critically inherited the classical economics, the main point was that it is the capitalist who saves and invests. In classical economics, land, by definition, does not increase in value when it is invested, and as long as it is invested, the landowner is acting as a capitalist in the economic sense. On the other hand, workers, for reasons that differ between the classics and Marx, cannot afford to save or invest because their wages do not exceed their subsistence level. Thus, the worker has no choice but to earn money by himself, using only the labor he has on hand, and the gap between the capitalist (and the landowner who is also a capitalist), who can invest and increase the source of his earnings, and the capitalist, who can invest and increase the source of his earnings, will open up more and more in the long run. In other words, in the perspective opened up by Marx from classical economics, economic disparity and inequality are dynamic phenomena, and cannot be understood separately from the problem of long-term economic growth.

I would like to add two more points to this discussion.

First, the Smith-Malthusian mechanism of population adjustment. The classical economists could not clearly argue whether labor is bought and sold in its entirety or only its services are leased. However, with the benefit of hindsight, from the perspective of modern economics, it can be interpreted as follows: In the world they envisioned, the subject of labor supply is not the individual worker, but the family. If we consider the family as a unit, we can understand the birth and raising of a new child as a kind of investment.

If we think of economics as a tool for policy evaluation and evaluate the welfare level of a society from a moral perspective, then we have to say that such an "investment" will only raise the welfare level of the "family" unit and not the individual. Therefore, we have to be cautious about such "investment" in procreation. However, from the perspective of objective empirical analysis, it is a possible viewpoint. In fact, modern economics tends to model ordinary people in terms of "households" rather than individual "workers" or "consumers."

Marx, the thinker of technological innovation

One more point about Marx. The greatness of Marx, as far as I can see, is that he tried to distinguish between labor and labor power, but he also focused on the importance of what he later called "technological innovation.

Of course, it could be argued that Smith and Ricardo were both discussing the innovations produced by the market economy as companions of the English Industrial Revolution. However, Smith's *Wealth of Nations* was still limited to handicrafts, and Ricardo, ahead of Marx, imagined that machines could create unemployment, and that economic growth would stop in the long run due to the constraints of land, perhaps because he could not imagine how to sustain a full-scale increase in productivity. Growth would not be driven by rising productivity, but solely by the further accumulation of capitalists, which would hit a wall as land became scarcer, land rents rose, and the fruits of growth were siphoned off to landlords. To such an outlook, which was shared by Ricardo and Mill, Marx gave a clear "no". And so far,

Marx's prediction has prevailed. The increasing efficiency of resource use through technological innovation continues to push the finite nature barrier further away.

At the same time, however, Marx rejected the possibility that a side effect of technological innovation would be to raise the standard of living of workers. Rather, in his theory of the industrial reserve army, he emphasized the tendency of unemployment caused by technological innovation to keep wages at a low level. What this means is that he underestimated the possibility of resistance to wage cuts by, for example, the bargaining power of labor unions.

On the other hand, in his analysis of working hours, Marx mentions in detail the resistance of workers and the regulation of working hours by the public policy of the state (factory law in the 19th century). Moreover, as can be seen in the idea of "the production of relative surplus value," this resistance by workers became an incentive for capitalists to rationalize and innovate. In other words, in Marx's view, workers' resistance is the key to technological innovation. In other words, Marx presents a somewhat inconsistent and ambivalent argument about workers' power. Marx inherited the Rousseauian motif, so to speak, that "in a society of private ownership and a market economy, isn't there something tremendously unjust going on? " However, he tried to critique inequality in a more sophisticated way based on Smith's recognition that "private ownership and a market economy have achieved an unprecedented increase in productivity." Therefore, he argues that workers continue to be structurally disadvantaged in capitalist society and must be powerless. This is reflected in the theory of the decline of the wage level in the theory of the industrial reserve army.

But on the other hand, since Marx's goal is to overthrow capitalism by siding with the working class, he cannot depict the workers as powerless too much. That is why, while he also pointed out the limits of wage increases in my conclusion, he could not accept the Smith-Malthusian theory of population and had to formulate his own theory of relative overpopulation, and he could not be straightforwardly negative about the achievements of labor movements and social policies in reducing working hours.

One core of distribution theory since the second half of the 20th century has been the possibility of savings-investment in the working class. In the next chapter, I will point out that the founders of so-called "neoclassical economics," which became the source of modern standard economics, paid attention to this possibility in their thought. Paradoxically (but not surprisingly, if you think about it), it was the founders of neoclassical economics, such as Alfred Marshall and William Stanley Jevons, who were more optimistic than Marx about the possibilities for the development of the working class in a capitalist society.

Chapter 3: Neoclassical Economics

The Age of Democratization and the Formation of Neoclassical Economics

Up to this point, we have been talking about "classical" economics, so to speak, but from here on, the discussion will become much more modern and complicated.

First, economics took a major turn after Marx's time, and from the 19th century to the beginning of the 20th century, the foundation of today's mainstream economics, so-called "neoclassical economics," was established. For a while after that, economics was divided into Marxist and non-Marxist factions centered on the mainstream school, and entered a period of lack of relation. In this book, I will focus on the neoclassical school, but also pay some attention to the Marxist school.

Second, the period after Marx's death was also a turning point, and since then, not only economics, but science and thought as a whole, and the real world as a whole, have entered a different phase from the period in which Smith and Marx struggled as contemporaries - the "Age of Revolutions," in the style of historian Eric Hobsbawm, the civil and industrial revolutions and the chaos that followed. It can be said that we have entered a different phase.

For example, if you look at the heroes of classical economics, such as Smith, Malthus, Ricardo, Mill, and Marx, only Smith can be called a "university professor" among them. Smith was so disgusted with the fact that the university was not a place of learning that he became a tutor for a great aristocrat. Ricardo was a stockbroker and a delegate, Mill had worked for the East India Company and had also been a delegate, and Malthus was a clergyman. In the 18th and 19th centuries, the emerging field of economics was not yet based in universities, and British universities at this time were not exactly centers of academic research. The situation was similar in France, where the heroes of the Enlightenment in the 18th century were based in the salons of the aristocracy, and the center of academic research was the Academy, an organization separate from the university. Germany was a relative exception, but Marx failed to become a university man in Germany and became a journalist and then a revolutionary.

However, in the last half of the 19th century, the situation changed dramatically. As a result of university reforms based on the so-called "Humboldtian model of higher education" in Germany and its spread to other countries, universities were reorganized from educational institutions for the children of the upper class to centers of not only education but also academic research. In these new universities, economics steadily gained ground, and from the latter half of the 19th century onward, the major economists were generally academics. William Stanley Jevons and Alfred Marshall of Britain, Léon Walras of France, Karl Menger of Austria, and John Bates Clark of the United States, the early leaders of the neoclassical school, were all university professors.

This was the era of democratization in many European countries, to put it crudely. The reason why the theory of violent revolution was so strong among socialists in the 19th century, including Marx, was because the politics of European countries at that time were not democratic at all - the common people, workers and peasants, who made up the majority of the population, did not have the right to vote or participate in politics in the first place. By the end of the 19th century, however, the growing presence of trade unions and socialist parties, including those influenced by Marx, and the pressure they exerted, led to a significant broadening of voting rights. Therefore, in Germany, the Social Democrats became the leading party in the parliament during the 19th century (even though the political power of the parliament itself was low in the German Empire at that time).

During this period, in response to the demands of workers and socialists, or in order to check them, a series of social reforms and policy changes that would later lead to the welfare state took place in European countries. Specifically, the legalization of labor unions, the establishment of social security systems such as medical insurance and old-age pensions, and the establishment of a free compulsory education system. The correction and alleviation of ieequalities has become a political issue.

Many of the neoclassical pioneers mentioned above were also reformists who were deeply committed to the social reforms of the new era. Marshall and Jevons were far from the "small government" theorists of the old classics; they advocated active policy intervention in labor and social issues, and Walras was an avowed "socialist. They were not indifferent to issues of inequality and distribution.

Despite this, or perhaps because of it, the neoclassical economics they developed was, at least in its appearance, more neutral and less sensitive to distributional issues than classical economics.

If we were to look back now and try to summarize what neoclassical economics has been like since the end of the 19th century, what should we focus on?   Let's take a detour and review the characteristics of classical economics once again (but from a slightly different perspective), and then contrast them with the characteristics of neoclassical economics.

Differences between classical and neoclassical economics
In classical economics, the three major factors of production, labor, capital, and land, had already been formulated, but the focus of analysis was more on labor and capital. However, the focus of analysis was more on labor and capital, because, as we saw earlier, capital accumulation was considered to be the driving force behind economic growth. It is capital and labor that accumulate and grow in accordance with the dynamic market mechanism, while land is treated as a passive entity in comparison. On the other hand, however, it is also land that imposes the final limit on the accumulation and growth of capital and labor, and although the land issue is not the focus of discussion in classical economics, its presence is not insignificant as it sets the limits of the entire discussion.

In principle, land was seen as a "nature" that could not be produced no matter how much capital or labor was invested in it. As such, even if land is traded in the market in the same way as capital and labor, the way it is traded, especially the mechanism of price formation, is clearly different from that of capital and labor.

For capital and labor, as well as for ordinary commodities, not only are immediate inventories traded in the market, but also short-term trends in trade affect their production in the long run. Goods that remain unsold due to lack of demand will not be produced in the next fiscal year, and the production line itself will eventually shrink, while production capacity will be increased if there is excess demand. Short-term capital and labor shortages lead to long-term investment, labor supply, and even population growth through higher interest rates and wages. In the case of land, however, it is not easy to increase the supply of land just because there is excess demand. In the long run, there is only so much land on the planet, and the ceiling will eventually come. In other words, classical economics is strongly based on the idea that the land transaction market cannot affect the total amount of land or the production of land itself, but can only exchange existing land.

On this basis, how was the price of land (whether land rent as a rental price or the total land value) supposed to be determined?   Land was typically viewed as a means of agricultural production. Moreover, investment in land for the purpose of land improvement, etc., did not change the nature or value of the land itself, but was considered as capital that was integrated with the land. If this is the case, then the pure value of the land itself, without the value of the capital invested in it, is the basis of the price of land. The value of the land before human intervention was considered to be derived from the natural properties of the land itself. Even in the untouched stage before irrigation and soil improvement, there are various differences among lands, some are suitable for cultivation, some are not, and some are expected to yield a good harvest. Such differences are the differences in the value of land. However, such differences in value are not created by human effort - labor or investment, so when they are brought into the human world and given a concrete economic value - a price, it is determined only by the balance between supply and demand. To summarize, in classical economics, the prices of capital, labor, and most human products are determined in the short run by market transactions and the balance between supply and demand, but in the long run by the cost of production - by the technological structure of production, which determines what resources, or more specifically, what amount of land, capital, and labor, are required to produce them. In the long run, it is determined by the cost of production - what resources are needed to produce it, in other words, how much land, capital, and labor are needed. (This is the basic idea of the so-called "labor theory of value.")

In simple terms, classical economics says that the price of land is determined only by supply and demand, while the prices of capital, labor, and most commodities are of course determined by supply and demand in the short run, but in the long run they are determined by the production technology of the economy as a whole and the structure of people's consumer lives (if we consider labor). So, the movement of capital and labor, or in other words, the human productive forces, will determine the movement of the economy, the accumulation of capital and economic growth in the immediate, human-scale time frame, although the factor of the finite nature of land on the earth will constrain the overall framework in human history. This is how it is.

If you think of it this way, then neoclassical economics is the opposite of classical ideas. If classical economics took a production-centered view of the economy, with capital and labor playing the leading roles, then neoclassical economics takes an exchange-centered view of the economy. In contrast to classical economics, which focused on capital and labor and tried to understand price formation in relation to production technology, in the neoclassical, the analysis of the market and the price mechanism is not based on the trade of products that require large amounts of capital and labor, but on the trade of land that cannot be produced by humans, as a paradigm, a frame of reference. As for the relationship between trade and production, if the classical school tended to regard production as primary and trade as secondary, the neoclassical school reverses this tendency. The neoclassical school is thorough in the sense that it believes that the reason why things, whether natural or artificial, are given economic value and prices is because they are wanted and traded by people, that is, because there is supply and demand.

And in another sense, neoclassical theories are based on land rather than capital and labor. This is the idea of "diminishing returns" and the method of "marginal analysis" to capture it. (The establishment of neoclassical economics is also called the "Marginal Revolution" in the history of economic theory.)

What is "diminishing returns"?

The term "diminishing returns" refers to the following in the context of agriculture. When you cultivate a certain area of land under a certain environment by investing labor and capital, the harvest will increase as you invest more and more, but the increase will gradually slow down.

This is called diminishing returns. If we keep the same ratio of input of labor and capital (liquid capital such as seed fodder and fertilizer, and fixed capital such as livestock, agricultural machinery, and land improvement), and add new land with similar conditions to be cultivated, the situation will not be like this, and the harvest will increase in proportion to the amount of input.

It was Ricardo who summarized the general framework of this argument, but from Ricardo to Marx, this "diminishing returns" was considered to be a fundamental feature of agriculture. Then the focus of economic thought shifted to capital and labor, and especially to industry (which does not have to worry so much about land as an input resource) in Marx. Apparently, Ricardo and Marx thought that this diminishing returns did not apply to industry. Under a certain production technology, the combination of inputs, whether labor or capital (more specifically, various materials and production equipment), is strictly fixed, so that, for example, increasing only the input of labor when capital is constant does not lead to anything (and vice versa for the various materials and equipment that make up capital). The same goes for the various materials and equipment that make up capital.

In other words, there is no substitutability between capital and labor. In other words, it is not possible to compensate for a shortage of labor by increasing capital input (or vice versa). However, it was assumed that there was no such fixity for land. If we consider only land and the other factors of production, capital and labor, there is a certain degree of substitutability between the technologically determined combinations of these factors (hereinafter referred to as the "capital-labor set," for lack of a better term), and even if one of them decreases, the same output can be maintained by increasing the other. However, the efficiency of substitution is not infinite. However, the efficiency of this substitution is not infinite, and it is expected to decline rapidly - even if at first one unit of capital-labor set can compensate for a shortage of one unit of land, a shortage of two units of land will require two or more units of capital-labor set (or vice versa), and so on.

The existence of such "diminishing returns on land" makes a big difference in agriculture compared to industry (as assumed by the classics). In other words, in the case of industry, where production increases in proportion to input, if there is no market constraint, demand is unlimited, and the more you produce, the more you sell (this is a situation of "perfect competition" in the later term), the goal of the manager is to increase production and sales at will. This is a straightforward way to increase profits. However, in agriculture, where there is diminishing returns, this logic does not hold true. In other words, the yield per input is decreasing rapidly. In other words, the yield per input is decreasing rapidly. This means that the yield, income, and profit per cost of production is falling rapidly. Of course, as harvest and production increase, total income will increase. However, if the cost of production increases more than the income, the profit (income minus cost) will start to decrease. In other words, in agriculture, unlike in industry, there is an optimal amount of production that maximizes profit.

The neoclassical school, however, generalizes this theory and believes that the theory of "diminishing returns" applies to industry as well. In other words, if you increase the number of people with a certain level of equipment, production will increase to some extent, but labor productivity will continue to decline (or if you keep the number of people at the current level and increase the input of equipment and inventory, production will increase to some extent, but capital productivity will decline). In other words, in most industries, including agriculture, there is a constant harvest with respect to scale and a diminishing harvest with respect to individual inputs. The "diminishing returns" that I mentioned earlier is more accurately defined as "diminishing returns on individual inputs (i.e., labor alone, capital alone, land alone, etc.). By "constant harvest on a scale," I mean that if you increase the amount of land, capital, labor, or any other input, the amount of production will increase in proportion to that increase. With this general assumption of "diminishing returns," the idea of "optimization" (as it is called later) can be applied to all industries, not just agriculture. In other words, "in order to maximize profits, we must aim for just the right amount of production = just the right amount of sales, not too much, not too little."

I won't go into details, but according to neoclassical economics, this logic actually applies not only to the producer, but also to the consumer. In neoclassical economics, the term "utility," inherited from utilitarian philosophy (see Chapter 10 for details), is used as a measure of the well-being of consumers, or rather, of people in the flesh, or to go a bit further, as a criterion to guide behavior (in modern terms, a "utility function"). In other words, people are regarded as maximizing their own utility. And that maximization is, of course, optimization. In the first place, no matter how much you like a food, you can't eat it indefinitely, nor do you want to. This is how the optimal amount of food consumption is naturally determined.

The neoclassical view of the possibility that everyone can be a capitalist

This has been a rather long discussion, but let me summarize it in retrospect. The characteristics of neoclassical economics in comparison to classical economics are as follows

1. The theory is based on exchange and trade, not production. The idea is not that "what is produced is traded," but that "trade causes production.
2. The behavior of both companies and people can be understood as "optimization".

That's about it.

To put it very crudely, neoclassical economic theory is more general and abstract than classical economics. On the subject of "inequality and growth," first of all, the difference between agriculture and industry is diminished, and both tend to be viewed more as subdivisions of the general "industry". Marx's rejection of the idea of a steady state - a limit to growth based on the limit of the total amount of land on the earth - is already strong, but the neoclassicals seem to avoid this idea even more strongly, though in a less obvious way. (Although, in fact, Jevons' seminal work, *The Coal Question*, is a very pioneering development of the "limits to growth" argument, focusing on coal as an exhaustible resource, quite apart from classical land constraints.)

Secondly, the generalization of the behavioral principle of "optimization" tends to de-emphasize class differences among people's behavioral patterns. In classical economics, the point of view that people belonging to different classes have different behavioral patterns was clearly introduced, perhaps ad hoc, such as capitalists accumulate but workers do not (or cannot). In the neoclassical school, however, all people are the same and such differences in patterns disappear. So why do capitalists save and workers don't?  Rather than introducing class differences in an ad hoc manner, it gives a coherent theoretical rationale of saving behavior, such as, "After all, the difference is whether they can afford to save or not".

The neoclassical school, while inheriting from Smith and Marx the argument that savings = investment is the engine of capital accumulation and economic growth, opens the way to a theory that does not assume a capitalist class as its subject. In other words, if the economy as a whole becomes sufficiently prosperous, anyone can become a capitalist. The economic and social theory with this possibility in mind was fully developed in sociology and journalistic social criticism, rather than economics, in the 20th century in "The Affluent Society" (J. K. Galbraith), but already in the early days of the neoclassical school at the end of the 19th century, a very interesting point was raised. The idea of "human capital".

Marshall's "human capital" perspective

Alfred Marshall is best known for his *Principles of Economics*, one of the first systematic neoclassical theories, but labor issues, overcoming poverty and improving the social status of workers remained central to his concerns. He asked himself, "Can the working class become gentlemen?" He tried to answer, "Yes! The key idea at that time was "personal capital".

Among the wide variety of workers who earn a living by earning wages for their labor, he focused on the difference between those who engage in high-wage skilled labor that requires relatively high levels of training and those who engage in simple, low-wage labor that does not require much experience or knowledge. He understood the difference to be the difference in human investment. In other words, they saw training to acquire skilled skills as an investment in time and effort, and in some cases, in direct monetary costs such as tuition and living expenses during the training period.

Even though the wage gap between skilled workers and simple workers is mainly due to their ability, the main cause of this ability is the presence or absence of training, and the presence or absence of training is the presence or absence of the burden of training costs. The children of skilled workers, thanks to their parents' high income and social connections, have a relatively better chance to be trained for high-paying skilled jobs, whereas the children of unskilled, low-wage workers do not have such a chance. This is how Marshall saw it.

Although he later retracted the term "personal capital," he did not abandon his position that not only vocational training in the narrow sense, but also everything from rudimentary literacy in schooling to the acquisition of more advanced science and technology could be viewed as "investments" that would improve workers' abilities and result in higher wages. In the latter half of the 20th century, this idea was fully developed in the form of considering vocational training and schooling as "human capital" (as it is commonly used today) and actually calculating its rate of return, or understanding corporate labor organization as a mechanism for human investment. But let's look at that later, and focus on the deeper level of Marshall's thinking that led him to this idea.

Marshall himself credits Smith's *Wealth of Nations* for inspiring him to view training as an investment. Indeed, Smith argues that the direct cause of the high wages of skilled craftsmen is to recoup the cost of long-term training (apprenticeship). However, Smith does not believe that the apprenticeship of skilled craftsmen in unions has the effect of increasing the skills and labor capacity of the craftsmen. In Smith's view, the most efficient way to acquire skills is to learn by doing while actually working, and apprenticeship and schooling have no such effect. According to Smith, apprenticeships in unions raise wages only because the unions' monopoly of the market raises prices and wages, but not the actual productive capacity, so it is a social waste. Marshall, on the other hand, believed that even in a perfectly competitive market where no such monopoly existed, training could raise the ability of workers to be productive and raise their wages as a share of that ability.

Another noteworthy difference, of course, is with Marx. Unlike Marx, who insisted on referring to "labor power" as a "commodity" and consciously rejected the term "labor capital," Marshall used the term "personal capital," at least for a time. In other words, Marshall believed that the wage level of skilled workers, at least, could and does clearly exceed the survival level. In other words, there is a surplus there that can be invested.

Now, from a purely formal point of view, we can say that this surplus does not necessarily need to be invested in human capital. Even if it is to be carefully saved and not fully consumed, there are many other ways to invest it besides human investment such as education and training. It can be deposited in a bank or lent to someone as a financial asset, or it can be converted into a real asset. In other words, you can buy your own land and production facilities and stand as a self-employed capitalist entrepreneur, albeit in a small way, or, similar to lending, you can invest in an appropriate capitalist business, or even buy shares in a stock company, etc. From here, we can see the possibility that anyone can become a capitalist (albeit a small one). Marx was also a theorist.

Marx also theoretically emphasized the "impoverishment" of the working class under capitalism with his "theory of relative overpopulation," but he was unsure whether this "impoverishment" was absolute or relative, and he also had a premonition of the rise of a new middle class that would later lead to the white-collar workers. Marshall, on the other hand, was more clear that the effects of general economic growth would be felt by workers, and that workers could be transformed into capitalists through investment, or become "gentlemen" as skilled workers with a high level of human capital. In other words, even under capitalism, the possibility of not only eliminating poverty but also alleviating inequality was foreseen.

In this way, neoclassical economics emerged against the backdrop of the establishment of a modern, university-centered scientific system on the one hand, and the development of democracy and social reformism on the other. The problem of inequality and inequality has become a major issue in economics. Issues of inequality and inequality are no longer hot issues in economics. Why is that?

Let's look at this issue again in the next chapter.

Chapter 4: Economic Growth in theoretical perspectives

After some twists and turns, by the end of the 20th century, neoclassical economics became the de facto standard in theory. In other words, the neoclassical theory was able to incorporate the situation assumed by the classical and Marxist theories, i.e., that the wages of workers are kept at a subsistence level, and that only capitalists can save and invest, as a possible special case, while the reverse is not true. Theoretically, everyone can be considered a capitalist, but it took longer than expected to settle the issue. Why is that?
Let's look at it a little more carefully below.

The "Typical Situation" in the Classical and Marx
First, let's look at the typical situation as envisioned by the pre-Marxian classical image, Smith and Ricardo.
Let's assume that there is no technological innovation, workers' wages are fixed at the subsistence level, landlords do not make productive investments, and only capitalists save and invest. The profit made by capitalists is invested and capital equipment is increased. However, according to the Smith-Malthusian assumption, the increase in wages causes an increase in the population, which in the long run puts downward pressure on wages, pushing them back to the survival level. If this pressure on population growth does not extend to capitalists and landowners who do not live on wages, their population will be suppressed regardless of the growth of the economy, allowing each capitalist business to expand its scale of operations and the consumption per capitalist or landowner to increase.
In the long run, however, the constraint of the absolute quantity of land comes into play. When all cultivable land becomes arable and no new land supply can be expected, the limit will become clear. New investment, and the additional labor supply that goes with it, can no longer be expected to be profitable, and a steady state is reached where investment is made only for the maintenance and reproduction of existing capital equipment. Population growth will stop there. The scale of capitalist management, the expansion of consumption among capitalists and landowners, and the improvement of living standards will also stop.
In contrast, what kind of image does Marx's *Capital* portray?  The basic structure does not change much, but the industrial sector adds the factor of technological innovation, or in Marx's words, "the production of relative surplus value. The image of this in Marx's mind is basically to increase the ratio of capital equipment through mechanization and to raise labor productivity.
The assumption that wages will be kept at a subsistence level is not very different from the classical one, but it has little to do with population change. In fact, the pace of the business cycle that Marx was experiencing in his own time was only a few years at most, so it makes sense that Marx thought it was nonsense to assume that the working class would adjust its own population to accommodate the change. Instead, Marx brought in unemployment as the dead weight that holds down wages, and the main cause of this is labor saving through technological innovation.

When capitalists invest their profits, accumulate capital equipment and expand the scale of their business, the demand for labor will naturally increase accordingly. Marx's strength lies in his detailed analysis of various possibilities, such as the extension of working hours, labor intensification, and labor saving through technological innovation, because this does not necessarily lead to immediate employment growth. However, his analysis of where these various factors end up settling down (or, to put it in neoclassical terms, where the "equilibrium" is reached) was, of course, unreasonable, but it was too weak.

However, in general, what Marx foresaw was a situation in which existing employment would always be reduced due to the labor-saving effects of technological innovation, and this would not be completely offset by the growth of new investment, the resulting increase in demand for labor, and the increase in employment, so that there would always be a certain amount of unemployment. If the economy overheats to the point that the industrial reserve army and the labor pool are exhausted, this will cause a depression, the economy will turn into a recession, and unemployment will increase at once.

The increase in labor demand due to capital accumulation and economic growth is financed not by the addition of labor supply due to population growth, but by the unemployed that capitalist management ejects due to labor saving caused by technological innovation. The problem is supply and demand. However, the problem is the mechanism for adjusting supply and demand, and in Marx's case, there is no smooth adjustment mechanism like Smith-Malthus's. Since we cannot expect wage fluctuations to adjust labor supply, the business cycle of labor shortage and wage increase during a boom, followed by a reverse recession due to a depression, takes its place. So the business cycle of labor shortage and wage increase during a boom, followed by a reversal of the depression, takes its place.

It is good to note that Marx rejected the Smith-Malthusian law of population (and the labor market theory that underpins it), but there is no population theory (distinct from the labor market theory) that can replace it. The argument implicitly assumes that there is some kind of survival mechanism outside the capitalist economy, whether it is the traditional rural community or the state's poverty alleviation administration. This is how I see it.

### The "typical situation" for the neoclassical

So, how should we think about the "typical situation" for the neoclassical school? The neoclassical school is characterized by its flexibility, and the above two cases can be discussed as special cases within that framework, so it is precisely for the neoclassical school that we have to think about the "typical situation," but it is tricky.

However, even if it is difficult to envision a specific "typical situation," you may already be aware of the difference in the awareness of the problem of growth and distribution between the classical, Marxist and the neoclassical schools. While the classical and Marxist schools focus on the qualitative and discontinuous difference between the class that possesses accumulating wealth (capital) and the class that does not (landowners and workers), the neoclassical school, in its early stages but as it matures as a school, focuses on the individual (or more precisely, the family and the corporation). In short, it focuses primarily on quantitative and continuous differences in wealth holdings among individuals (more precisely, families and businesses, which are individual agents of specific economic behavior). The presence or absence of property is not seen as a qualitative difference, since even a poor worker with no property may accumulate property as a result of luck or hard work and move on to become a capitalist or landowner. (I will ignore the issue of "human capital" for a while, or discuss it only as if it were essentially the same as physical capital.)

I have discussed this issue from time to time, so it may seem a bit late, but I would like to confirm a basic point here. When thinking about the economy, it is essential to distinguish between flow and stock. In "production," where new goods and services are created, or in "consumption," where existing goods and services are used up and disappear, newly produced goods and services or goods and services that are consumed and disappear are "flows," but goods and services that are produced but not used for a while and stored as inventory, and even goods and services that are used but not consumed are "stocks". In addition, fixed capital equipment and durable consumer goods (including housing) that do not wear out immediately but last for some time are also called "stock". In economic terms, it is not too much of a mistake to say that "income" refers to "flow" in terms of money. In economic calculation, at the level of flow, income = production = consumption + investment (savings). In economic calculation, saving and investment are the same thing: increasing "capital" as stock. The special economic term "wealth" is the counterpart of "income" and refers to capital and land as a stock.

What I have referred to in this book so far as "property" (as distinct from consumed goods) is actually this "wealth". It also refers to the factors of production, excluding labor. (Except in the case of slavery, "labor power" or "human capital" is not priced by itself and cannot be bought and sold in its entirety, so it is excluded from "wealth" here.) A capitalist economy is a type of market economy in which not only consumer goods and services as flows, but also capital and land as stocks are priced and bought and sold, that is, they are commodities in a broad sense. Then, what is the basis for the price, value of capital and land as stock?

From the classical point of view, capital is also a product of production, so it is the total cost of production, but the value of land cannot be understood by this logic. On the other hand, from the neoclassical point of view, the value of capital is not determined by its cost, that is, the value invested in it, but on the contrary, by the value it is expected to produce from now to the future. In other words, wealth, such as capital and land, is assigned a value because it is expected to be a source of economic value - or, more directly, income - and is therefore priced high, and is routinely the subject of time-based leases rather than whole transactions.

Whereas the classicals assumed that qualitative and discontinuous differences in the possession or non-possession of wealth in this sense and the kind of wealth (capital or land) would lead to class differences, or in other words, to differences in the logic of behavior among people, the neoclassicals assumed that the amount of "wealth" in monetary terms would be the same as the amount of "wealth" in monetary terms. In the neoclassical school, the focus is on the quantitative and continuous differences in how much "wealth" people possess in monetary terms, and people are assumed to be subject to basically the same principles of behavior.

What difference does this difference between the classical Marxian view of class society and the latent one-class (or no-class) view of neoclassical society make to the way we view the relationship between growth and the distribution of income and wealth?

It cannot be said in general that the one-class view of society is directly related to indifference to inequality. However, we can say the following. Under the classical-Marxian assumption, it is possible to argue that the concentration of wealth in the hands of capitalists (and, conversely, the deprivation of wealth from workers and landowners) is positive for the expansion and growth of the productive forces of the economy as a whole, whereas under the neoclassical assumption, this is not necessarily the case.

Under the classical-Marxian assumption, the behavioral logic of landowners and workers is not investment-oriented, whereas under the neoclassical assumption, whoever they are and whatever class they originally belong to, they will invest if they judge that investment is in their interest. In other words, under the classical and Marxian assumptions, changes in the way income and wealth are distributed may have an effect on investment and thus on economic growth, whereas this is unlikely to be the case in neoclassical theory. Regardless of the distribution of income and wealth, if the total amount of wealth (i.e., factors of production) in the economy and society as a whole remains unchanged, and if the market is sufficiently competitive and people act rationally in their own self-interest, wealth will be used efficiently, and the resulting amount of production, investment, and economic growth will remain unchanged. It makes sense to think this way.

This idea of "distribution does not affect production" can be seen in one sense as an idea that allows a high degree of freedom in the distribution of income and wealth, or as an idea that broadens the range of options. In other words, it is possible to argue that "if you want to eliminate inequality through redistributive policies, you can do whatever you want to some extent. (In fact, there is a big trap opening up there, but we will look at that later.)

But on the other hand, these ideas may have some effect of distracting economists from the distributional issues. In other words, the number of economists who think that distributional issues are not an academic subject for economists, but should be left to political scientists, sociologists, and welfare researchers, will increase. (Of course, this does not mean that all economists will think that way.

The Neoclassical Steady State

Another important point is that under the neoclassical assumption, we should assume the possibility of diminishing returns not only for land but also for capital and labor, in other words, for all factors of production.

Let us first consider the case where there is no technological innovation and no accumulation of human capital. As in the Marxian situation, population is assumed to be constant (or, more precisely, population change has no meaning in this analysis). Since capital and labor are to some extent substitutable, even when technology is constant and unchanging, the combination of capital and labor, the amount of capital per unit of labor, and the capital-labor ratio will change. Therefore, if capital accumulation is low at the starting point, it will continue to grow in this economy. The motivation that drives people to do so is that, at least initially, if they accumulate capital and increase the capital-equipment ratio, the productivity of labor, and thus the quantity of output, will increase, and profits will increase. However, this growth rate will decline as capital accumulation progresses. Eventually, the optimal point will be reached where further capital accumulation and increased production will result in a decrease in profit (the difference between income and cost). Therefore, it is best for people living in this economy to stop accumulating at this point and continue to maintain the capital-labor ratio and output at that level. This is a situation very similar to the classical steady state that Ricardo and Mill envisioned. In the initial phase, as in the Marxian situation, labor productivity will continue to rise, but since there is no innovation in the true sense of the word, the dynamism will eventually stop. Of course, this does not explain the mechanism of sustained growth since the Industrial Revolution, so it is necessary to somehow incorporate the mechanism that triggers technological innovation into the theory, but this has not been very successful.

Nevertheless, this vision of reaching and maintaining an optimal capital-labor ratio in the long run is still influential. The main reason for this is that it seems to match the global trend since the end of the 20th century of catching up with the least developed countries and converging economic disparities among countries.

I will come back to this point later, but until about the 1980s, Asian and African countries that had largely shed their colonial status and become independent by the 1960s did not continue to grow as expected despite the enormous aid from developed countries. The ever-widening gap between the developed countries that were growing and the less developed countries that were left behind was regarded as one of the central problems of the world economy. However, in the 80's, several East Asian countries, followed by the People's Republic of China in the 90's, started to grow rapidly under the open economy, and in the 21st century, India, which had long been a symbol of stagnation, entered the era of high growth. Sub-Saharan Africa, which continues to suffer from political instability due to corrupt dictatorships and the turmoil that followed their collapse, remains stagnant in general, but in recent years, not a few countries have been booming. In the 1980s, the term "middle developed countries" was coined to describe this process, and in the 21st century, many of these former middle developed countries are now considered ordinary developed countries.

In the 60's and 70's, it was believed that the key to the development strategy of the least developed countries was a certain amount of active government intervention and official assistance from the international community. However, the economic growth of developing countries during that period (especially in the 1960s, when developed countries were enjoying unprecedented high growth) was not good. Rather, if it is the transition of many developing and former socialist countries to free market-led open economic systems, along with the collapse of the socialist economic bloc and the transition of the system after the 1990s, that is causing this global catch-up and convergence, then it is certainly not surprising that there is an argument for seeking the mechanism behind this convergence to the neoclassical steady state.

In other words, this argument can be interpreted as "under certain conditions, free competition in the market may naturally lead to a more equal distribution of wealth." The question, of course, is what exactly are these "conditions" and what do they look like?

Why have neoclassicals lost interest in growth and distribution issues?

Let me summarize what I have seen so far.

The broad theoretical orientation of separating distributional issues from production (use of resources) issues has reduced interest in income and wealth distribution issues among economists who take the neoclassical position.

The expectation of a possible convergence of capital-labor ratios, and hence labor productivity and living standards, as a consequence of diminishing returns also spread the idea, at least among some economists, that free markets not only maximize production regardless of distribution, but in some cases even contribute to the equalization of distribution. This is what I mean.

That's about it. At least, the debate on the relationship between growth and distribution became somewhat muted in economics in the second half of the 20th century.

The above summary is based on the line of reasoning that "the rise of neoclassical economics led to a decline in interest in the question of the relationship between growth and distribution." I don't mean to deny that line of reasoning itself. (I don't have the guts to assert that it is the most important cause.) However, as I have confirmed several times, it was only in the late 20th century, around the 80s, that we could say that we had reached such an "end" for the time being, and this was not necessarily the case in the 50s and 60s.

Simon Kuznets' discussion of the "inverted U-shaped curve," which was brought back into the spotlight by Piketty's work, was such that "in the initial phase of economic growth (the period after the start of the Industrial Revolution), distribution becomes more unequal with growth, but this tendency eventually reverses and distribution becomes more equal with growth in affluent societies". However, Kuznets' research was primarily statistical in nature and did not lead to theoretical clarification of the mechanisms behind such phenomena.

However, during this period, when the high postwar growth had not yet ended in the developed countries, the influence of neoclassical economics grew stronger in the economic world. In the meantime, there were many researchers who built theoretical models of growth and distribution starting from the classical and Marxian assumptions, mainly orthodox Keynesians (those who inherited this position are now called post-Keynesians).The characteristic of the arguments of the older generation of Keynesians is that they model the situation of unemployment and recession, which John Maynard Keynes analyzed in his *General Theory of Employment, Interest and Money*, as a malfunction of the market mechanism, especially a situation where prices do not change smoothly and supply and demand cannot be balanced. This is the point. The first real mathematical model of economic growth was also this kind of model (the Harrod-Domer model in economics textbooks).

The emergence of neoclassical growth theory, in which the protagonist is an agent that makes optimal decisions about saving and investment, as well as substituting between capital and labor as appropriate to reflect price changes, was somewhat delayed. The Solow-Swan model, which is famous as the originator of neoclassical growth theory, takes into account the possibility of changes in people's behavior in response to price changes. In subsequent textbooks on economic growth theory, the Solow-Swan model is regarded as the starting point for modern growth theory, while the Harrod-Dominger model tends to be treated as a "prehistory.

However, the "victory" of neoclassical growth theory over classical and old Keynesian growth theory was mainly due to its theoretical consistency, not to the level of empirical research explaining empirical facts. Compared to Old Keynesian, Neoclassical had fewer arbitrary descent assumptions - for example, why prices become rigid, which are suddenly assumed without any theoretical basis. Therefore, I cannot deny the sense that the above "interim summary" is "hindsight" looking back from the end of the 20th century. Moreover, since the 1990s, within the neoclassical framework, there has been a renewed interest in the relationship between growth and distribution (and the theoretical work of the young Piketty can be positioned within this trend).

The Core of Economic Growth

In concluding this chapter, we must also note that the classical, Marxian, and early neoclassical theories of growth that we have seen so far are within certain clear limits. That is, with the exception of Marx, they describe a process of growth toward a zero-growth steady state and do not discuss technological innovation in the proper sense. In other words, it is not possible to understand economic growth in the sense that we are concerned with today, i.e., sustained increases in productivity and living standards. It is easy to see that this is the case with the classical framework, but it may come as a surprise to some that it is the case with the neoclassical framework as well, but I will take the trouble to confirm it.

To begin with, there are various types of "productivity," and we can distinguish between productivity per individual factor of production, i.e., land productivity, capital productivity, and labor productivity. In addition, there is the concept of "total factor productivity" that needs to be considered.

If we assume the classical situation that I had in mind when I explained "diminishing returns on land," there is no substitutability between capital and labor, but, in reality, there is some substitutability between labor and land, or capital and land. To be more precise, there is substitutability between the combination of labor and capital in the appropriate ratio - the "capital-labor set" mentioned above - and land. Therefore, even under the same production technology, there is not just one combination of land, capital, and labor required to achieve a certain yield or output, but countless combinations, and these combinations can be drawn on a graph. This combination can be plotted on a graph (called an "isoquant curve"). And for each particular combination, there can be various combinations of land productivity and capital-labor set productivity under the same technology. Under the assumption of "diminishing returns on land," as long as the technology is the same, in order to raise the same output, the capital-labor set must be increased by more than the amount by which the land input is reduced. If we use the term "productivity" to describe this, it means that for the same output, as land productivity increases, capital and labor productivity decrease (capital and labor are a set and move proportionally). If this were the "typical situation" in neoclassical theory, substitutability would hold between land, capital, and labor, but in practice it is not so complicated, since land is often omitted and only capital and labor are considered. To put it simply, the typical neoclassical image of growth is that if the capital-labor ratio is increased, that is, if the capital input per worker is increased, labor productivity will increase. Of course, as long as technology remains unchanged, the rate of increase will gradually become worse.

What we call technological change or innovation, and what we usually have in mind when we say "productivity increases," is not this increase in labor productivity caused by a change in the combination of labor and capital under the same technology, but an increase in both labor and capital productivity (and, in fact, land productivity). If you consider a particular set of capital-labor (and land), it means that the output per unit of that set will increase. This is called "total factor productivity".

While labor productivity and capital productivity (and land productivity) are relatively easy to calculate, total factor productivity is more difficult to calculate because it can only be calculated after assuming the structure of technology (production function) based on certain assumptions of economic theory (neoclassical economics). But theoretically, this is the most standard way of thinking about "productivity".

However, of course, this is a "theoretical" conceptual construct, so technological change and innovation in the real world will not necessarily make the concept of "total factor productivity" easier to understand. In other words, technological innovation in reality and the resulting increase in productivity do not necessarily occur in a smooth, "increase in output per unit of capital-labor (land) set.

Relationship between technological innovation and productivity

In most cases, the direct motive for technological innovation by companies operating for profit is to save the factors of production that happen to be scarce and expensive in the economy - whether individual materials or large "intrinsic" factors of production such as capital and labor. To put it bluntly, the reason is "wages are too high, so I want to reduce manpower". Under the neoclassical assumption of technology with high substitutability between factors of production, it is possible to reduce manpower if wages are high, and to compensate for the reduced manpower by increasing capital to maintain production at the same cost, but there is a limit to this because of "diminishing returns on individual factors of production. This is why they embarked on technological innovation.

However, in the case of technological innovation aimed at reducing manpower and saving labor, at least in the initial phase, it appears to the casual observer that "keeping the technology the same but making it more capital-intensive" and "conducting R&D activities and investing in new equipment for technological innovation" are both "investments". Even if the new technology saves capital input and improves not only labor productivity but also capital productivity and eventually total factor productivity, if the focus is on saving labor, the improvement in capital productivity may be treated as an "appendix," so to speak, and may not actually improve as much as labor productivity. On the other hand, even if it worsens slightly (the capital input per unit of product has increased! ), but from the micro perspective of the company, as long as the total cost is improving (the savings in wage costs outweigh the increase in capital costs), it is fine for the time being.

Therefore, it is important to note that laymen outside the field of manufacturing and development tend to think of technological innovation as a way to increase productivity, and that "productivity" means "labor productivity," and tend to forget about "capital productivity" and "total factor productivity. Increasing labor productivity is not the only way to innovate.

For example, there is a term "Industrious Revolution" that is still not well known outside of historical circles. The term was originally coined from the study of the expansion of agricultural productivity in Japan during the Edo period, but it is now a concept that is also used in the study of early modern history in Europe and China. To be more specific, it has been known for a long time that a major revolution, the Agricultural Revolution, took place in early modern agriculture prior to the Industrial Revolution and industrialization. The key to this can be understood as an increase in yield per unit area of land, an improvement in land productivity, and behind this was innovation in agricultural technology, but these innovations were labor-intensive as a whole, increasing land productivity but not labor productivity. Rather, they have increased the efficiency of land use by increasing the labor input per unit of land (e.g., the introduction of new crop rotation systems to stop fallowing, etc.). Japanese rice paddies are also a system that eliminates the need for fallow, and multiple cropping is also an improvement in land use efficiency), and increase the harvest. Thus, in the early modern period, the harvest per unit of land increased, and the supply of agricultural products increased, but the demand for labor also increased, and the increased crops were used to support the increased population, so the standard of living of the common people did not improve much overall - it is said that the wage level in the early modern period was stagnant. (Actually, we can think that something similar happened in the prehistoric Neolithic Revolution, at the beginning of agriculture - at a slower pace, of course. The transition from hunter-gatherer to agrarian life is thought to have worsened the average nutritional intake and health status of the total population, even as it increased.)

If we don't call this "Industrious Revolution" and its results "economic growth" and "economic development," then we are being somewhat biased in our use of the term. In fact, such an increase in agricultural productivity and population has created affluence among the ruling classes such as aristocrats and upper class citizens, fostered culture, and enabled the military revolutions of early modern European countries and the development of modern administrative and financial structures to support them. However, it is also true that "economic growth" and "economic development" for us today must be based on the increase of "labor productivity". This is because without it, it is impossible to increase per capita income and improve the average standard of living.

Economic historian Eric L. Jones has proposed a distinction between "extensive growth" and "intensive growth," taking these circumstances into account (*Growth Recurring: Economic Change in World History*). The former is a mere expansion of the economy in terms of scale, without technological innovation and with an increase in population, while the latter is a qualitative change with an increase in productivity and an improvement in living standards. According to Jones, if even "extensive growth" is counted as a type of economic growth, then economic growth is not a phenomenon unique to modern times, but has been occurring frequently since the beginning of history. However, this kind of extensive growth always ends up running into environmental constraints at some point. It is only by dodging these environmental constraints through technological innovation and improving the efficiency of land use that "intensive growth" becomes possible. The Industrious Revolution, the Agricultural Revolution, and the Industrial Revolution are the events that marked the transition to intensive growth.

However, as far as we have seen in this chapter, neither the classicals nor the neoclassicals have anything to say about the increase in total factor productivity. Only Marx barely addresses the issue head-on, but he is so concerned with the exploitation of labor that he fails to fully explore the difference between labor productivity and total factor productivity. And, as it turns out, we will have to wait until the end of the twentieth century to see a neat economic theory of such a rise in total factor productivity, a full-fledged technological innovation. Interestingly, the emergence of this new growth theory, called the "endogenous growth" model in the sense that it endogenizes the mechanism of technological innovation, has also helped to rekindle interest in the relationship between growth and distribution.

But before I introduce it, let me turn back the clock and look at a trend that probably prepared the way for the revival of interest in theories of growth and distribution since the end of the 20th century.

It is the subgenre of labor economics.

Originally, the focus of labor economics was to analyze the class conflict between capitalists and workers as part of the study of labor problems, but in the latter half of the 20th century, such interest faded away, and the main focus of labor economics became the study of workers as profit-seeking economic agents, and how their behavior affects corporate management and the economy as a whole. In other words, the focus has shifted away from the distributional issues between capital and labor and toward the potential contribution of labor to production and growth.

However, the story is not so simple when you look at it in detail. In fact, in labor economics, there has been a steady debate about the relationship between production, growth, and distribution in a secretive way, perhaps foreshadowing the resurgence of interest in the relationship between growth and distribution since the end of the 20th century.

Chapter 5: Human Capitan and the Hierarchical Structure of Labor Markets

The Neoclassical Theory of the Financial System: Why Workers Are Supposed to Be Capitalists
First of all, I would like to talk a little bit about finance before I get into the topic of labor economics. The first reason is that, as I said just before, while the focus of the study of labor issues used to be on the class conflict between workers and capitalists, in the latter half of the twentieth century, such an issue has become obsolete, and. the development of the financial system is not unrelated to the obsolescence of the class conflict.

In classical and Marxist economics, the problem of income and wealth distribution among classes was considered inseparable from the problem of capital accumulation and economic growth, but in neoclassical economics, the problem of distribution and the problem of growth have come to be understood relatively separately, and the interest of many economists has been drawn more toward the latter.
In the classical and Marxian schools, it was assumed that the class to which people belonged, and thus the principle of behavior, was determined by what kind of property (stock) they possessed or did not possess, so the growth problem and the distribution problem were inseparable. In neoclassical theory, on the other hand, the concept of class in this sense has disappeared, but what does this mean?
I have already touched on the possibility that as the wage level rises and far exceeds the level of subsistence, the opportunity to accumulate capital or acquire land becomes too great to ignore, even for those who initially own no property and have no choice but to become laborers hired by others (or tenants). However, we can also consider some other implicit historical assumptions that support the neoclassical assumption. The most important of these is the development of the financial system.
The term "finance" in fact encompasses a wide variety of systems, but the focus of this discussion is the system that supports credit transactions. If you understand that "credit transactions" refers to transactions between different points in time, you are not far off the mark. For example, if A lends money to B on the condition that B will repay the loan after a certain period of time with a certain amount of interest, there is an irreversible time difference between the time of the loan and the time of repayment, between the beginning and the end of the transaction.
Contrast this with what we think of as a standard trading transaction. Strictly speaking, even in such a transaction, the process of, for example, a buyer offering to buy an item in a store, the seller accepting the offer, and the item being delivered, naturally takes time, but it is not inevitable. What is important is that the transfer of ownership of the goods takes place as quickly as possible, and the ideal limit of the time taken is zero. In contrast, when lending or borrowing money, it is inevitable that there will be a certain amount of time where the borrower will do something with the borrowed money. However, since there is a passage of time, the lender will take a certain risk - the possibility that the borrower will not repay as promised - in order to earn the reward of interest. If they dare to "trust" the other party and do business with them, even at the risk of taking such a risk, but if the other party is not necessarily trustworthy enough as it is, they will devise various ways to deal with it. If you are going to do the same business with someone, you may want to choose a partner that you can trust, and you may also want to monitor them to make sure they don't betray you. Even so, there is still the possibility of betrayal, or even if there is no betrayal due to malicious intent, there is still the risk of the transaction failing due to force majeure, so preparations are made to mitigate these risks as much as possible, i.e., insurance is taken out.

Of course, credit transactions are not limited to the lending and borrowing of money, but also involve time differences in the buying and selling of goods. It is common to have a time difference in the buying and selling of goods (think of credit card payments or monthly installment sales). In fact, even in "buying and selling," except in the case of selling daily necessities and mass consumer goods to consumers, it is rather exceptional to have an immediate cash transaction for a finished product that already exists. Consider goods that take a long time to produce, such as custom-made buildings and fixed capital equipment. Or in the case of business-to-business transactions, new products such as equipment, new materials, and new kinds of parts are usually ordered, including the development of the product from the beginning, which also involves time differences. Since "finance" in the narrow sense of the word, which centers on the lending and borrowing of money, is also involved in such transactions (credit companies make advance payments to manufacturers on behalf of consumers and then lend the proceeds to consumers), it is a mechanism that supports the central axis of credit transactions, which are not limited to the direct lending and borrowing of money, but also include sales and purchases.

Having taken a slight detour, let me return to my main point. The bottom line is that the nature of investment changes with the development of finance. Under the typical classical and Marxist assumptions, the only actors of investment were capitalists (and landowners) who already had capital and could use their profits to finance new investments. However, with the development of financial markets, even the indigent could theoretically borrow money from financiers and make investments. Of course, the borrowed money would have to be repaid with interest, so such investments would not be made unless the profits were expected to exceed the borrowed money.

### The Stock System

Another thing that is thought to have contributed to the popularization of investment is the stock company system and the stock market, which are also part of the financial system.

In the classical and Marxist conceptions of the typical corporation, the owner of the company is at the same time the capitalist, and the corporation (although it already existed historically) was considered to be an exception. So, there, "being a capitalist" and "owning capital" only means either owning the whole company, or else having enough money to lend to others (including the company).

Smith considered the epitome of a joint-stock company to be a heavily mercantile state-run company like the East India Company, and rejected them as undermining the functioning of the free market. Marx, on the other hand, had a keen sense of their future potential, but was unable to incorporate them into his system of *Capital*.

However, through the system of stocks, one can become a capitalist without owning the whole company (if one does not own a company now and wants to acquire a new one, one must either already have enough money to buy the whole company or borrow enough money to do so), one can become a capitalist. By owning shares, one can earn dividends from the profits made by the company and participate in the management through shareholders' meetings.

Thus, under the neoclassical worldview, in addition to the path of accumulating a small amount of income as I have already suggested, an proletarian worker can also become a capitalist through the contrasting paths of borrowing money from the money market and starting his own business, or purchasing a small amount of stock and participating in the business of others. In a sense, this is a contrasting path that opens up the possibility for the laborers to also become capitalists. The meaning of this, as I have suggested before, is ambivalent. On the one hand, it opens up the possibility of thinking about the problem of income and wealth distribution not as a disparity between classes, which is inseparable and insurmountable, but as a continuous inequality between individuals, without fault. On the other hand, the idea that "no matter how wealth is distributed, as long as there is a well-organized market system, it will be fully utilized and the maximum productive power will be achieved" may lead to indifference to the distribution problem.

In any case, I believe that the development of the financial system has played no small role in changing the setting of the distribution problem from a class problem to a problem of income and wealth distribution among individuals - or, more precisely, among the subjects of concrete economic behavior and property rights, including corporations. I think that you have understood my point.

I will now turn to the study of labor economics, that is, the analysis of labor markets and employment relations. As I mentioned at the end of the previous chapter, here too we can see a shift in interest from distributional issues to production and growth issues, but there is more to it than that. Earlier, in my discussion of finance, I mentioned that even in cases where the distribution of capital and wealth is an issue, the way in which the issue is addressed has shifted from class inequality to individual inequality. The same can be said for the field of labor economy. In this context, the concept of "human capital" has also resurfaced.

Fading "labor problems": Changes in labor economics

Originally, labor research was not a monopoly of economics, but an interdisciplinary field of study involving law, which is deeply involved in the formation of institutions, sociology, psychology, and medical and health sciences, etc. For this reason, each country has its own unique characteristics (law, in particular, is inevitably domestic). However, it is possible to find some common patterns among the developed countries of the West.

Until the first half of the 20th century, the focus of labor economics was rather on industrial relations, labor-management (labor-capital) relations as political relations between labor unions, labor unions, capitalist management, and the state. In other words, it can be said that labor economics at that time had a strong color as a political economy of class relations. However, the "human capital revolution" of the latter half of the 20th century promoted the de-"political economy" of labor economics, and at the same time, it came to focus on the issue of the inequality in wages and working conditions, the distribution of income among workers, rather than the competition between wages and profits, that is, the distribution between labor and management.

In the first place, labor research was "labor problem research," which meant studying the "social problem" of laboring people. Furthermore, it can be said that until the first half of the 20th century, the dominant perception was that "labor problem" is "the" social problem, rather than simply being part of "social problems". In other words, the "social problem" is the "problem" of capitalist society, and its core is the class conflict in capitalist society as an unequal class society.

Traditionally, the "social problem" was mainly the "poverty problem," but in the mid-19th century, it was made part of the "labor problem" because "in capitalist society, the poor are part of the working class." The problem of rural poverty was also considered a "labor problem" if the poor peasants were highly dependent on hired labor, and even if they were not, it tended to be discussed in relation to urban labor problems, such as the problem of population outflow.

Women's issues and racial/ethnic issues tended to be neglected until the first half of the 20th century, and issues of racial and sexual discrimination in the labor market tended to be positioned as part of "labor issues" as "class issues."

This kind of composition is not the exclusive property of Marxism. As we have already seen, the basic structure was established in classical economics, and Marx and other socialists simply denounced it as a moral evil and made it a political issue. This was shared by moderate social policy theorists, trade unionists and social democrats who aimed for social improvement rather than revolutionary overthrow. They aimed to "institutionalize" the "class struggle" into a legitimate framework of parliamentary politics and collective bargaining, not necessarily to deny it. There was a shared understanding that if left to the free market, wealth would only accumulate more and more under the capitalists (and landlords), while the workers would remain poor, and that political means were needed to deal with this.

Along with this "class struggle" between workers and capitalists - the conflict between labor and capital over wages and working hours - another central "labor problem" in the past was the problem of unemployment. As we saw earlier, the Marxists tried to recover this issue into the framework of the "class struggle" through the theory of the "industrial reserve army. In other words, the existence of unemployment is one of the factors that make the struggle between workers and capitalists inevitably dominant in the latter. In the case of trade unionism and social democracy, in order to prevent unemployment from becoming a dead weight in the labor market, a response through labor movements and socioeconomic policies is required, and the main actors in this response are also considered to be trade unions and workers' (trade unionist and social democratic) political parties as representatives of the interests of the working class. This is in line with the aforementioned "institutionalization of class struggle."

If we assume that the composition of "labor studies" from the end of the 19th century to the middle of the 20th century was to depict the macroscopic rivalry between "total capital and total labor" with the labor unions as the main players, then from the latter half of the 20th century, the microscopic development became the leading force in labor studies, and at the same time At the same time, labor research became less and less about studying "labor problems" as "social problems.

As is well known, after the period of rapid economic growth, the weight of "labor problems" and "poverty problems" in "social problems" has been decreasing, especially in developed countries. The question that guides labor research has become "What are the factors that determine the good or bad performance of workers in the workplace or production site?" Until the 1980s, the focus of global attention on the "good performance of the Japanese economy" was also on "Japanese labor-management relations" and "Japanese employment practices". I do not think that such developments are unrelated to the neoclassical disconnection between the "distribution problem" and the "production problem," and it is needless to say that they are related to the disappearance of the "poverty problem" in the developed countries - or rather, its transformation into a problem of the exceptional few rather than of the working masses as a whole. It goes without saying.

In the midst of such a landslide of shifting interest in "labor," the leading role in collective labor-management relations shifted from labor unions to individual companies. This is because, after all, in capitalist society, it is the enterprise that has the presence as the subject of concrete production.

The Transformation of Industrial Relations

In the past, the main actor in collective labor-management relations was the labor union as a solidarity organization of workers, but the employer-capitalist side of the bargaining table was sometimes a collective trade association, sometimes an economic organization, and sometimes an individual company.

In particular, until the 19th century in the Western world, it was the unions of highly skilled workers, such as mechanics, that led the trade union movement, and they were able to work anywhere, and their skills, independent of management, allowed them to base their lives outside of the company. In other words, if they didn't like their current workplace, they could easily move elsewhere, and if they accumulated their own capital, they could turn to management themselves. The unions of these skilled workers focused on job placement and mutual aid functions to support their free movement of labor. In other words, rather than imposing restrictions on the labor market, they were themselves the infrastructure of the labor market outside the company. When dealing with such craft unions, it was common for employers to unite and organize to deal with them.

However, the relationship between workers and employers (capitalists) is usually not one-to-one, but rather multiple workers are usually employed by one employer - and in modern societies, that "one" is often a "corporation" that is itself an organization. In such a society, the bargaining power of the capitalist employer becomes stronger due to his monopoly position, and the workers prefer to unite in a union to strengthen their bargaining power as well.

Since the end of the 19th century, when heavy industry was developing and the number of large corporations was increasing, the leading role of the labor movement was played by the type of unions that united the workers inside the huge corporations, who could no longer easily go outside the corporations (if they did so carelessly, they would be at a disadvantage).

In the giant corporations that became the main players in heavy industrialization, technology was also no longer something that workers could control as their own skilled profession, but was occupied by corporate managers and specialized engineers, and was monopolized as "intellectual property," or became inseparable from capital equipment. Workers' skills are no longer highly skilled and generalized, but rather depend on their specific experience in a particular workplace, and are no longer transferable beyond the workplace. Moreover, as production facilities become huge and complex systems, each worker's job becomes only a fragment of it, simple and not requiring much skill.

In such a situation, the threat of "quitting if you don't like it" is no longer something that each worker can hurl at his or her employer. It is here that the image of today's strikes (collective strikes) and collective bargaining conducted under the threat of strikes is established, where workers united in the same workplace use their strength in numbers to say, "We will not work until our demands are met."

However, it is not necessarily true that such unions were the "company-based unions" that became typical in Japan in the latter half of the 20th century. In other words, perhaps it is because the historically leading labor movement was based outside the company and was a job-based union that attempted to control the labor market outside the company, that in Western Europe and the United States, such workplace-based unions were not "company-based unions" but rather "industrial unions" or "general unions" whose main body of organization was still outside the company. In the image of post-war Japan, industrial union organizations are just federations of company-based labor unions in industrial units, but in this case, on the contrary, branch offices of industrial unions are located in individual companies or factories.

However, with economic development, especially in the latter half of the 20th century, the ratio of factory workers in the manufacturing industry, which had traditionally been the base of labor unions, decreased, and the ratio of commercial and service industries and office and customer service workers increased, and labor unions also changed. Most importantly, as a result of the failure to expand organizing into emerging industries, the organization rate has been declining in general, and even in the unions that have been able to stay afloat, there is a tendency for the center of gravity of the movement to shift to organization at the company level. On the other hand, this can be interpreted as a process in which the leadership in collective labor-management relations shifts from the side of labor unions to that of employers (capitalists and the state), or as a process in which the personnel and labor management of individual companies and the state have more influence over the labor market than labor unions.

The term "leadership" here does not refer to the strength or weakness of power. In terms of simple bargaining power, it is the norm in employment relations that the employer is stronger than the employee. Rather, it is a question of who or what side provides the basic framework within which the concrete framework of labor-management relations, social practices and legal systems are developed. If employers are to unite against the professional unions of artisanal skilled workers, then it is the workers who are taking the leadership here, because the relationship can be understood as one in which management moves to counter the actions of the workers' unions. It is the union activities of the skilled workers who control the labor traded in the labor market, and thus (in Marxian terms) labor power, and (in neoclassical terms) human capital, that determine the scope of the labor market to be controlled by unions, and the capitalists act on this acceptance.

In contrast, in the era of giant corporations, when technical knowledge is enclosed on the side of capitalists and managers, the power to control the internal reality of skills and capabilities has shifted to the side of corporations and employers (this is sometimes called "internalization of the labor market"). This is called "internalization of the labor market".

Another important point is that industrial unions, general unions, and company unions, unlike job-based unions, focus on workers who are currently employed by a particular company and cannot include the unemployed. The workers of these unions, unlike the skilled artisanal workers organized by vocational unions, rely heavily on the company and workplace where they are currently employed to form and maintain their professional skills, and if they leave the workplace, their value as a guarantee of bargaining power will be greatly reduced. Therefore, if you are laid off or otherwise leave the company, you will not be able to contribute to the organization as a union member in any significant way. If you quit or are forced to quit and leave the workplace, you can no longer participate in collective bargaining to exert pressure, and if you lose your job, you can't even continue to pay union dues. It is also important to note that wages are lower than those of skilled tradesmen and professionals, and it is difficult to become self-employed and independent. The only way for these unions to protect their members is to focus on maintaining their current employment, and they do not have much power to protect their laid-off colleagues.

Thus, these new age unions cannot organize the unemployed and protect them as their own members. On the other hand, if they leave the unemployed unattended, or if they confront them as adversaries rather than as members of their own ranks, the unemployed will appear in the labor market as "dead weight" that will lower the wages of organized workers. Therefore, in order for the unions of the new era to confront their employers as labor unions and secure their bargaining power, they need to do something about the unemployed who cannot protect themselves well, at least without getting in their way. The measures taken in many countries were to make the state and government responsible for employment policies, to prevent the increase of unemployment by securing jobs through economic policies, and to ensure the livelihood of the unemployed through social security so that they would not become dead weight in the labor market. In other words, labor unions will have no choice but to deal with the problem of unemployment by finding or creating their own political parties to represent their interests.

Explaining Historical Change through the "Developmental Stage Theory

The framework within which labor economics and the study of labor issues in economics have attempted to understand this historical shift has been, very roughly speaking, through the "developmental stage theory". In other words, since the end of the nineteenth century, the capitalist economy led by heavy and chemical industry has shifted to the "monopoly capitalist" or "imperialist" stage in Marxist terms, where free competition is distorted by giant corporations.

Under these circumstances, as we saw earlier, the skills of the workers in the giant corporations are unique to each company, and they inevitably tend to seek long-term employment (and management also seeks long-term employment for key employees). On the other hand, the market adjustment function is weakened, making recessions more likely to occur and be prolonged. Under these circumstances, giant corporations will not be able to simply respond to economic ups and downs by increasing employment when the economy is booming, and shortening operations and reducing employment when the economy is in recession. At least the key workers who are hired on a long-term basis and play a central role in the field - or in the case of the management department, the reserve army of managers - will not be immediately laid off because of a downturn in the economy. Rather, many of the tasks at the end of the line are performed by workers hired from the beginning with a different status and employment status than the core workers, or subcontracted to other - often smaller - companies, and then used as a safety valve by firing or cutting deals when the economy worsens. This is a safety valve. Such a hierarchical structure in the labor market - regular employees and part-timers, main workers and period workers, parent companies and subsidiaries and subcontractors - will be created in the 20th century capitalism when the presence of large corporations will increase.

For example, Shojiro Ujihara, a leader in the study of labor issues in postwar Japan, developed this framework at the end of the 1950s, based theoretically on Marxian economics and the labor union studies of Beatrice and Sydney Webb, Lujo Brentano, and others, and empirically on surveys of actual conditions in the manufacturing industry and the labor market for new graduates on the eve of rapid economic growth. In the late 1950s, it clearly took shape. (The basic framework of Kazuo Koike's argument, which became internationally influential against the backdrop of the growing interest in "Japanese-style management" after the 1980s, was also based on Ujihara's inheritance.)

In the United States, where the influence of Marxism was smaller than in Japan and Western Europe, labor economists of the so-called (old) institutional school developed the "internal labor market" theory based on similar ideas. Michael Piore, a leading figure of the "internal labor market" theory, wrote *The Second Industrial Divide* (co-authored with Charles Sable) as a grand theory to support it. Although the historical theory depicted in the book is slightly different from that of Marxian economics, it is based on the idea that a structural shift in the industrial economy took place from the end of the 19th century to the 20th century - from a system of small-lot production of a wide variety of products, in which skilled multi-skilled workers played a leading role, to a system of mass production, in which mainly single-skilled workers were assigned to production lines for flowing work.

Again, under this kind of problematic interest and framework, the study of labor problems, or rather labor research, becomes less and less of a "social problem" study. As long as the focus is on the potential for economic growth in developed countries, or even in developing countries, the focus of labor research shifts from the poverty and hardship of workers to the improvement of workers' abilities, the production organization of workplaces that utilize workers, and the management strategies of companies. Under such circumstances, when the "labor problem" as a "social problem" is dared to be highlighted, the focus is not on the conflict between labor and capital, but on the disparity and conflict of interest among workers. In other words, the "hierarchical structure of the labor market.

In post-war Japan, the "dual structure" has long been a problem. In the midst of industrialization, from pre-war to post-war high growth periods, many studies were conducted based on the awareness that there was a clear disparity between "large companies" and "small and medium-sized companies" in Japan. In addition, a variety of economic and social policies have been implemented to support small and medium-sized enterprises (SMEs) as "the weak".

Two types of disparity

However, there are two types of disparities that can be roughly described. One is the disparity at the level of the company or management. Not only is there a stark disparity between large companies and small and medium-sized companies in terms of name recognition and influence in sales, but also in financing, small and medium-sized companies often have to accept loans on unfavorable terms. In general, there is an image that large companies dominate the market with their superior technology, while small and medium-sized companies are placed in an inferior position and have a higher risk of bankruptcy.

However, the focus of the "dual structure" as a "labor issue" is, of course, the disparity in wages, treatment, and working conditions. While large corporations offer higher wages and lower risk of unemployment, workers in small and medium-sized enterprises (SMEs) are forced to settle for low income and precarious employment - such a disparity exists between workers employed by large corporations and those in SMEs.

The question, of course, is why such a double or hierarchical structure is created. If the market were sufficiently competitive, small and medium-sized enterprises (SMEs) that are not competitive in the first place would be eliminated from the market and only efficient large enterprises would remain. (Why is it that the size of a company is directly related to its inefficiency and weak competitiveness in the first place?) Such market pressures should come not only from the product market, but also from the money market and the labor market. Why do workers in small and medium-sized enterprises (SMEs) continue to be underpaid?  Why can't they move to big companies with higher wages, or why can't they be forced to raise their wages to those of big companies through collective bargaining?

The leading growth industry of the twentieth century, the heavy chemical industry, requires huge fixed capital equipment that cannot be moved around. It takes time to set up the equipment and train a large workforce to run it, and once the equipment is up and running, it can't be shut down just because the economy is in a bit of a slump. Momentum makes such companies slow to act in the market.

However, on the other hand, such companies are forced to hold on to huge capital equipment and a large amount of labor, which allows them to exert a great deal of influence, or in other words, monopolistic control, in the market. Such market dominance makes up for the sluggishness of the market and allows the giant corporations to survive. However, under monopoly capitalism, the self-adjusting ability of the free market economy becomes sluggish, and recessions and unemployment are more likely to occur. However, in order to pass on the damage of such recessions and unemployment, giant corporations deliberately let small and medium-sized companies survive without destroying them, and outsource some of their operations. (i.e. subcontracting). Or, they keep some of their employees as part-timers to perform only auxiliary and peripheral tasks.

If the power to divide workers along lines such as "large companies vs. small and medium-sized companies," or "regular/stable employment vs. irregular/unstable employment," which are not exactly the same thing but overlap, is acting from the demand side of labor (companies), what about the supply side (workers)? What about the supply side?

In fact, the professional unions that originally led the early trade union movement had the exclusive aspect of being an organization of a select group of skilled workers. However, on the other hand, the unions were based on a broader network beyond individual workplaces, on the external labor market, and thus had a mechanism of solidarity between employed and unemployed workers. In contrast, under monopoly capitalism, workers who have lost control of technology and leadership in skill formation at the production site to management, workers who are not only commanded but also nurtured by the corporations, have lost their bargaining power without uniting at the workplace level beyond differences in skills and proficiency. In other words, the main actors of this era, the workers, could no longer have bargaining power without uniting as one. In other words, it can be said that the industrial unions that played a leading role in this period, and later the company-based unions in Japan and other countries, were free from the logic of exclusion based on job skills, as embodied in the job-based unions. But of course, on the other hand, it has become difficult to cross the fault line outside the workplace, outside the company, especially with the precariously employed and unemployed, such as part-timers, small and medium-sized enterprises, etc., whom the labor unions have not been able to reach.

Disparities among workers are disparities in "human capital.
— The logic we have seen so far is that of "traditional" labor studies, strongly influenced by Marxian economics, the (old) institutional school of economics in the United States, or the historical school of economics and social policy in Germany. As a whole, labor markets and labor economics have been understood as areas where "market failures" are fundamentally strong. In the labor market and in the management of labor in the workplace, the "logic of the market" as depicted by economics does not work well, and people are constrained by fixed social conventions or are driven by logics other than the logic of economic interests, such as morality, political power, and religion. Such an emphasis on the constraints and restraints that various non-market social contexts, the natural environment, and technology place on the workings of the market was the basic idea of traditional "applied economics" until the mid-20th century, including not only labor economics but also agricultural economics, finance theory, and development economics.

In contrast, from the 1960s, there was a growing momentum to reorganize development economics, agricultural economics, and labor economics, which had been regarded as issues centered on "market failures," with a consistent logic from the standpoint of neoclassical economics, rather than the eclecticism of the (old) institutional school. The leading figures in this movement were Theodore Schulz and Gary Becker, based at the University of Chicago in the United States.

To put it simply, they assumed that even in areas of human society and human life where markets are not established and markets have not penetrated, people must often make

"rational" decisions in the sense that economics assumes, and they thoroughly implemented this idea. For example, domestic work, which is not calculated in monetary terms, is in fact (consciously or unconsciously) compared and weighed against alternative choices of goods and services that have a monetary value and can be purchased for a price on the market outside the home - such as eating out, domestic help, etc. If this is the case, then theoretically it should be possible to assign a market value to it.

Based on this idea, Becker, in particular, boldly uses the blade of economics to cut into all sorts of social phenomena that have traditionally been considered outside the scope of economics because markets have not overtly penetrated them. His research on the "economics of housework," the "economics of marriage," the "economics of childbirth and childrearing," the "economics of crime," and the "economics of suicide" was at first the subject of laughter and indignation, and was criticized as the embodiment of "economics imperialism". However, after the initial enthusiasm (both positive and negative) wore off, it became clear that the Beckerian approach could be very useful as long as it was viewed as "a new way of looking at things." It is clear from a little bit of history that the era when housework was mainly done by housewives as free laborers was rather unique, that domestic servants as wage laborers were not uncommon, and that marriage was driven not only by love but also by economic calculation.

In this spirit, Schulz, Becker, and others boldly advanced the idea of "human capital" since Marshall, and used it not only as a theoretical model but also as a tool for empirical analysis. In other words, the difference in human investment creates the difference in wages, income, and working conditions afterwards. Human investment can take the form of schooling and formal vocational training (such as in public training institutions, professional unions, or apprenticeships), or it can be informal, such as learning skills by imitation in the workplace.

In the case of formal training, it is easy to see that there is money invested in it. In the case of schooling, tuition is usually required, and in the case of apprenticeship training, the apprentice is usually unpaid during the period of apprenticeship, so it can be said that the cost of free labor is borne by the apprentice. From an economic point of view, even though he could have found other work that would have earned him more money during his education, he dared to give it up and spent his time and effort on his education and training, and this "income that he should have earned" that he dared to give up is also the cost of his education and training. This kind of "cost" is called "opportunity cost" in economics.

If we keep this idea of "opportunity cost" in mind, we can understand that the acquisition of skills in an informal and imitative workplace, which is not clearly called "job training," can also be an "investment. The common wage curve in which wages are low in the early years of employment and rise as the worker continues to work - the "seniority wage" that is often referred to as the "Japanese wage" is an example of this - is also a wage curve in which the cost of training is deducted from the worker's original contribution to production in the early years of employment. When a worker gains experience and no longer needs training, his or her wage will rise.

Thus, the idea of "human capital" reintroduced by Becker et al. provides a logic to explain the hierarchical structure of the labor market by the difference in the ability of workers to bear the cost of training (human investment). Whereas the Marxian theory of "monopoly capitalism" was based on the demand side of labor, and lacked an explanation of how to sort out the core and peripheral workers, this theory makes up for it from the labor supply side. In other words, companies are not just hiring arbitrarily and discriminatorily, but they are also selecting workers based on their educational background and job training history.

Another important perspective introduced by Becker is the distinction between "general training" and "specific training. In this case, the criterion for distinguishing between "specific" and "general" is not the content of specific labor and skills - knowledge and physical practices - but rather the difference between those that are specific to a particular company or workplace and those that are not. The training required to acquire a professional skill - such as that of a doctor or a lawyer - that is supported by an official qualification system is "general," even if it is an extremely advanced and complex skill. And, roughly speaking, Becker argues that companies (employers) have no incentive to bear the cost of "general training" and that the cost tends to be borne by the workers, while "specific training" for skills specific to a particular company or workplace, on the other hand, workers have no incentive to bear the cost and the companies (employers) must bear the cost. This can also be used as a complement to the hierarchical structure of the labor market theory mentioned above.

However, these two approaches to the hierarchical structure of the labor market are not merely complementary to each other, but may be better viewed as being in a kind of tension. The former approach, from the labor demand side, focuses on the "market failure" - the market mechanism did not work well in the industrial economy led by heavy industry in the 20th century, and this had a great impact on the labor market. The latter theory of "human capital" does not necessarily adopt such an idea. Schulz and Becker, who revived the idea of "human capital," are considered to be representatives of the so-called "Chicago School," but they rather expanded the frontiers of economics by finding a "quasi/virtual market" even where there is no market. In other words, they are "market fundamentalists" and "economics imperialists". From their point of view, it is possible to explain the wage gap only by the logic of "human capital", and to portray it as a penetration of the "logic of the market", irrespective of the "failure of the market" - because it is more natural.

Does "human capital" belong to the team?

The review of labor economics up to this point has followed the direction of the separation of "distributional problems" and "production problems" since the hegemony of the neoclassical school in the 20th century. The focus of labor studies has shifted to production rather than distribution, and even when dealing with distribution problems, the focus has shifted to the inequality among workers - whether they are organized workers or not, large or small firms, regular or irregular, etc. - rather than the "labor versus capital" structure. And the income gap between workers is said to be due to differences in innate ability and human investment. The implication of this is that, firstly, it is efficient in terms of resource utilization, and secondly, it makes no difference whether the inequality is left unaddressed or whether the distribution of human investment is made more equal in order to correct the disparity, as long as the capital market is sufficiently efficient.

However, if we take a closer look at the content of labor research, the story is not so simple. The idea of "human capital" is certainly not without an individualistic slant. At the very least, employed workers form a team with their employers, and since one employer usually employs many workers, they also form a team with their colleagues. On the other hand, the theory of "human capital" tries to reduce the factors that determine the income of each worker in a company to the human investment borne by the worker (or his/her parents) as much as possible. However, this is not the whole story. As we have seen earlier, and as the architects of the "human capital" theory have made clear, workers as individuals have no incentive to bear the cost of training in skills specific to a particular workplace or company. So what exactly are these company- and workplace-specific skills? Why are there such special skills that cannot be externalized, that is, cannot be freely procured in the outside market?

There are many possible reasons. In the case of primary industries such as agriculture, it is typical, but in the manufacturing industry as well, natural factors such as the environment of the workplace and the physical characteristics of the equipment cannot be ignored. However, at least part of it is that the workplace team becomes a concrete and individualistic group of people, and the nature of the group influences the productivity of each individual within it. In other words, a team is more than just a collection of individuals. The simplest example of a team is a group of people carrying a heavy load that cannot be lifted by a single person. In the case of a task one can do alone, sometimes it is more efficient to work in a group than to work individually.

The reasons for the increase in productivity through group work and collaboration are of course complex. A very popular example, which is different from the case of heavy lifting mentioned above, is the effect of the division of labor, as explained by Adam Smith in his famous case of pin production in *The Wealth of Nations* ( Book 1, Chapter 1). Rather than gathering ten craftsmen who perform a series of processes all by themselves, a team that divides the process into ten simpler work units and dedicates one craftsman to each unit will be more productive. The emergence of skills that are unique to each company and workplace can typically be understood in this way.

Complementarity is stronger than substitutability between the work of each member in such a team. Let's say the member in charge of a certain task suddenly quits and is no longer there. Is there anything else that can be done in this case other than hiring a new person to take over the job?  If there is some degree of substitutability between each task, you can make up for the loss by increasing the amount of the other tasks. We have already seen that the neoclassical way of thinking is to base things on "some degree of substitutability between input elements," but this kind of substitutability does not work very well between the tasks and jobs that are responsible for each process in a single workplace or a single production line. It is easy to see that in large-scale, complex production technologies, there is not much substitutability, at least in the short term, between the work of workers in individual sub-processes, nor between labor and raw materials or capital equipment, but rather complementarity comes into play. In such a situation, there is often a phenomenon of increasing returns to scale, i.e., the larger the scale of operation, the higher the productivity. This is another reason for companies to retain their own trained workers for a long time and stabilize their employment.

"Increasing returns to scale" or "economy of scale" is not an uncommon phenomenon. In the case of technologies that cannot be produced at all without first building a fixed facility on a large scale, but once the facility is set up, it can be used for a long time, this "economy of scale" - that is, productivity increases as production volume increases and product unit price decreases - is always observed. The more the production volume is increased, the higher the productivity and the lower the unit price of the product. In addition, when a new technology is adopted, productivity increases as a result of the accumulation of experience among the people involved as the operation continues compared to the initial start of production - this includes, of course, the maturation of the work group as a team - a phenomenon called the "experience curve" in business administration and production engineering. In this case, as time passes and experience is gained, the scale of operations often increases, so at least apparent "economy of scale" can be observed.

However, this increasing returns is not permanent, and it is often implicitly assumed that almost any production technology will eventually reach a certain level of constant returns to scale. The division of labor described above may require a team to work together in one place at the start, but once the method is established and the procedures standardized, it does not necessarily have to be done in a group under a single management, and each work unit can be broken up and subcontracted through the market. In this way, company and workplace specific skills are transformed into general skills that can be procured in the external market. The question is, which of these should be assumed to be the more standard situation? In other words, should we think of human labor as a team entity that cannot be decomposed, or as a fragmented entity that can be pieced together through the market-based division of labor? The neoclassical school - or rather, the mainstream of economics, including the classical school - has been thinking in the latter style for many years, while the former was the domain of business administration, sociology, or engineering.

However, since Marx and Joseph Schumpeter, capitalism has been viewed as a "market economy with continuous technological innovation," and both sides of the equation cannot help but be taken into account. That is to say, the closed and continuous "organization" (as opposed to the open and fluid "market") has become the new focus of analysis at the workplace level within companies, and even between companies, not only in terms of competitive relationships in the market, but also in terms of cooperation, technological alliances, contracting relationships, etc. I mentioned earlier that labor research in the latter half of the 20th century ceased to be research on labor "problems" and shifted its attention to production rather than distribution, but such changes can also be seen in this context.

While the stimulus for the development of new technologies can only be provided by fierce competition in the marketplace (which is why technological innovation stagnated in the centrally planned economy), the development and commercialization of such new technologies requires more than just the division of labor among independent businesses connected through the marketplace. It is necessary to have a close communication network and a team that involves physical and face-to-face contact. The skills of the participants will be highly specific to the network or team, at least until the new technology is established and standardized. Companies with employment relationships at their core are a typical way of organizing such networks and teams. Even though there is fierce competition among companies, there is a need for closer cooperation than competition among workers employed within a company - let's put it this way for now. So what happens if we do that? At the level of the firm, or the level of the collaborative network or team, the issues of production and distribution become inseparable.

The inextricable link between production and distribution

I mentioned earlier that hired workers have no incentive to invest in firm- or workplace-specific skills, and that employers must bear the cost of training. In order to come to such a conclusion, it was necessary to make the assumption that the workplace is only a temporary place for hired workers, and that they have no particular commitment to it. There is no reason to be excited about the workplace, or the company, as long as the employer pays their wages well and without regret, and they are not interested in acquiring skills that have no value away from the workplace.

In other words, the point here is not whether you are a worker or an employer, but whether you are committed to the workplace or the company itself. The reason why employers bear the cost of company/workplace specific training is because they are committed to maintaining the value of the company as the owner of the company (a position similar to that of a capitalist). In other words, if the hired workers have a reason to commit to the corporate value itself, they will have a reason to bear this cost themselves. The question is, how does such a motivation arise?

A company in which the capitalist or his agent is the employer is a system in which the capitalist bears the cost of organizing the team, but the capitalist also monopolizes the profit of organizing the team. This means that the level of productivity that could not be achieved if each worker worked separately is achieved by having them form a team, but the capitalists keep most of the profits. By paying each worker the same as, or slightly more than he or she would have earned if they had worked separately, the capitalist can secure the necessary labor force.

Now let's think about how they can realize the benefits of team production without capitalists or entrepreneurs. If we do so, we will find that this is quite a difficult task.

In order to realize worker cooperatives, workers must bear the cost of organizing and managing the teams themselves, which in the employment relationship has been taken care of by the capitalist employer. This naturally includes the cost of training for the formation of skills specific to this team.

The problem is that these team-specific training costs, as well as the effort required to manage the team, must be borne by the team members in a coordinated manner to be effective - if only one person works hard while everyone else slacks off, the effort will be largely wasted. This is the nature of a team. If the size of the team is large enough, there is a temptation for each member to cut corners and become a "free rider" who does not properly bear the costs. This is because if the team is large enough, the damage caused by one person slacking off is minimal. However, if everyone thinks the same way and everyone slacks off, the team will not be able to survive.

To prevent this from happening, it is effective to assign someone to monitor each member to make sure they don't slack off, but in order for that person to do their job properly, they need to be paid accordingly. But what if that reward is paid out of the profits of the team's production?

In the most extreme case, this watchperson - or hired manager - would have a monopoly on most of the profits of team production. The watchdog would then work hard to maximize his profits and achieve the best possible results, while the ordinary workers would be left with only the same level of compensation as they would have earned outside the team. The reality of the team would be almost the same as the capitalist management. The hired manager, even if he is elected by the team members, is no longer a member of the team, but rather an outsider. On the other hand, if the manager's compensation is too low, he will not work diligently and will not be able to successfully prevent the general workforce from slacking off. In reality, the negotiations take place between these two extremes. In other words, the distribution of compensation among the members of the team is directly related to the production performance of the team.

In other words, the distribution of reward among the team members is directly related to the production performance of the team. The image of the employer who takes all the profits of the team production is, so to speak, the ideological extreme image of the all-powerful tyrannical leader who plans every step of the way what work is to be performed by each worker, gives strict instructions, and imposes sanctions when necessary. In the real world of capitalist management, the capitalist employer needs a large number of staff and middle managers to assist him in his administrative tasks in order to manage a large team, and he has to pay these administrative workers a share of the profits of the team's production in addition to the money they would have earned outside the company. So, even in capitalist management, there is a need to make sure that the people who are in charge of these administrative tasks are given a share of the profits of team production in addition to what they would have earned outside the company. Thus, even in capitalist management, if it is a "team", the relationship between distribution and production becomes irreconcilable.

In practice, the problem of the distribution of the return to capital is added to this. In other words, capitalist corporation is not really just a mechanism for realizing team production of a large number of workers. The typical capitalist corporation is not owned entirely by a single capitalist manager, but also by a cooperative team of capitalists. In the ideal extreme of capitalist corporation, the capitalists soak up most of the profits of the team's production in the form of dividends or capital gains, but in reality there is a distribution problem with the board of directors who are delegated the management of the company.

On the eve of the "Inequality renaissance

If we think about it in this way, we can sum up that the retreat of the "labor problem" in labor economics, the retreat of interest in distribution problems at the macro level, and the shift of focus to production problems also caused attention to the inseparability of production and distribution at the level of individual management and the firm as the focus of production problems. Thus, in the 1980s, "economic analysis of organizations" concerning employment relations and corporate governance at the firm level, business alliances and subcontracting, and inter-firm networks such as regional industrial clusters, became a popular field in economics. In fact, this trend provided a foreshadowing for the resurgence of interest in disparity and inequality at the end of the 20th century. This new stage of the inequality debate will be referred to in this book as the "inequality renaissance. This "inequality renaissance" meant a resurgence of interest not only in inequality per se, but also in the relationship between distribution, production and growth.

# Chapter 6: Inequality Renaissance I – after the Kuznetz curve

What is the "Inequality renaissance?

Let us now return to the "inequality renaissance," or the revival of interest in the relationship between growth and distribution in the context of macroeconomic growth theory.

As I mentioned in the previous two chapters, the "Kuznets curve" (inverted U-shaped curve) is an empirical (quasi-)law or a trend in economic history that Piketty has also taken into account and criticized. In other words, economic inequality among people - for the time being, inequality in income/wealth distribution - increases rapidly as the economy develops, but the pace eventually slows down and eventually reverses. Kuznets conducted this study in the 1950s, and as a defector from Russia (Ukraine, to be exact) to the US, it may have implied a critique of Marxism, but it is basically empirical. What is presented is not so much a theory as a very rough view of history by today's standards of economics, but it is something like the following....

In the traditional economy, which was mainly agricultural and limited to local areas, people's livelihoods were homogenous and inequality was not high. This is because incomes are higher in the emerging urban-centered industries and the gap with the traditional rural sector, which is left behind, is larger, as is the diversity and disparity within the emerging urban industrial (manufacturing and commercial services) sector. However, this inequality is a transitional one during industrialization, and as industrialization progresses, the population will shift to urban areas and modern industrial sectors, and this transitional disparity will disappear. As industrialization progresses, the population will shift to urban areas and the modern industrial sector. Furthermore, the effects of social security and redistribution policies are not insignificant. Thus, in the process of industrialization, inequality temporarily increases, but after a certain level is exceeded, inequality begins to decline again.

Kuznets' original paper was published in 1955, on the eve of full-fledged post-war reconstruction, or in other words, on the eve of full-fledged high growth, beyond the recovery to pre-war levels. However, the United States, whose land had not been affected by the war, had already entered an era of rapid growth and mass consumption. The subsequent high economic growth in Western Europe and Japan reduced the "double-structure" gap between urban and rural areas, and between large corporations and small and medium-sized enterprises. Of course, the disparity will not be completely eliminated, but the high economic growth will, so to speak, "raise the bottom," and realize a general rise in income, including the bottom part, making the poverty problem, which in the past was a survival crisis like the "food problem," a problem of the unfortunate few rather than "mass poverty" in developed countries. Rather, the "poverty problem" has come to be seen as a problem of developing countries, a problem of global scale rather than a problem of a single country, and the "North-South divide". Poverty and inequality were then understood as problems of the "developmental stage," that is, as transitional problems until economic growth is achieved. Although it was undoubtedly "mass poverty," it was not seen as a distributional problem within a single society of the entire global human race (or "global society" as it was later called), but as a domestic problem of each country, and not as a distributional problem, but as a production and growth problem of insufficient absolute productive capacity. This trend is in line with the neoclassical worldview (not the theory itself, but the ideology that tends to derive from it) that I mentioned earlier.

However, this situation changed in the last half of the 20th century, and a "inequality renaissance," as it were, began to take place. Specifically, it has been pointed out since the 1990s that domestic inequalities have been increasing in advanced countries, or more broadly, in OECD countries including newly industrialized countries that have caught up with the rest of the world since the 1980s - a situation that deviates from the predictions of the Kuznets curve.

Interestingly, this seems to be in some ways contrary to the reduction of national and international inequalities at the global level. Yes, as evidenced by the increase in the number of OECD countries themselves, especially since the 1980s, some developing countries have emerged from the phase of "mass poverty" and have undergone rapid industrialization and urbanization. China and India, with populations in about one billion each, have had a particularly large impact on the marketization and openness of their economies. These countries now have some of the worst domestic inequalities in the world, and it can be assumed that these inequalities are basically similar in nature to Japan's former "double structure". In other words, these countries are now in the ascending phase of the Kuznets curve, the phase where inequality increases with growth, and we can expect that it will be some time before they enter the descending phase.

However, if the recent increase in domestic inequality in these countries is only such a phenomenon, it does not mean that the Kuznets curve or the Kuznets hypothesis is disproved in any way, but rather that it is an example of the opposite. Rather, the problem is that inequality seems to be growing even in the "developed" countries that have already passed through the era of the "dual structure". In other words, in these "advanced" countries, income inequality has been easing under the process of high growth, as if to confirm the prediction of the Kuznets curve, but apparently, income inequality has been increasing again since the 1980s, when the high growth clearly ended.

However, what I would like to call the "inequality renaissance" here is not only the discovery of such facts but also the flourishing of theoretical research to explain them.

From production to distribution and from distribution to production

As I argued in the previous two chapters, it is possible that for orthodox neoclassical economists, the existence of inequality itself is not a big deal, whether from the standpoint of a scientist or a policy advocate. In fact, there are plenty of researchers who claim that.

If the characteristic of neoclassical economics is to separate the problems of production and growth from the problems of distribution, to concentrate the subject of economics on production, and to regard the market as a mechanism for the effective utilization of socially dispersed resources (wealth, skills, knowledge, etc.) rather than as a mechanism for the distribution of income and wealth they could say it like that. Regardless of the distribution of income and wealth among people in a society, if the market is efficient, the resources will be used efficiently anyway. If we think that the question is how efficient the market is in that sense, then whether or not inequality is actually growing is not a very important question for neoclassical economics. Even if it were an academic or policy issue, it would be a problem for sociologists and political scientists, who are primarily concerned with distribution rather than production. This is a topic for sociologists, political scientists, and social workers, not for economists.

Let's check a little more what is meant by this "separation of production and distribution issues (and the concentration of attention on production issues)". It should be noted that this does not mean that production and distribution are unrelated.

Neoclassical economists do not deny that the degree and pattern of production activity, capital accumulation, and economic growth centered on firms competing in the market will affect the distribution of income and wealth among people. It may preserve the existing distribution of wealth (disparity structure of wealth) almost as it is, as is typical in the case of the classics and Marx, or it may become even more so, or conversely, it may have an equalizing effect, as argued by those who focus on the catch-up of some developing countries in recent years.

However, at least so far, as a large number of neoclassicals, we have accepted this kind of "production to distribution" causality, but not the reverse "distribution to production" causality. Where a free market economy exists, whatever the distribution of wealth, knowledge, and abilities among people, they will be put to maximum use - regardless of how the newly created wealth is distributed among them.

Therefore, to criticize the "separation of production and distribution problems (and the concentration of attention on production problems)" and to insist on the "inseparability of production and distribution problems" means that "distribution patterns may affect production, capital accumulation, and economic growth in the market," that the "distribution-to-production" causal path cannot be ignored. In the "inequality renaissance," there is a strong interest in this "distribution-to-production" causal path. In a sense, it can be said that this is a revival of the classical interest in the issue, but in a reversed direction.

In other words, current empirical research points out that "there is a negative correlation between the degree of inequality and the economic growth rate of each country, i.e., the worse the domestic inequality, the lower the economic growth rate of the country tends to be. The question is then posed as to whether the former causes the latter.

In other words, economic inequality within a country may be stunting its economic growth, and conversely, more equal distribution may have the effect of raising the growth rate. The question of whether there is a causal relationship from distribution to production and growth, and if so, what kind of mechanism it is, has become increasingly active, and several theoretical models have been developed.

The Unclear Theory of "Growth and Distribution

I'm getting a little ahead of myself. Let's go back to the situation before the advent of this "inequality renaissance.

In the first place, the usual interpretation of the Kuznets curve is that it remains a "story" rather than a clear theory, but it is generally understood in the order that the stage of development of production technology and productive power affects the distribution of income and wealth, that is, production and growth are "causes" and distribution is "results. However, the issue here is only the absolute level of productive capacity, not the speed of dynamic changes such as the economic growth rate. There is no particular awareness of the question of how changes in the rate of economic growth affect the distribution of income and wealth.

However, to go a little further, to take the interpretation of the Kuznets curve as I mentioned earlier, that is, to regard the phase of rising inequality associated with economic growth as a transitional period in which the traditional economic sector, which is not fully marketized and business-oriented, coexists with the modern sector, which is market-oriented and uses the latest technology, is a very important point. This means that

\***The market economy itself, contrary to appearances, is actually an equalizing force.**

Or...

**\*The inequality that occurs within the market economy is small compared to the inequality that occurs between the market economy and the outside world.**

As we have already seen, to put it very crudely, classical economics emphasized the possibility of raising incomes in general in a market economy, up to and including the bottom, but not the possibility of reducing the gap between the bottom and the top, between the poor and the rich. They believed that a market economy would lead to unequal distribution, and even that inequality - the concentration of wealth in the hands of capitalists - would lead to more growth. Neoclassical economics, on the other hand, was characterized by a more neutral stance, leaving room for discussion of various possibilities.

The dominant interpretation of the Kuznets curve, however, goes one step further and suggests the existence of a tendency toward equalization under a market economy - or after industrialization. However, Kuznets' thesis itself is, again, not very theoretical. So, is there a stronger and clearer theoretical consideration of "How does competition in the market affect existing patterns of wealth distribution? Isn't there a theoretical consideration of this question?

The starting point for this trend is actually the 1969 article "The Distribution of Income and Wealth among Individuals" by Joseph Stiglitz, a young economist ( needless to say, who become a Nobel Prize-winner and an enthusiastic endorser of Piketty's *Capital in the 21st Century* ). In this paper, Stiglitz came to the rather clear conclusion that "in a perfectly competitive market (including capital markets), the distribution of wealth and income will level off in the long run," based on the Solow-Swan model, the originator of neoclassical growth theory, which I mentioned earlier.

Stiglitz's subsequent fame is based on his analysis of market failures due to uncertainty and incomplete information, as well as his theoretical analysis of various non-market institutions and practices that cover such problems, and his harsh criticism of "market fundamentalism" from such a standpoint, so this conclusion is somewhat surprising. In fact, there is no information that Stiglitz himself wrote a follow-up study to this paper.

This 1969 Stiglitz paper seems to be nothing more than a theoretical justification of the conventional interpretation of the Kuznets curve. In fact, it was written before the end of the rapid economic growth, so it may have reflected the atmosphere of the time. It is also possible that Stiglitz himself had such an intention. But on the other hand, Kuznets' name itself does not appear in this paper in the first place. More importantly, at the time it was written, it did not seem to have been widely read (for Stiglitz's work). It was not until the 1990s, when we seem to have entered an "inequality renaissance," that work began to explicitly address, critique, or take over the thesis.

From a layman's point of view, a glance at the history of research shows that there is still no clear conclusion on "whether distribution becomes more or less equal in a perfectly competitive market" - there seems to be no common understanding in the entire academic world. This is because, taking into account the long-term growth process, there are multiple ways to understand a "perfectly competitive market" - a situation where so many actors participate in the market that no one can manipulate the market as a whole, let alone the actions of a particular person - it is possible to set up a variety of models. Even if the real market works so well that it can be approximated by a model of perfect competition, the question is, "Which" model of perfect competition is closer? Of course, this is especially true if the reality is a world of imperfect competition. The concept of a "perfectly competitive market" is important as a benchmark. In other words, if we want to create various models of imperfectly competitive markets that are closer to reality, we must first assume a perfectly competitive market, and then create a model that deviates from it and corrects it.

The problem is that in the theory of growth and distribution, the "benchmark" itself in this sense is not yet clear.

To put it crudely, for example, if a perfectly competitive market causes equalization of income and wealth distribution among its participants, it would be better to attribute the inequality mainly to non-market factors - various conditions that hinder the smooth functioning of the market - and to remove such factors in order to achieve equalization. If, on the other hand, we believe that perfectly competitive markets lead to inequality, then we cannot achieve equality without active market intervention. In any case, we cannot safely discuss "inequality under capitalism" without at least this level of clarity. At present, however, I suspect that even this level of agreement has not yet been properly reached.

As we saw in Chapter 4, in macro growth theory itself, the Harrod-Domer model is already in the museum, perhaps because it does not incorporate price changes and their adjustments. However, the Solow-Swan model, the first neoclassical growth model, is still in use, probably because it incorporates the adjustment of the capital-labor ratio in response to changes in prices, wage rates, and interest rates, and is still useful, at least from an educational perspective. However, even here, the assumption that the savings rate is constant regardless of the subject and regardless of time is introduced into the descent equation, so the theoretical foundation is still weak. At the researcher's level, there are two major types of models: the Ramsey model, which depicts how subjects with infinite life expectancy (called dynasty) save and invest to maximize their lifetime utility, and the overlapping generations (OLG) model, which depicts how subjects with limited life pass on their inheritance or debt to subsequent generations indefinitely. They are established as the "benchmarks" for depicting long-term growth under perfect competition, and modified as necessary when depicting imperfect competition, including Keynesian situations. (The Ramsey model is also briefly explained and criticized in the text and online appendix of Chapter 10 of Piketty's *Capital in the 21st Century*.) From there, for example, some "secondary benchmarks" have been created for specific problem areas, such as, as we will see later, some more specific standard models in "endogenous growth theory" for analyzing technological innovation. However, the standard model in the "theory of growth and distribution," as it were, has not yet been established.

Therefore, in the next chapter, I will briefly attempt to create a benchmark, but only for this book. Specifically, I will construct the Ramsey model and a generational overlap model consisting of multiple entities with disparities in wealth ownership, and investigate both behavior. Here I will basically explain the model in words, but behind the scenes, I have created simple mathematical models to check the behavior. If you are interested in mathematical models, please see mathematical appendix in this volume.

Chapter 7: Inequality Renaissance II – Toy models of growth and inequality

Consider the model of the "Theory of Production and Distribution

To simplify the discussion, let's assume that capital is the only factor that remains inequality. However, let's assume that everyone has enough capital from the start to save and invest at least a minimum amount, especially if they are unable to trade capital.

We also assume that there are no differences in capabilities among people. (Or, if there is, it is due to differences in human capital.) And let's lump "labor" as the wealth that people have equally from the beginning and cannot accumulate any more. (Let's ignore "land.") Furthermore, let's assume that the population is constant, the working hours remain the same, and the amount of labor per person is the same.

Let's assume that the same production technology is used throughout the economy. The only inputs to that technology are labor and capital, and the single product that results is consumed in people's lives or used as capital. In this technology, there is a constant return to scale and a diminishing return to labor and capital. In other words, doubling the total amount of labor and capital equipment at once will double production, but doubling only labor or capital will increase production but not double it.

With the above production technology as a premise, we do not assume the existence of companies as units specialized in production activities, distinct from households as living units, but assume a world with only household-level self-employment. The labor market does not exist (or is excluded from the analysis). People use their own labor on their own. As for capital, if there is a capital market, people borrow from others or lend extra to others. One reason for setting up the model in this way is that I want to analyze the problem of inequality by narrowing down its cause to the distribution of capital (I also want to treat the wage gap as a problem of human capital), and another reason is that I want to compare the "world with capital transactions" with the "world without capital transactions," as we will see later. If we assume the existence of corporations, we cannot assume a "world without capital transactions" - not to say that we cannot, but it would be very unnatural.

If we assume such an "economy consisting of self-employed people with no difference in labor capacity under a certain production technology with constant return to scale," one very interesting result emerges. That is, "the maximum social output is achieved when the amount of capital at hand is exactly equal among all agents. (Later that assumption will be relaxed a bit.)" First of all, the maximum amount of production is achieved by using up all the capital and labor that already exists in society - full employment. The question is whether it is possible to achieve this in a free market economy with a system of private ownership, where each individual voluntarily engages in production activities for his or her own benefit, rather than in a centrally planned economy.

For this to be achieved under this assumption, the capital-labor ratio of each business must match the ratio of total capital to total labor in the economy as a whole. This means that each business must have an equal amount of capital on hand that can be invested in production activities. At least in the initial conditions, this can only be achieved by a lucky chance. With the exception of businesses that happen to be in the average position, everyone else falls outside of it. Of course, those who own above-average capital here are able to produce above-average output. Of course, the output of those with below-average capital will be less than that of the average business owner.

And the key point here is that thanks to "diminishing returns on capital," the degree to which the output of the former exceeds the average is not enough to offset the degree to which the output of the latter falls below the average. This is because the more capital per capita, the less productive it is. If we take the extra capital from the above-average capital owners and transfer it to the below-average capital owners, we can achieve the maximum social output, but there is no way to achieve this except by coercion by those in power from above. Without transaction, there is no way to achieve this except by coercion by those in power from above, because it is privately optimal for the wealthy with above-average capital to spend all their capital on themselves, no matter how far outside the social optimum. These inefficiencies are generally greater the more unequal the distribution of capital is.

In other words, in this model, distribution determines production, at least in the short run. However, even in the short run, if capital markets are established and free capital transactions are conducted smoothly, the task of "achieving the maximum social output by utilizing all wealth without waste, regardless of its distributional status" can be achieved. If a capital market is established, transactions of capital lending and borrowing are possible between those who are undercapitalized and those who are overcapitalized. In this way, the amount of capital on hand and available for use can be equalized among all people. But of course, even if everyone's output is equal, it does not mean that everyone's income is equal. Those who lend capital become richer than average through interest income, while those who borrow become poorer than average. This is exactly the neoclassical idea of "separation of distribution and production" that I mentioned earlier.

How will capital be accumulated?

A further question is how this will unfold in the course of time. As capital is accumulated in this way, the amount of capital owned by each economic agent, that is, each household, as well as the total amount of capital in the economy and society as a whole, will increase, and the ratio of capital to labor at any given time will change. What we have just seen is that in the case of a free capital market, the amount of total capital in the economy as a whole, and the average capital-labor ratio in society as a whole that results from that, automatically becomes the "optimal capital-labor ratio" that individual economic agents should adjust to, and how people adjust to that in a market economy. This was the mechanism. In the long run, however, the story is not exhaustive. In fact, there is an "optimal capital-labor ratio" at the level of the entire economy and society, and the question of how to adjust to it for the entire economy and society emerges.

The point is that "our model of economic growth assumes a long-term life plan and actions based on that plan for subjects living in the flow of time". Since we are concerned here with economic growth, especially the increase in per capita productive capacity, it is impossible to conceive of it without capital accumulation, or investment. In the first place, saving and investment means not consuming goods now, but saving them for the future, or converting them into something that can be used in the future. In other words, economic growth is seen here as a process in which an entity that lives for a certain length of time takes action to maximize its profit over its entire life cycle, rather than just for the immediate benefit of the day. Therefore, this subject not only compares the two options in the present and ponders which one to choose, but also compares the two options in the present and at some point in the future, such as whether to spend the money they have now or to save it and spend it a year from now.

The way to model this choice between different points in time is to consider the "present value of future options" and make a comparison of the present options and their values possible in the very present moment. The key concept in calculating the present value is the (subjective) discount rate. This is simply a number used to express the behavioral pattern of "morning, three, but evening, four", that is to say, to place more importance on the near future than the distant future. For the monkeys in the story, the combination of four acorns for breakfast and three acorns for dinner had a higher present value than the combination of three acorns for breakfast and four acorns for dinner, at least in the morning.

If the present value of an acorn in the morning and the present value of an acorn in the evening are the same for the monkeys, then the combination of three in the morning and three in the evening have exactly the same value, and either one can be chosen (in economics terms, it is "indifferent"). However, if the latter is chosen, the present value of an acorn for dinner is diminished compared to that of an acorn for breakfast, even if only slightly. This decrease in value is called the (subjective) discount rate.

There are many reasons why a positive discount rate is possible (the future is "discounted" from the present). In simple terms, it may be natural to give more weight to the certainly feasible options of the present than to the uncertain options of the future. Even if they are exactly the same, it is natural that the former would be chosen over the latter.

Intuitively, this is the reason why interest is charged on debts. In this case, the discount rate corresponds to the interest rate. For example, if money is lent and borrowed at an interest rate of 3% per year, it means that the value of 1,000,000 yen in your hand now is equal to the present value of 1,030,000 yen one year from now.

If you think of it that way, the discount rate acts as a "cutoff line" for your investment plan. If the rate of return is less than the subjective discount rate, then the investment is not worthwhile for that entity. To give you an idea, let's consider a situation where you can get a 3 percent interest rate if you deposit money in a bank. If you simply deposit money in the bank without thinking about it, you are promised a 3 percent rate of return, so any business plan that offers less than that is worthless.

Now, "diminishing returns on capital" means that the rate of return on investment is high in the beginning, but declines rapidly as capital is accumulated. For example, if the first 1,000,000 yen yielded 100,000 yen, or a 10 percent rate of return, but the next 1,000,000 yen yielded only 50,000 yen, the overall rate of return would drop to 15/200, or 7.5 percent. At the time of the initial investment of 1 million yen, the rate of return on that capital increase is 10 percent, even if a small increase in capital is made in addition to the initial investment. However, when the accumulation progresses and the total value of capital reaches 2 million, the rate of return on a small increase in capital is 7.5 percent.

This "return on a small increase in capital" is the "marginal productivity of capital," and "diminishing returns on capital" means that this "marginal productivity of capital" is declining rapidly. Although it is declining, the assumption here is that it will never fall below zero. But that doesn't mean that we should just keep on accumulating. So, to what extent? Until the rate of return - and the marginal productivity of capital - is equal to the discount rate, the extra investment still promises more profit. But once that line is crossed, further investment no longer generates (subjective) profit, but rather reduces it. In the previous example, if the rate falls below 3 percent, it is better to do nothing and deposit the money in the bank.

To put it very roughly, the optimal point is where the marginal productivity of capital and the subjective discount rate coincide, and the capital-labor ratio at that point is the optimal ratio. The ratio of capital to labor at that point is the optimal ratio. Increasing capital beyond this point will result in a loss. (Strictly speaking, of the two models we will look at later, the Ramsey model shows a match between the marginal productivity of capital and the subjective discount rate in the steady state, but the overlapping generations model does not. Nevertheless, even there, there is a clear and unambiguous correspondence between the two.)

The model assumes that there is no (economically significant) difference between people except for their ownership of capital, so the discount rate for all people is equal. The question we want to ask is, "Can it be that people have equal discount rates but different long-term saving and investment behavior because of differences in capital ownership, with different end results? This is a very important question. Needless to say, this idea follows fundamentally the same logic as the traditional steady state of classical economics. If the initial state of the economy starts from a point where capital is still low and insufficiently accumulated, then the growth process toward this steady state can be derived. Since the model does not incorporate the mechanism of the increase in total factor productivity due to technological innovation, the steady state is zero growth. The "steady state with growth" that incorporates technological innovation will be discussed later.

How does the presence or absence of a capital market affect economic growth?

Two representative theoretical models of how people accumulate capital toward a steady state where the marginal productivity of capital matches the discount rate are the Ramsey model and the overlapping generations (OLG) model, which I have already briefly introduced. To review, the Ramsey model focuses on dynasties that live infinitely long lives, while the overlapping generations model focuses on generational succession of people with finite lifespans (although, for simplicity, this generational succession is also depicted as if a single "family line", or dynasty, is perpetuated). In the former model, each subject is assumed to be purely selfish, while in the latter, each subject is assumed to be altruistic to some extent, specifically, to the extent that he or she is willing to leave assets only to his or her direct successors.

If we look at how each of these models portrays the savings and investment behavior of each individual subject under these assumptions, we see an interesting contrast. Since the nature of the people who make up this economy is exactly the same from the assumptions, differing only in the amount of capital they own at the starting point, one would expect that the differences in their saving and investment behavior patterns would basically stem from differences in the amount of capital they own. Interestingly, however, in the overlapping generations model, everyone's savings rate is equal and invariant over the long run, regardless of the amount of initial capital ownership.

Incidentally, the Solow-Swan model, which is featured as the "first neoclassical growth model," is a model that assumes a constant savings rate from the beginning and leaves the reason for this assumption unanswered. In contrast, the constant savings rate in the overlapping generations model is derived as a result of the choice behavior of economic agents, and thus has a more solid theoretical basis.

In the standard overlapping generations model, the savings rate - the ratio of savings to income - remains positive and constant even after reaching a steady state of zero growth, as each generation leaves a legacy to the next. In the case of the Ramsey model, on the other hand, the savings rate is usually high during low accumulation, declines gradually as accumulation continues and steady state is approached, and reaches zero at steady state. The difference seems to come mainly from whether the decision makers are replaced one after another or remain the same throughout.

As you may have guessed by now, for each of these two models, let's further consider the case where the capital market is perfect, and the case where there is no capital market at all. This is an unrealistic and extreme contrast, but it is a thought experiment to see how the capital market affects investment and growth, as well as distribution.

As I mentioned earlier, if capital markets are perfect, anyone can borrow any amount of capital they lack or lend out capital they have no use for, as long as they pay the appropriate interest determined by the market. Thus, inequality in capital ownership does not hinder the efficient use of capital in the short run (each business or individual operates with an equal amount of capital). In contrast, if there is no capital market at all, capital cannot be lent or borrowed, and people will try to make full use of the capital they have. However, as mentioned earlier, this situation is inefficient from a social point of view. This is because a decrease in production under an undercapitalized business compared to the average level cannot be compensated by an increase in production under an overcapitalized business compared to the average level.

Of course, when capital is used efficiently and optimally in the short run, it does not necessarily mean that it is optimized in the long run. As we saw earlier, the optimal level of capital accumulation in the long run is determined by the point where the marginal productivity of capital and the subjective discount rate coincide. However, the process of adjusting to this point takes the form of a capital market, in which the interest rate established in the market approaches the discount rate. As we saw earlier, the rate of interest mediates capital transactions to equalize the marginal productivity of capital among people for short-term optimization. However, as long as the interest rate deviates from the discount rate (which is originally equal among people), there is a need to adjust investment in society as a whole and in the long run. In this case, the rate of interest = the marginal productivity of social capital is still higher than the discount rate, so it is in our interest to invest more and more until it eventually matches the discount rate. And the growth rate is proportional to the difference between this interest rate and the discount rate. (In special cases, they match.) Since both the interest rate and the discount rate are social numbers, the growth rate is also the growth rate of consumption for all people as well as for society as a whole. So the original asset disparity is preserved.

In contrast, in the absence of a capital market, each individual will perform the task of investing more and more to the point where the marginal productivity of capital matches the discount rate, in a discrete manner, without the intervention of a capital market. Each business invests on its own, without borrowing from others, until the marginal productivity of its own capital matches the discount rate. Since the growth rate varies from individual to individual, and the less capital ownership at the starting point, the faster, the asset disparity will decrease as the economy as a whole grows, and will disappear in a steady state of zero growth.

### Will the gap be preserved or will equalization be achieved?

First of all, I tried to think about how the presence or absence of a capital market affects the growth process, regardless of the two models, but in fact, this is not necessarily the case when we try to fit this discussion into the two models.

It is in the case of the Ramsey model that the story fits exactly as I have described above. In the Ramsey model, if there is a perfect capital market and smooth capital transactions, equalization of capital use is achieved without affecting the disparity in capital ownership, and the disparity is preserved throughout. In contrast, in the absence of a capital market, equalization of capital ownership is realized as a byproduct of long-term optimization at the expense of short-term optimization, so to speak.

However, in the case of the overlapping generations model (and, as Stiglitz's 1969 paper argues, in the case of the Solow-Swan model as well), with or without a capital market, the long-run optimum is achieved as well as the long-run equalization of capital ownership. The reason for this is probably the effect of a constant savings rate.

What I mean by this is that the savings rate is the percentage of one's income that is saved. This means that the higher the income, the higher the savings, so at first glance, it may not seem to have anything to do with the reduction of inequality. However, things look different when you think in terms of capital (stock) instead of income (flow).

In this case, there is no difference in people's labor capacity (if human capital is included in capital, it is the original capacity before it is increased by investment), so the difference in income is solely due to the difference in capital. However, if capital ownership is doubled, income is not doubled. This is because income can be divided into labor income and capital income, and there is no difference in labor income between people. Even if the ownership of capital doubles, the total income will not double, but will be a smaller amount. This means that under a constant savings rate, even if capital ownership doubles, savings will not double, but will only be less. In other words, a constant savings rate means that the ratio of savings to income is constant, which means that the ratio of capital accumulation to capital falls rapidly as capital increases. This effect seems to cancel out the gap-preserving effect of capital markets, even when they exist.

In summary, we can see that there are roughly four patterns to keep in mind.

(1) The reduction of inequality depicted by the generational overlap model in a perfect capital market situation. (This is an upgraded version of Stiglitz's 1969 paper using Solow-Swan model.)
(2) The same tendency of reduction in inequality is also observed in the overlapping generations model in the absence of capital markets.
(3) The maintenance of inequality in the Ramsey model with perfect capital markets.
(4) The reduction of inequality in the Ramsey model without capital markets.

These are all rather abstract models, but if we dare to relate them to reality, we can say that models (1) and (3), which assume the operation of capital markets, should be applied to disparities at the domestic level, while models (2) and (4) should be applied to global disparities. (2) and (4) can be interpreted as a model of a world where free trade is to a large extent practiced and product markets are globally integrated, but labor markets and capital markets (especially those that lend and borrow long-term funds necessary for capital investment) are not so integrated. It looks like a good explanation for the catch-up of some developing countries, the start of rapid growth and the transition to middle-income countries. But it is only apparent, and it is not clear why this is so. In contrast, (1) and (3) can be interpreted as modeling the level of a single economy with active capital movement. However, they are also in sharp contrast to each other. While (1) seems to support the classical interpretation of the Kuznets curve by inheriting the problems of Stiglitz's 1969 paper, (3), on the contrary, maintains inequality. This is extremely difficult to interpret.

The models we have discussed so far are all models with no technological change and no technological innovation, in which "growth" becomes zero growth once a steady state is reached, and the same state is repeated endlessly. Let's take a look at what this steady state looks like. In the absence of a capital market, full employment of capital is achieved for the first time, as well as maximum equalization. On the other hand, in the case where capital markets exist, full employment of capital is achieved in the short term at each point along the way to this point. Therefore, when the capital market is fully functioning, the "separation of distribution and production" always exists, but even when the capital market is absent, the "separation of distribution and production" exists in the steady state, that is, in the long run. As for the nature of the steady state, we must pay attention to the full employment of capital and the maximization of social production, rather than the limit of equalization (in the first place, if the capital market is perfect, this equalizing tendency will not be possible).
In light of the above, let us now consider an economy that can sustain growth over the long term.

### Incorporating Technological Change into the Model

Ultimately, the focus of our interest is on "intensive growth" (see Chapter 4), in which technological change leads to sustained increases in total factor productivity, and under which income and wealth are distributed. As a matter of fact, in our capitalist market economy, most of the technological innovations are realized through trial and error by profit-seeking companies and individuals. For a long time, technological innovation has been given exogenously to models as if it were falling from the sky.

If I had to put a realistic interpretation on it, I would say that technological innovation comes from the academic research world, where people engage in purely intellectual pursuits regardless of economic interests, and is offered to the business world for free. And the mechanisms of the academic sector are left as a black box. The Solow-Swan model was a typical example. In this model, growth stops in the steady state. Therefore, when incorporating technological innovation and productivity growth into the Solow-Swan model, it was common to assume that the growth rate would continue to rise, even though we did not know why, as a given without explanation, as if it were a blessing that had fallen from heaven. Of course, since Marx and Schumpeter, it has always been emphasized that the main driving force of technological innovation is the capitalists and other actors who participate in the fierce competition in the market economy. However, again, it has been very difficult to incorporate this into a coherent theory.

The so-called "endogenous growth" model, which attempts to theorize that technological innovation is the result of rational choices made by economic agents seeking private profit, has been actively studied since the late 1980s. (Jones' "inteisive growth" is naturally caused by this "endogenous growth," but the two concepts are not equal. Under the traditional "exogenous model," productivity growth is given descent without explanation.) And while there is still no established theory, we have gone beyond ad hoc conjectures and have developed some "standard" models that can be found in textbooks as the common property of the academic community. And much of the theoretical work in the "inequality renaissance" referred to in this book focuses on the impact of distribution on this "endogenous growth". In other words, a non-zero-growth steady state is derived in which the per capita productivity and total factor productivity increase is sustained, and at the same time, the distribution of capital and income is affected by the growth rate in such a steady state - not only temporarily and in the short run, as in the classical "zero-growth" steady state described above, but also in the long run and permanently.

The simplest and most primitive of the "standard" models incorporates what have been called "Marshallian externalities" and "network externalities". More straightforwardly, it is the spillover (leakage, diffusion) of knowledge and technology.

For example, technical knowledge enables individual economic agents to increase their productivity and get ahead of their rivals who do not know it by acquiring and using it. If this is the case, individual people and companies will want to keep their knowledge and skills to themselves. However, if such knowledge and technology cannot be kept monopolized and becomes known and shared throughout society, will the original monopolist suffer a loss? Not necessarily. Of course, it may no longer be possible to gain a position of relative advantage over rivals. But what if your knowledge and skills are so versatile that they can be used widely outside your industry?   You may no longer be able to monopolize the knowledge and sell it better and cheaper than your competitors. In that case, as a producer or seller, the former monopolist would have lost money. However, let's say that the leaked knowledge has been used in other industries beyond the narrow industry and has contributed to making better products at lower prices. Then, the former monopolist, who was losing money as a producer or seller, will gain money as a consumer or buyer. And if the versatility of the knowledge and technology is very high, the gain will outweigh the loss. At the most basic level, such spillover effects are emphasized in the context of the "abacus of reading and writing," that is, literacy and basic academic skills at the primary and secondary levels, but it is easier to understand when we think about various technological standards.

If we assume that such investments not only benefit us but also spill over to the outside world, we can construct a model of a "steady state of positive growth" with "intensive growth" relatively smoothly. What is even more interesting is that when such spillover effects, or "external economies" as they are called in the jargon of economics, are present, there is a blatant misalignment between individual and social optimality, as opposed to when they are not. In the standard case where there is no "external economy," the maximum effort of each subject for its own benefit results in the maximum social use of resources, but in the case where there is an "external economy," not all the results of the effort are attributed to the subject of the effort, and there is spillover to others. However, when there is an "external economy," not all the results of the effort are attributed to the subject of the effort, but rather spill over to others, creating a motivation to "ride for free" on the achievements of others. In other words, in a market economy model with knowledge and technology spillover, unlike a model without it, a steady state of positive growth is possible, but that steady state is not the maximum growth theoretically possible in this economy. This is an example of what is called a "market failure" because it leads to such an inefficient outcome under perfect competition.

But let's put this "free ride" issue aside and return to the "steady state of positive growth. We need to look at how the various distribution patterns in the "zero-growth steady state" we discussed earlier change when this spillover effect is added.

If we create a model in which new investment is accompanied by new knowledge and technology, and such knowledge and technology spill over to people other than the subjects of the investment, total factor productivity increases in society as a whole, causing positive growth as a steady state, then investment takes on a strange character. In other words, while "diminishing returns to capital" are observed at the level of individual economic agents, "constant returns on capital" are observed at the level of total capital in society as a whole. If this is the case, then the marginal productivity of capital in society as a whole will remain constant and will never decline, and therefore will never match the discount rate. In conclusion, the difference between this marginal productivity and the discount rate determines the growth rate. This contrasts favorably with the case of technological constancy, where marginal productivity and the discount rate coincide to form a steady state of zero growth. Also, the "constant return to scale" is broken, and a kind of "economy of scale" - a situation in which productivity increases as scale increases - is established.

I will now briefly introduce the results of replacing the production technologies with those with spillover effects for the four models described above.

(1)'If we look at the model with perfect generational overlap in the capital market (and the Solow-Swan model), it is not so far off from (1) at first glance. The growth process generally converges to a "steady state of positive growth," and inequality shrinks in the process. (In fact, the economy depicted in the generational overlap model has the interesting property that even in a zero-growth steady state without spillovers, the maximum technically possible growth cannot be achieved under perfect competition, but I do not have time to go into that here.)

(2)'The situation is similar for a generational overlap model with a missing capital market, but the convergence to the steady state is slower and the level of production is permanently lower than in the case of (1)'. The more unequal the distribution at the starting point, the slower the convergence and the more severe the decline in the level of production.

(3)' The Ramsey "dynasty" model, in which capital markets are perfect, is similar to (3). If the capital market is perfect, everyone faces the same rate of interest, has the same amount of capital equipment on hand through capital borrowing and lending, produces the same amount of output, and earns the same gross income, but there is a gap in the level of net income due to interest payments and receipts. The mechanism works here as well, and everyone experiences the same rate of growth, thereby preserving the initial disparity indefinitely.

(4)' The Ramsey model lacking a capital market also converges to steady-state growth, as in (3)', but, as in (4), the lack of a capital market leads to a process of equalization in the process of convergence. Also, as in the case of (1)' and (2)', the convergence to the steady state is slower in Mal than in the case of (3)', resulting in a permanently lower level of output. The more unequal the distribution at the starting point is, the slower the convergence will be, and the lower the production level will be.

Here are two graphs to give you a rough idea. Needless to say, the detailed logic behind these graphs can be found in the mathematical appendix.

Figure 1 shows the difference between(1)' and(2)' . The vertical axis is capital per capita and the horizontal axis is time (generations). Under the same production technology, even if the starting state (the amount of total social capital and its distribution) is the same, there will be a difference in the time it takes to reach the final steady state between a perfect capital market and a lack of a capital market. As a result, even if the steady-state growth rate is the same, there is a difference in the absolute amount of capital, and thus in total output. The graph shows the two paths taken by the three groups of households (not individuals, as they change from generation to generation): hosehold A, household B, and household C, in that order, have a higher amount of capital at the starting point. Household B has just the average position. The same starting point - starting from a state where there is a certain amount of disparity between households and eventually there is no disparity, but compared to the case of (1)', the convergence in the case of (2)' is delayed and the final level of income and wealth is lower.

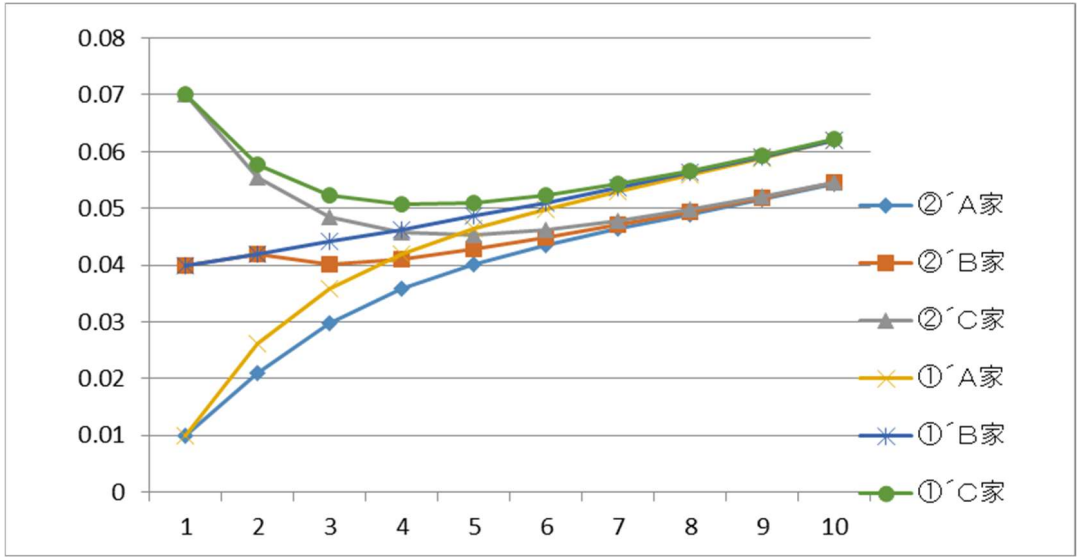

Fig. 1: the OLG intensive growth model; comparison of (1) and (1)'

Figure 2 shows how inequality in the distribution of capital leads to slower growth in the case of (2)'. Although the two groups are equal in terms of the total amount of social capital at the starting point, and hence the average amount of capital per household (household B in the large gap group and household B' in the small gap group), the former is more unequal than the latter. Both groups eventually equalize their capital ownership and reach the same steady-state growth rate, but the former cannot catch up with the latter.
(it is impossible to show this in a graph that is as easy to understand. For those who are interested, the only way is to look at the formula in the mathematical appendix).

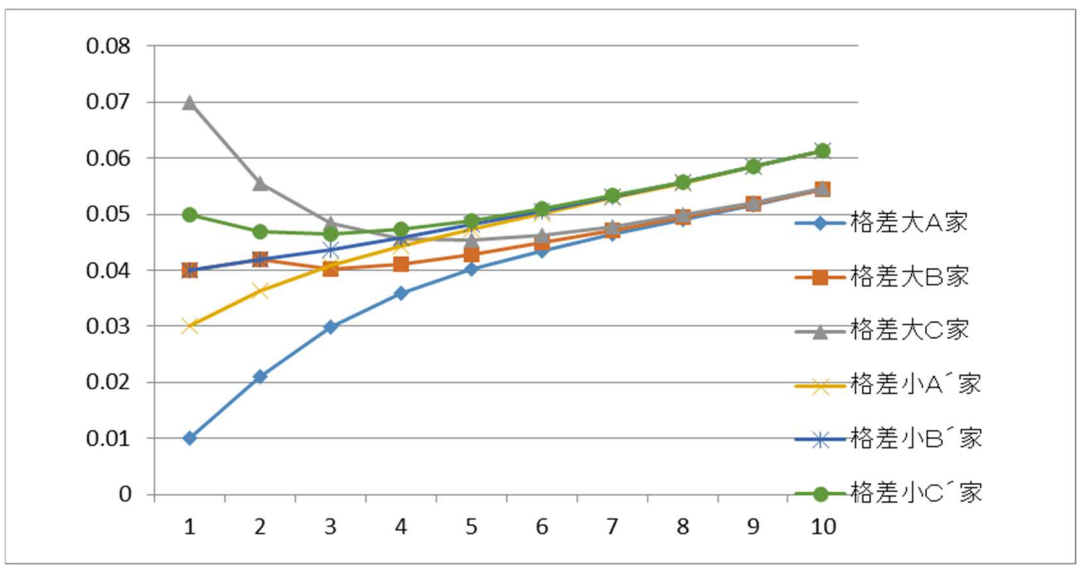

Fig.2: the OLG intensive growth model; the effect of inequality in (2)'

The above is a list of eight "benchmarks" for zero-growth steady state and positive growth steady state, four patterns each, focusing on the presence or absence of capital markets and whether the savings rate is constant. Of course, this is not an exhaustive list of all possible patterns, but merely a list of easy-to-follow landmarks for a quick intuition of the overall structure of possibilities. However, the above list should give you a vague idea of how the existence of capital markets, the nature of the savings rate, and factors such as diminishing and constant returns on capital can affect growth and distribution. So, let's look at the implications of this in a new chapter.

Chapter 8: Inequality Renaissance III – capital market completion or redistribution?

How to resolve the uneven distribution of capital

In the model so far, I have made the rather extreme assumption that the minimum capital requirement is not set, that the scale of business can be reduced as much as possible, and that there is no particular disadvantage even if it is broken down to the level of a sole proprietorship. However, even if we don't go to such extremes, it would be more efficient if the ratio of capital to labor is the same in each business unit when technology is shared socially. However, in reality, it is difficult to think of such a situation as fortuitous, where such conditions are in place from the beginning. Capital is unevenly distributed, being available to those who have it and not available to those who do not.

There are at least three ways to solve this problem and bring the capital-labor ratio to a socially optimal level, if you have been following the discussion above.

The first is to let things happen as they will, and let each individual make a concerted effort to accumulate capital. This is an easy way to let nature take its course, but it takes time, especially if there is no capital market.

The second is to establish a complete capital market. Once this is in place, it will have an immediate effect, but it is actually quite difficult to create proper institutions such as banks and stock markets. Also, if the real economy is closer to the Ramsey model than to the generational overlap model, this will not help to improve inequality.

In addition, there is of course a third way, namely, fiscal redistribution policies that involve the mobilization of state power. There are two types of redistribution policies: one is to redistribute capital and assets themselves, that is to say, the stock at once - agrarian reform is similar to this - and the other is to redistribute the flow of income, consumption expenditure, etc. The former is so revolutionary that I will exclude it from my discussion for the time being. Even if we limit ourselves to the latter, there are reasonable costs associated with taxation and the income transfers and free service transfers that it finances. This is the disincentive effect of taxation, i.e., the possibility that income taxes will discourage work and business, and consumption taxes will discourage consumption.

There are a number of problems, but the first one to consider is that the first "let nature take its course" approach is surprisingly unreliable. This may be fine if you can think about a steady state of zero growth, with capital markets but no technological change, but that is a far cry from the real world we live in. In a world of technological change and imperfect, if any, capital markets, "letting nature take its course" is not only too slow, but also unreliable. If this is the case, then the second thing to focus on is the development of capital markets and the third is redistribution through public finance.

Having thought this far, the focus of theoretical interest in the "inequality renaissance" naturally emerges. To put it very crudely, it can be said that the leading role is played by arguments based on the idea of a generational overlap model that lacks capital markets in the previous list. As a strategy to overcome this inequality and low growth, two directions are suggested: capital market reform and fiscal redistribution.

In textbooks, the paper by Oded Galor & Joseph Zeira in 1993 is often referred to as a milestone. Piketty's paper of 1997 also assumes a constant savings rate and basically follows the same generational overlap model as above.

Let's review. In a steady state of zero growth with no technological change - or new technology falling from the sky for free - the initial inequality and inefficiency will change in the growth process. In a steady state, the initial inequalities and inefficiencies will change during the growth process. Without a capital market, the capital-labor ratio will become more efficient and equalized through self-sufficient saving and investment by each individual. (Even if there is a market economy, people trade mainly in final products, consumer goods, not in capital, productive goods. When the capital market is perfect, the use of capital can be made more efficient more quickly, and this does not require equalization of the distribution of capital itself. In any case, in the long run (in the steady state), the distribution does not affect the growth rate there.

However, this is not the case when there are no or dysfunctional capital markets in an economy with endogenous technological change (due to the self-help efforts of individual agents with spillovers). Again, in the absence of capital markets, improvements in capital efficiency and the resulting equalization of the social distribution of capital are not carried out through capital transactions and the transfer of capital between economic agents, but through their own isolated, self-sufficient saving and investment activities. Investment continues to the point where its efficiency can no longer increase, which in a zero-growth steady state is where the marginal productivity of capital, which continues to decline thanks to "diminishing returns on capital," coincides with the subjective discount rate.

However, under the steady state of positive growth, the irony is that the marginal productivity of capital does not match the discount rate, but ends up just short of it. In the case of the model assumed here as a benchmark, a "constant return to capital" is assumed, so the marginal productivity of capital is constant and unchanging regardless of the amount of investment and regardless of the total amount of capital stock, and it never even comes close to the discount rate. And the absence of a capital market has a distinctly negative effect.

To summarize, in an economy with persistent technological innovation and a steady state of positive growth, but lacking or very weak capital markets, the tendency of the distribution of capital to equalize is rather weak, and the more unequal the distribution in the beginning, the lower the output level in the long run. The more unequal distribution of capital and the lack of capital market lead to the less efficient social use of capital and the lower total social output here, and adding to that, unlike an economy with a zero-growth steady state, the negative effects of these initial inequalities persist even in the steady state. In other words, the "separation of distribution and production" does not hold, not only in the short run but also in the long run. Moreover, contrary to the former classical assumption, the conclusion is that inequality leads to lower productivity and lower growth rates.

Of course, this kind of debate has been stimulated by the observation of unequal distribution of income and wealth in the developed countries since the end of the 20th century, and it has become active in order to clarify it theoretically. In a sense, it appears to be a revival of the classical concern with the relationship between distribution, production, and growth, but in a reversed form at the conclusion level. This is not to say that these arguments deviate from the traditional neoclassical framework. For example, the fact that the absence or incompleteness of capital markets is focused on as an important factor causing inequality leaves room for the argument that if capital markets become sufficiently efficient, the "separation of distribution and production" will be established, and the problem of inequality will be less severe, if not eliminated. If you think about it that way, the new theory of inequality is not as anti-Kuznets curve as it seems.

If we assume that the mechanism behind the Kuznets curve is not only the increase of productive forces but also the further penetration of the free market economy (in the overlapping generations model, the capital market may be more of an equalizing factor), then the inequality trend since the end of the 20th century can be explained by the fact that the development of the capital market has not sufficiently caught up with technological innovation, in other words, as the result of the immaturity of the market economy.

From this point of view, when it comes to the question of "capital market reform or fiscal redistribution," the former is probably the better choice. This is because fiscal redistribution is a policy intervention that distorts the market from the outside and is a bad idea from the perspective of maximizing social production. If efficiency rather than fairness and maximization of production rather than equality are important policy goals, the best response would be to develop capital markets. If we want to alleviate inequality, or at least improve the situation of the poor, we should not redistribute at the level of human capital, such as education, but at the level of final consumption, such as livelihood assistance, according to this line of thinking.

However, many of the theorists in the "inequality renaissance" - including the young Piketty - did not necessarily take such a position. Many of them call for redistribution through public policy, especially the public provision of human capital, as a practical policy guideline. In this sense, this trend is not a departure from neoclassicism, but it is a revival of classical or perhaps even Marxian concerns. Why is that?

Does "human capital" fit into the capital market?

The reason is that most of these theorists, first of all at the empirical level, regarded the wage gap and the labor income gap as the main part of domestic income inequality in developed countries, and the gap in income from physical and financial capital (interest, dividends, etc.), the gap in capital ownership itself, and the gap between those who own capital and those who do not own capital as rather secondary. At the level of theory, human capital was naturally found behind the disparity in labor income, but human capital, that is, investment in human labor capacity, knowledge and skills, was considered to be an order of magnitude more difficult to trade in the market than physical capital.

Although there are various patterns of market transactions for physical capital, the "capital market" depicted in previous theoretical models has basically been a market for lending and borrowing capital, not for buying and selling capital. The reason why it is based on lending and borrowing rather than buying and selling is that capital is too expensive to buy and sell in its entirety. In the modern era, the typical "capital" of a company is divided into partial ownership in the form of shares, and when a company or a large piece of equipment is sold or bought in its entirety, banks and other institutions usually intervene to lend and borrow funds over a long period of time.

On the other hand, the situation is very different when we consider the market transactions of human capital. First of all, human capital is a person's ability, and as knowledge and information, it is something that a person can use, so it cannot be separated from its owner and distributed as a "thing". Also, in a world where slavery is illegal, human capital cannot be traded, even with the people who invested it. In other words, there is no corresponding "stock market" for human capital. It is not enough to say that the market for human capital is a lease market, as in the case of land and capital. There is usually no market for buying and selling the whole thing. (Unfortunately, this book omits the issue of slavery from its consideration.) The market for the lending and borrowing of human capital itself is a "labor market" in the most ordinary sense of the word. As I have suggested before, wages and salaries are not the price of the commodity labor power, but rather the rental price of human capital. In addition to this, a market for financing the investment cost of human capital can be considered. In other words, a market for financing the cost of education and training, i.e., a market for student loans (including interest-bearing scholarships).

Of course, the cost of education and training can be paid as a short-term transaction, as in the case of parents who have the financial resources to pay their children's tuition immediately, but when the individual bears the cost, there is a limit to what he or she can earn by working while undergoing education and training, and thus dependence on loans is inevitably high. Since we in the developed world live in a world where primary and secondary education is usually financed by the public, it is difficult for us to be familiar with such a concept. However, if we think about higher education, the model of "tuition paid for by student loans" is not so far from reality of today's developed countries. This is why the focus of the following discussion will be on the student loan market.

Of course, market imperfections should not be a problem only for the student loan market. Investment that contributes to intensive growth, that is, to technological innovation, whether in physical capital equipment or in human knowledge and skills, is a bet on an uncertain future, and to that extent it must be beyond the reach of private markets. However, even in the case of established knowledge and skills that are not new technologies, the uncertainty of whether or not they can be acquired by a real person is much greater than in the case of physical capital investment. Even in the case of seemingly simple physical labor, which can be done by anyone with a human body, there are certain aptitudes, including not only physical strength and health but also temperament. Some of these aptitudes cannot be adequately assessed by preliminary examinations alone, but become apparent only after a certain trial period.

In addition, when borrowing money to finance a physical investment, the capital goods of the target investment can be used as collateral, but the person's body cannot be used as collateral under modern law. For example, it is not possible (beyond a gentleman's agreement that has no legal effect) to bind a former employee who quits after studying abroad with the company's money to change jobs or start a new business, saying, "You must work for the tuition I paid for you." This is, in essence, a appropriate way to prevent enslavement, but it also raises the risk of loan losses in human capital financing. In order to avoid this, lenders have to be very restrained.

With the above in mind, let's think about school loans. Suppose that school education exists, but is run on an independent profit basis, and that the cost of education must in principle be borne by the person receiving it. In that case, the student has no choice but to either pay immediately or borrow money from the capital market. However, because of the risk of default, the financial institutions that provide the loans have strict screening procedures to determine the ability and sincerity of the borrower students. This will naturally be costly, and even if they do, it doesn't mean that they can make perfect predictions. Moreover, unlike in the case of real investment loans, the investment object cannot be used as collateral, so human investment loans, that is, school loans must demand a higher interest rate than real investment loans.

Let's think of a world of perfect information, where there are no unknowns and where we can predict the future with certainty. In our imperfect information world, the supply of funds is insufficient compared to this ideal perfect information world, and human resources are not fully utilized - people who could have earned more through education have to give it up and settle for a lower income. This means that inequality is not being improved through education, and human resources are not being used effectively.

The question is, how close can we make the bumpy real world, and the use of its resources, to this ideal world? Of course, this comes at a cost. These are mainly what we call "transaction costs," the cost of information processing, which is either non-existent or negligible in an ideal world of perfect information, and the effort and resources required to gather information and make optimal decisions. The capital markets for innovation and human capital investments, no matter how sophisticated, are plagued by these negligible transaction costs and cannot be very efficient.

What about fiscal means, i.e., the government financing the cost of this human investment through mandatory taxation and free transfers - i.e., public education - rather than through voluntary transactions in a free market? From the perspective of textbook economics, taxation and income transfers based on taxation distort the price mechanism (taxing wages has the effect of lowering wages for workers, which affects their willingness to work, and taxing certain commodities, such as alcohol and tobacco, has the effect of raising prices, which reduces their demand. On the other hand, the taxation of certain products, such as alcohol and tobacco, will also have the effect of raising their prices and reduce their demand. (Conversely, the labor and consumption behavior of those who receive transfer payments and increase their income will also change.) However, these disadvantages and negative effects are only in comparison with an ideal world of perfect information. What we should be comparing and contrasting here is the private-sector-led human investment finance market that we saw above. It is also significantly inefficient compared to an ideal perfect information economy. The question is, which is more inefficient (compared to a fully information economy)?

In addition to this, there is the issue of external economies of human investment ("economies of scale" in team production), as mentioned in Chapter 5. In an economy with a steady state of positive growth accompanied by sustained technological innovation, the existence of knowledge and skill spillovers has been a key factor. And it has long been argued that these externalities are more pronounced in human capital than in physical capital. Typically, it is the effect of team production that has been the focus of attention in labor economics. Of course, a team is typically a group of people who work together in a single workplace without a market. However, essentially the same effects of team production often work beyond the walls of the workplace or even the company. This is often referred to as "regional industrial agglomeration," which has been the focus of attention in management science and geography. This is a situation in which a large number of companies in the same or related industries are clustered in a neighboring region, competing with each other as rivals, while at the same time engaging in insurance and mutual aid activities and technical cooperation on an informal and routine basis. In the context of the "inequality renaissance," attention is focused on such industrial clusters, or on general purpose technologies such as basic literacy and information and communication technologies, and the external economy of human investment is not limited to team production at the corporate or workplace level. The external economy of human investment is not limited to team production within a company or at the workplace level, but is also considered to be a "Marshallian externality" or "network externality" that operates at the level of external markets and the macroeconomy. These "externalities" are considered to be typical examples of "market failure" as well as so-called "public goods" - goods and services that are difficult to establish strict ownership rights for and are easily available to those who do not pay for them - and environmental destruction ("pollution" is the opposite of "public good(s)" - public bad(s)). As we have already discussed, it is also known as a classic example of "market failure". In a situation where we can both "ride for free" on the fruits of other people's efforts, no one is going to work hard enough on their own.

Will effective redistributive policies be politically chosen?
Because of the strong uncertainty and externality in human investment, it can be argued that the strategy of "providing education and training to those who cannot afford it through mandatory redistribution by the government" is more effective than the strategy of "building a more sophisticated capital market that overcomes the uncertainty inherent in human investment. It was found that the strategy of "providing education and training to those who cannot afford it through mandatory government redistribution" would result in more efficient use of human resources, maximization of social productivity, and higher growth.
On the other hand, it is even more interesting to note that in the 1990s, the theorists of the "inequality renaissance" went beyond the policy argument that redistribution and public education could be effective, to the political argument that effective redistributive policies could be politically chosen to generate growth.
One of the milestones in the early theoretical examinations of inequality and growth in the mid-1990s was the 1994 paper by Alberto Alesina and Dani Rodrik. In this paper, a simple mathematical politics - a simple mathematical model of democratic voting - is used to make the following cogent argument.

In a democracy, a one-person, one-vote system, the "median voter" often holds the casting vote and determines the outcome. Let's consider the simplest case. If, for example, tax rates are the target of policy choices, then the menu of policies surrounding them, and the voters' support for each policy, can be ordered on a single line. When we order the voters from those who support the lowest tax system to those who support the highest tax system, the "median voters" are right in the middle. Suppose there are two political parties competing for the support of the voters from opposite sides. Let's say there are two parties vying for the support of voters from opposite sides: the tax hike party and the tax cut party. If the party with the most support wins the government by a simple majority vote, it is clear that both parties can win a majority by getting the middle voters to support their party. This is called the "median voter theorem.

If the Median Voter Theorem is valid, the question is, "Where does this median voter stand socially in terms of the distribution of income and wealth? Of course, there are many possible measures of distributional inequality, but here, we will simply assume that when people are lined up on a line according to their income (or wealth, but for now), the income of the " median-income citizen" - the person who is in the median of the income order - corresponds to the "median voter" mentioned above. If the average income is higher or lower than the average income, then the average income is higher or lower than the average income. If the average income is higher than the middle income, then inequality in the country or society is high, and if it is the opposite, then inequality is low. In other words, if the average income is higher (or lower) than the middle income, then the income of the top earners is higher (or lower) than the national income. And as is well known, the pattern of the distribution of income and wealth is basically higher for the mean than for the median.

On top of that, let us consider a tax system that is not for public goods, but purely for income redistribution. For simplicity's sake, consider a system in which taxes are levied on those with higher than average incomes, and subsidies are given to those with lower incomes. What happens if we try to determine the tax rate for such a tax system by democratic vote?  If we apply the Median Voter Theorem in a straightforward manner, the tax rate desired by median-income earners will be the one that the party seeking to win the election promises to implement. In this sense, the more unequal the income is, the higher the tax rate and the larger the redistribution will be. This means that in a country with a well-run democracy, the more unequal the original (pre-tax) income, the higher the tax rate, the more distortive the effect on the economy, and the lower the growth rate.

This argument is interesting as it is, of course, but it is rather in the minority of the trend we are calling here the "inequality renaissance". In the Alesina and Rodrik paper, the mechanism that leads to inequality and low growth is not inherent in the market economy, but in democratic politics and state power (which is, to use a cliché, "neoliberal"). The majority in the "inequality renaissance" would rather see the mechanisms of inequality in the market economy, as we saw earlier. They argue that the "lack or imperfection of capital markets" is "natural" rather than the result of distortions caused by government intervention, and that it is the development of markets to overcome this that requires government intervention and policy support. However, no matter how rational such policies are from an economic point of view, whether they will be politically chosen and implemented is another matter. As for the Alesina and Rodrik paper, regardless of its conclusions, I must commend it for its focus on the political process that defines policy. (For the record, the authors' own policy recommendation in this paper is not to "stop democratic redistribution from getting in the way of economic growth," but to "promote redistribution at the asset level, since asset inequality is negative for growth."

In 1996, Roland Benabou wrote another often-referenced paper that used a similar political model with the Median Voter Theorem, but also adopted the previous model (2)' for the basic structure of the economy (in contrast to the model(1) in the Alesina & Rodrik,), and asked, "In what cases does democratic redistribution increase the growth rate and in what cases decrease?" After all, the point here is that in an economy with a positive steady-state growth due to the spillover effect of human investment, the situation is more complicated than expected in the Alesina and Rodrik paper. Benabou's own conclusion is that, under some conditions, redistribution through democratic decision-making can result in both decreased inequality and improved growth.

Is Inequality Evil? From Rousseau and Smith to Piketty

At this point, let's review what we have seen so far.

Let's review the confrontation between Rousseau and Smith in the latter half of the 18th century, when Rousseau asked "Is it morally acceptable that the system of private ownership and the division of labor under it create and even maintain and reinforce inequality?" I have interpreted that Smith's reply was such that it is acceptable because, when combined with a market mechanism that allows people to freely trade their property with others, it can raise the level of overall wealth, if not eliminate inequality. In this interpretation, Smith did not dare to read Rousseau's question as assuming that inequality itself is evil, but rather that he thought that inequality is evil because it is the result of the exploitation of the poor by the rich, and answered that, if inequality makes the poor richer, then the criticism that inequality is evil because it is exploitation is unjustified.

However, in the view of Smith and his successors in the so-called classical economics school of the 19th century, the poor, specifically wage earners who have no production facilities, that is, capital goods of their own to operate, are limited in their ability to become rich under a market economy, and the possibility of the working class as a whole becoming rich to the point of owning its own capital is largely unforeseen. Marx's critique of the market economy, crystallized in his *Capital*, can be said to have exploited this point. Since the working masses, the majority of society, cannot accumulate property and become capitalists by themselves, that is, they cannot work on their own initiative and leadership, the capitalist market economy is a civil society that treats all people equally in legal form, but in reality it is a class-ruled society where only a few capitalists can be called "free citizens", Marx criticized. Furthermore, Marx believed that capitalism not only aggravates inequality, but also wastes the productive forces of both capital goods and labor power by repeating the business cycle of boom and bust, weakens competition by creating a small number of monopolies in the market, distorts the market mechanism, and stagnates economic growth. This is why Marx advocated the abandonment of the capitalist market economy itself.

Neoclassical economics, which was established at the turning point between the 19th and 20th centuries and later became the mainstream of economics, responded ambivalently to Marx's challenge. On the one hand, neoclassical economists responded to the possibility that inequality would reduce productive capacity and stagnate the economy, as Marx had argued, by saying that distribution and production could be separated if the capital market functioned properly, thus denying the limits of growth under a capitalist market economy. In a sense, this was a reiteration of Smith's answer to Marx's argument, which was a repetition of Rousseau's critique. At the same time, it also shifted the issue in a sense, as Smith's answer to Rousseau did, i.e., it sidestepped the question, "Is not inequality itself an evil?" Both Rousseau and Marx were not only concerned with "affluence" and the well-being of consumer life, but also with the opportunities for active participation in civil society - in Rousseau's case, political participation as a sovereign, and in Marx's case, in addition to that, the active participation of workers in economic life, labor and business management. If this is the case, "the working masses can be prosperous even under the capitalist market economy" alone is not a sufficient answer.

In the first half of the 20th century, there was a counterattack against Marxism from the neoclassical school in the form of the "Socialist Economic calculation Debate," which I did not introduce in Chapters 3 and 4. The Marxists rejected the capitalist market economy and proposed as an alternative a "centrally planned economy" in which the central government formulates a plan for the management of the entire economy and orders the people to follow it by administrative means. In fact, after the Bolsheviks came to power in the Russian Revolution, this concept was put into practice. In response, some neoclassical economists, most famously Ludwig von Mises and Friedrich von Hayek, boldly challenged the idea and criticized it, saying that planned economies do not work in practice. This criticism was later shown to have hit the nail on the head. It should be noted, however, that while this argument was certainly correct as a critique of some forms of socialism, including Marxism, it was a critique that "the alternatives to capitalism that socialists offered were worse than capitalism" and it does not necessarily mean that "the socialists' criticism of capitalism and its flaws were wrong." (There is room to argue that socialists' critique of capitalism is wishful thinking.)

But if we look at neoclassical economics as a whole, it was not the only response to Marx's critique. Marshall's typical view was that if we look at "human capital," it is possible for workers to accumulate capital, and that this is positive for the growth of the economy as a whole. Of course, there is a danger that this will degenerate into a slavish affirmation of the current state of capitalism. In other words, it could lead to the argument that those who are still poor are merely poor for lack of effort (or the exceptional bad luck of a few), because everyone has the opportunity to accumulate human capital, not to mention physical capital. However, since the end of the twentieth century, a new theoretical search in the face of the renewed growth of inequality in developed countries has led to the assertion that there are structural limits to the market financing of human capital, and that policy intervention is essential.

The young theorist Piketty came into the world on the back of this last trend. However, Piketty himself dared to step out of this trend and embarked on empirical research, searching for materials, compiling databases, and applying statistical analysis rather than tinkering with mathematical models. In other words, he became a leading figure in the empirical side of the "inequality renaissance." The question is, what was the significance of this development? It does not seem to be a simple "shift from theory to evidence". Piketty's empirical research since at least the 2000s has not been about verifying the theoretical findings obtained until the 1990s in light of data, but rather it has been about demanding further explanations based on new theories. However, Piketty himself, who was once a promising theorist, has yet to explicitly present the new theory that is required.

Why is that?　Let's look at it again in the next chapter.

## Chapter 9: Thomas Piketty's Capital in the 21st Century

What did Piketty discuss in "Capital in the 21st Century"?

There are already many excellent introductions to the great book *Capital in the 21st Century*, so I will not give a glimpse or summary of its contents here. I will simply present the main points.

First, Piketty concentrates his attention on the gap in physical capital and the gap between those who have it and those who do not, rather than the gap in human capital. In the "inequality renaissance" we saw earlier, the focus was on the general disparities within the developed world, especially the decomposition of the middle class and the impoverishment of the lower classes. This is why the focus was on wages and human capital inequality. If the vision of the theorists of human capital and inequality in the 1990s, including Piketty himself, was similar to the "generational overlap and endogenous growth model with incomplete capital markets", (2)' in this book, the vision of *Capital in the 21st Century* seems to be closer to (3) "the Ramsey model with perfect capital markets" or (3)' "the Ramsey endogenous growth model with perfect capital markets.

Also, as a matter of empirical technique, it is difficult to statistically capture the reality of the very small minority of asset owners. Consider, for example, conducting a social survey in a country with a population of about 100 million. With careful random sampling, one hundred is not enough to get a clear picture of the minority groups that make up about 10 percent of the population in the country. However, a sample of about 1,000 should be sufficient.

However, the large group of wealthy people who do not work and can live a luxurious life only on their asset income is one percent or less of the population, and cannot be successfully picked up by a random sampling survey. It is expensive for social scientists to conduct their own questionnaire surveys to collect samples, rather than using existing statistics from government agencies. The information and communication revolution has revolutionized the processing of data, but collecting data one by one is still very physical work. A sample size of 1,000 is almost the upper limit for a proper survey that can be conducted by individuals or small groups. In such a situation, Piketty searched for data from existing government statistics, especially tax statistics, that could capture the trends of wealthy people, and processed them for statistical analysis (matching units, removing obvious errors, and estimating missing parts) to produce the international comparative study that resulted in *Capital in the 21st Century*.

Piketty thus argues that "if we focus on the top 1 percent, the rapid increase in the share of asset income is remarkable," where conventional researchers focus on the top 10 percent and argue that "the main cause of the increase in income inequality is the expansion of labor income, as seen in the compensation of the top managers of large corporations".

Second, the trend to focus on human capital was generally friendly to the "Kuznets curve" view of history. The most common view is that "the mechanism of the market economy itself cannot be said to increase or decrease inequality in general, and there are various possibilities depending on the circumstances - the external environment, production technology, and the institutional framework of the economy and society". The focus on the widening gap since the end of the 20th century, which appears to be a deviation from the Kuznets curve, did not force a major rethinking of the earlier trends that Kuznets pointed out.

Piketty, on the other hand, urges a major rethinking of the Kuznets curve itself. While he does not deny the Kuznets curve as a phenomenon, he does not believe that there is a consistent economic mechanism or development trend of production technology behind it. If anything, Piketty believes that the market economy itself tends to preserve or even increase inequality. As for the trend of decreasing inequality in the developed countries in the mid-20th century, as suggested by the Kuznets curve, many of the "inequality renaissance" theorists focus on the trend of human capital. As we have seen earlier, many of them attribute the increase in wage inequality since the end of the 20th century mainly to the gap in human investment. Specifically, they say that the information and communication revolution has widened the wage gap between college graduates and high school graduates (and below). Whatever the case may be, all other factors being equal, an increase in the supply of a particular commodity, including labor, should lower the price of that commodity. If this is the case, then the wage advantage of college graduates over high school graduates in the second half of the 20th century should have declined, and indeed some such cases have been observed. However, since the 1980s, or more clearly since the 1990s, it seems that the skills to use personal computers and the Internet as "general-purpose technologies" mentioned in the previous chapter have led to a considerable wage gap. For at least the last 20 to 30 years, not only has the need for information and communication technology created a demand for college graduates, but the increased supply of college graduates has not immediately lowered wages, but has created a virtuous cycle of further information and communication technology innovation, its spread throughout the economy, and the resulting need for people who can adapt to information and communication technology. This is a virtuous cycle that continues. (This is called "skill biased technical change".)

Of course, there are implications as to how this should be interpreted, but this skill biased technological change is also understood as a phenomenon of intensification of human investment, as evidenced by the focus on "general-purpose technologies. In other words, the wage gap under the information and communication revolution is mainly a side-effect of the "free ride" phenomenon caused by the externality and spillover effects of human investment, and it can also be a chance to reduce the gap and increase the growth rate at the same time, if appropriate public policies are implemented.

The most obvious example of this type of argument is the series of work being carried out by Oded Galor. In the 1993 paper I mentioned earlier. Galor has committed himself to a project that he calls "unified growth theory," in which he has, somewhat anachronistically, developed a theoretical model of economic development that has been around since the beginning of human history (!)..

Galor integrates the Industrial Revolution and the population transition (from high fertility to low fertility) as major milestones in economic history. Prior to this, economic society had been trapped in the same mechanisms as depicted by Malthus's theory of population. Since the beginning of hisotory, the results of productivity improvements, such as the development of new resources and new technologies, have been absorbed by population growth and have not led to higher living standards per capita or continuous technological innovation. This is commonly referred to as the "Malthusian Trap". However, it is argued that in the 18th and 19th centuries in Western Europe, productivity growth began to be channeled into continuous productivity growth, including improvements in living standards and human investment in education. This is the first stage of Galor's research plan, so to speak.

The second stage is about the changes that have occurred since industrialization began in earnest. Galor roughly divides the means of production, or wealth, into three categories: land, which cannot be increased in value by investment; physical capital, which can be accumulated by investment but has no external effects; and human capital, which can be accumulated by investment and has external effects. In the process of economic development, the main players of growth shift from land in the early stage to physical capital and human capital, and the distribution of income and wealth is equalized along with the shift.

However, Galor's research is still in progress, and while the first stage is somewhat complete, the second stage is still in its infancy. In the case of the vision of "shifting from physical capital to human capital," we can say that it is (or should be) more rational from the perspective of production and growth, but we have not yet reached the point where we can prove that it is actually becoming so.

But, Piketty in the 21st century clearly distances himself from these trends that he used to be committed to in the 1990s. He does not believe that the reduction in income inequality in the mid-20th century was the result of an increase in the weight of human capital in growth. He believes that the two world wars of the first half of the 20th century and the prolonged recessions in between resulted in the full-fledged welfare state system, as compensation for the wartime mobilization of the working masses and their cooperation in the war, and as part of the recovery from recession and war. Piketty calls it "social state". Piketty calls it the "social state," not in the English-speaking or French-speaking way, but in the German-speaking way. In addition to livelihood assistance for the poor, public medical insurance and old-age pensions for all citizens were established during the interwar and wartime periods in terms of systems, and in terms of substance during the high growth period after the war. Not only compulsory and free primary and secondary education, but also public support for higher education can be included here. (A good example is the support for demobilized soldiers to go to college in the United States after the war.

More importantly, labor unions and workers' political parties were officially incorporated into the political decision-making centers that supported such welfare state policies, and labor unions established their official status as a party to negotiations over wages and working conditions in the field of civil society. In this understanding of the 20th century, Pikkety is more like a political scientist, sociologist, or some kind of Marxist rather than a neoclassical economist.

It should be clear, then, what Piketty thinks is the main cause of domestic inequality in developed countries since the end of the 20th century. In other words, the so-called "welfare state crisis" and the rise of the "neoliberal" policy line since the 1970s, the fiscal crisis and the austerity macroeconomic policies to cope with it, and the drastic reduction of the influence of labor unions through privatization and deregulation of state-owned enterprises and public utilities are the main causes of the widening gap. In other words, it is only because the control of the welfare state has been removed and the "ground metal" of the capitalist market economy is peeking out again. Piketty clearly rejects the optimistic scenario that "the main driver of growth shifted to human capital, hence the narrowing of inequality until the middle of the 20th century, and the reversal of this trend with the information and communication revolution."

Piketty emphasizes the positive effects of inflation

Third, and closely related to the above-mentioned view of the history of the welfare state, Piketty attaches great importance to the development of inflation in the 20th century. While institutionally it had its beginnings in the turn of the century when neoclassicism emerged, the welfare state as a real political system was not established until after World War II. It is clear that the welfare state has been considered as an inseparable part of Keynesianism, as evidenced by the fact that the term "Keynesian welfare state" was later established as a key term in political science and sociology, if not economics. What is noteworthy about Piketty's argument is that he does not take a negative view of inflation, which has traditionally been regarded as a "side effect" of the Keynesian welfare state, considered a chronic disease for the welfare state due to various factors such as the government's economic stimulus policies, deficit-based fiscal spending, and pressure from labor unions to raise wages. In fact, since the fiscal crisis of the 1970s, inflation has been the main target of neoliberal policies. However, as I will explain later, Piketty does not pay much attention to the macroeconomic policy debate from Japan's "lost 20 years" to Abenomics, or the debate over monetary easing in the United States after the Lehman Shock. He does not seem to have a definite opinion yet. Regardless of such economic policy debates, Piketty gives a not necessarily negative assessment of the development of inflation in the 20th century in the context of the long-term history of income and wealth distribution. To put it bluntly, inflation has contributed to a significant improvement in distributional inequality through a significant depreciation of asset prices.

Simply put, inflation is a phenomenon in which a large supply of money reduces the purchasing power of money and conversely increases the prices of goods - not just individual goods, but almost everything. This means that people have come to expect that prices will continue to rise, not just in the short term, but at least for some time to come, and that there is no telling when it will stop.

If inflation were a general rise in prices in the strict sense, it would not seem to be a serious problem in the real economy, since the relative prices between things would remain the same, but this is not the case. The exchange ratio between all things and money, money, changes. As long as inflation continues, the value of money will decline in the long run, and holding on to cash will mean losing money from time to time. Therefore, inflation increases the willingness to buy, whether for consumption or investment.

More importantly, inflation has the effect of redistributing income and wealth. In the debt-credit relationship, inflation causes a decrease in the real value of debt, which reduces the debtor's repayment burden. In other words, there is an income transfer from the lender, the creditor, to the borrower, the debtor.

What does the income transfer from the creditor to the debtor mean? Of course, this is not simply a transfer from the rich to the poor. Roughly speaking, asset holders of low-risk assets, typically land, and financial assets such as fixed face value bonds, especially government bonds, will lose. The same is true for pensioners and retirees who live off their savings. The same is true for employed workers - full-time employees of large corporations and full-time public servants - who have a stable position with no fear of unemployment and no decrease in salary. In contrast, the beneficiaries are debtors in general, including companies that issue bonds or borrow money by other means to keep their businesses running. Small and medium-sized businesses and farmers are usually the same. The same would be true for former students who are still repaying their scholarships. Even workers can benefit from this if they have a mortgage on their home. (Of course, if you have paid off your mortgage or inherited it from your parents, you lose.) The question of who owns stocks as a risk asset is a difficult one. It is difficult to say because, on average, the value of stocks does not correspond to prices per se, but to the growth of the economy as a whole, adjusted for prices. However, it is clear that it is rather easier to invest under inflation because debtors have an advantage, that is, it is easier to borrow money. If investment becomes more active, then if there are unemployed people, they will have a better chance of finding a job.

Roughly speaking, under inflation, landlords, interest earners, retirees, and stable workers will lose, while business owners, whether they are large corporations or small self-employed farmers, and those with scholarships and other loans will gain. This means that even a salaried worker with stable employment, if he or she is burdened with mortgage or scholarship repayments, the net result could go either way. And above all, the unemployed and precariously employed will benefit insofar as they will have more chances to find or change jobs.

This is enough of a generalization about inflation for now, but rather than this generalization, Piketty focuses on a particular period in the first half of the 20th century, the period of the world wars and the Great Recession. In this period, a series of major events that placed a heavy burden on the fiscal and financial resources of the state, such as the wars and the Great Recession, changed the nature of the state itself - both the political participation of the masses from below and the mobilization and control of the government from above were greatly deepened - as well as the civil society and the private economy. In addition, the civil society and the private economy were also changed. In addition to the recognition of labor unions and their integration into the system that I mentioned earlier, the expansionary fiscal spending and the expansionary monetary policy to support it were accompanied by severe inflation (on the other hand, the Great Recession was deflationary), which depreciated the value of previously stable assets such as cash, land, and government bonds. According to Piketty, this inflationary depreciation of assets and the hidden transfer of income to debtors is a key factor in the dramatic reduction of income and wealth inequality in the 20th century.

Historically, "r > g" has been the norm.

Finally, there is another kind of growth pessimism that has been popularized by the example "r (interest rate) > g (growth rate)". As Piketty himself notes, this inequality is not a theoretically inevitable law, but a common empirical trend. However, if you think about it for a moment, you can easily see that this is a very likely situation. After all, money lending has existed since before the Industrial Revolution, before the start of full-scale sustainable economic growth, and the term "loan sharking" has been around for a long time. In other words, interest rates were already normally positive in a zero-growth world, so "r > g" is completely normal in historical terms.

The exception to this is the 20th century, when the growth rate exceeded the interest rate for a relatively long period of time, the high growth of developed countries after World War II. These are the cases of the so-called NIEs (newly industrialized economies, from the "Four Little Dragons" of Hong Kong, Singapore, South Korea, and Taiwan to Thailand, Indonesia, Malaysia, etc.), which rapidly caught up with developing countries and became middle-income countries after the end of the 20th century, and in the 21st century, the BRICS (Brazil, Russia, India, China, and South Africa). These high growth rates, together with inflation, have contributed greatly to the reduction of relative inequality and, of course, more importantly, to the overcoming of poverty in an absolute sense, but in Piketty's view, this is a transitional phenomenon. The high growth of Western Europe and Japan after the war included the initial phase of post-war reconstruction, that is, the return to the original state before the war destroyed them, so it is natural that it would be faster. (Goods and lives were lost, but knowledge and technology were not.) What is particularly important for the NIEs and other less developed countries is the change in the international economic environment since the 1980s (transition to a floating exchange rate system, etc., as we will see later) and the shift to an open economy and the lifting of controls following the fall of socialism.

Until the 1970s, many developing countries followed a relatively closed, self-reinforcing path of economic development with strong state control, based on the development philosophy of the early postwar period, which was heavily influenced by socialism. The strategy of "import-substitution industrialization" was influential, based on the fear that under an open economy, the country would be exposed to imports from developed countries and multinational corporations, and would not be able to industrialize on its own with national capital. However, the "welfare state crisis" in the developed countries, the progress of neoliberal deregulation, and the fall of socialism led to a shift from an "import-substitution industrialization" strategy to an "export-led industrialization" strategy. The nationalistic vision of "creating a full set of industrial structures at the national level to make the country self-sufficient" was abandoned, and the country somehow found a place in the international division of labor network - according to the principle of "comparative advantage," even if a country is absolutely unproductive in every sector, there is always a sector where it can secure a trade advantage. - the aim is to develop export industries. Multinational corporations from developed countries, which were once avoided, became actively encouraged to enter the country, aiming not only for short-term job creation, but also for the long-term benefits of technology transfer (through the development of human resources and related local companies). Even under the inappropriate policy of "import-substitution industrialization," in the case of countries with a certain level of education and infrastructure, such open economies might be surprisingly effective and achieve remarkable growth in a short period of time, enjoying the "benefit of lateness" of adopting the latest technologies of developed countries.

However, this rapid growth will end when the war-devastated countries return to their original state and the developing countries become middle developed countries - catching up to the advanced countries. In other words, this kind of rapid growth cannot be a steady state, but merely an adjustment period until it is reached. Therefore, this kind of extreme high growth cannot be expected, at least in developed countries. As a general rule, it is not possible to expect such an extreme high growth in an economy where the "dual structure" I mentioned earlier has been dissolved. Then the economy would return to the very "r > g" normal.

Piketty does not deny the possibility or desirability of growth. In fact, Piketty believes that growth is necessary to prevent the widening of inequality. However, since the 21st century, the potential for rapid growth through catch-up has been shrinking, and countries that have achieved a certain level of industrialization and market economy no longer have much hope of achieving rapid growth beyond "r > g". After the two major waves of the world war, the transition to the welfare state, and the rapid catch-up of the war-devastated and less developed countries have passed, the tendency to preserve or increase inequality as the "ground metal" of the capitalist market economy, as shown by "r > g," will come to the fore.

As the theorists of the "inequality renaissance" of the 1990s predicted, if the gap in capital ownership itself were to narrow and the contribution of human capital to growth were to increase more and more, the disparity between the two would disappear. Piketty's empirical research in the 21st century raises serious questions about this possibility.

(Incidentally, one can read into this argument Piketty's interpretation of the converging trends of global inequality in Capital in the 21st Century, which focuses on domestic inequality in developed countries as a whole.)

Four Essences of Piketty's *Capital in the 21st Century*

Of course, it is impossible to narrow down the essentials of a large book filled with a wealth of issues to the four points listed above, but I will leave an adequate summary of *Capital in the 21st Century* to the many commentaries that have already appeared. I will pick up these four points in the context of this book.

The first is the focus on physical capital, not human capital (although, unlike the classics, *Capital in the 21st Century* treats both land and physical capital collectively as "capital," "assets," and "wealth"). Piketty does not neglect human capital and education from a policy perspective, but he does not consider them to be the key to success. The main cause of domestic inequality in developed countries is rather the distribution of physical capital, and as a countermeasure to this, the emphasis is on the redistribution of capital income, as seen in the well-known global asset taxation. In *Capital in the 21st Century*, Piketty evaluates human investment, investment in knowledge and skills, as a power to reduce inequality, but he is critical of the use of the term "huma capital", saying that there is a tendency to overvalue them as "property" that can create inequality, and as power that can compete with and surpass physical capital.

The second, the development of the "Keynesian welfare state" in the 20th century and its decline with the rise of so-called "neoliberalism," is in line with the current perception of political scientists and sociologists (including those with strong Marxist influences). However, the third point, which is closely related to this, is the focus on the redistributive effect of inflation, which is something that only economists can see.

If we pay attention to these three points, the meaning of Piketty's shift from the 1990s to the 21st century becomes somewhat clear. First of all, what has been consistent since his days as a mathematical economist in the 1990s is a return to a classical awareness of the problem in the broadest sense, that is, to take up the issue of the relationship between distribution, production, and growth head-on. In addition, his concern with not only competition in the marketplace, but also the policies of the centralized state and the political mechanisms that determine them, is a legacy of classical and Marxist thought.

On the other hand, Piketty, along with his mathematical models, seems to have given up on a simple vision of the relationship between distribution and growth. In contrast to the classical school, the theorists of the "inequality renaissance," of which Piketty himself was a part, argued that the equalization of distribution could be positive for growth. Such theoretically clear-cut arguments seem to be something that Piketty, a statistical empiricist since the 2000s, has rather tried to avoid. Of course, not only theorists but also empirical researchers have spoken out about the relationship between equality and growth, and in recent years international organizations such as the OECD have also attracted attention by publicly stating that inequality reduces growth. In the midst of such a trend, which he himself promoted, Piketty can be said to be rather ascetic. One of the reasons, I suppose, is that as the importance of physical capital rather than human capital as a cause of inequality has become apparent, it can no longer be said that the equalization of its distribution contributes to growth in the same way as human capital, which is expected to have spillover effects. Of course, as I mentioned earlier, Piketty in his book *Capital in the 21st Century* also believes that the spread of knowledge and skills has the power to reduce inequality. However, he does not expect it to have the power to surpass the disparity caused by inequality in the distribution of (physical) capital as "human capital.

The third point is also an important hint. The economic theories of inequality, growth, and redistribution of human capital that reached a certain level in the 1990s, including Piketty's, are basically supply side, real side, and real economy theories, lacking any analysis of the monetary, or even demand side of the economy in the Keynesian sense. Therefore, they cannot analyze recession, involuntary unemployment, and the relationship between recession/unemployment and inequality. This is not just the case with the inequality theories of Piketty and others, but with endogenous growth models in general, which remain at that level. Today's macroeconomic models, including not only the endogenous growth theory but also the real business cycle model, which is the mother of the endogenous growth theory, are called dynamic stochastic general equilibrium (DSGE) models. It is quite difficult to incorporate money in a theoretically consistent manner. (Incidentally, the analysis in Chapter 7 uses a "dynamic general equilibrium" model without the "probability" component.)

This macroeconomic theory based on the DSGE model started with the "Rational Expectations Formation Revolution," which is closely related to the neoliberal policy trend, but it has been getting a very bad reputation as "useless" in recent years, especially with the resurgence of the Keynesian world after the Lehman Shock. Prior to the "rational expectations revolution," mainstream (old) Keynesian economics, as we will see later, faced with the challenge of recession, looked to money and the unique character of the monetary side of the economy for its cause. However, they failed to provide explanations that go back to the assumptions (called "microfoundations") of economic theory, i.e., the usual market mechanism and rational choices of economic agents moving in search of profit. Of course, in science, theory is for understanding reality, and if there is a discrepancy between reality and theory, it is the theory that needs to be corrected.

The problem here, however, is that Keynesian "theory," which seemed to have more explanatory power than the unrealistic existing classical and neoclassical theories, did not connect well with the existing orthodox economic theories, which were significant and realistic enough once you got away from phenomena such as recessions and money. Rather than being a tool that can be used by anyone based on common sense and logic, I suspected that it was more of an ad hoc idea, a "masterstroke" that could only be used by researchers with special talent and experience. Then, as we will see later, if Keynesian policies seem to have lost their effectiveness, the equilibrium-theoretic macroeconomic theory of "rational expectations" that is consistent and coherent with existing economic theory would be more scientifically productive. So now, even when academic economists try to take a Keynesian direction, they try to build a Keynesian model somehow within the framework of the DSGE after the "rational expectations formation revolution". This trend is called "New Keynesian," but it is still in its infancy.

Very intuitively, in the Keynesian world, the adjustment of supply and demand through prices does not work so smoothly. As a result, it takes a long time to achieve equilibrium in the market, and imbalance in the transition period lasts for a long time. (This is called "liquidity preference." In Marxian terms, this is the "fetishism of money.) This is a familiar argument in philosophy and sociology, but it is quite difficult to explain it on the basis of the rational choices of economic agents, and although some proposals have been made, they do not seem to be decisive enough to be established in textbooks. Of course, there are also many researchers who deny the reality of the Keynesian situation itself (and therefore consider this research direction itself sterile and meaningless).

Piketty is aware of the importance of such Keynesian issues as inflation (and deflation) and the resulting disturbances in the market economy, as well as the distributional effects of macroeconomic phenomena due to the existence of money, and he discusses them with great emphasis in his final policy recommendations. For example, he discusses the redistributive effect of inflation, which I mentioned earlier, not only on the correction of inequality, but also on the reduction of public debt, which has become an urgent issue in Europe since the Euro crisis. However, he is not yet ready to go into it in earnest, and is probably at the stage where a rigorous theorization of the issue is very, very difficult.

Piketty, who has been a staunch critic of the current austerity policies that push for tax hikes and spending cuts (not progressive capital taxation, which is his theory, but labor income and consumption taxes) as an active fiscal theorist (it is still fresh in his mind that he severely criticized the consumption tax hike when he came to Japan), is, frankly speaking, not a strong advocate of macro monetary policy. Also, although Capital in the 21st Century frequently mentions unemployment, it does not make a full-fledged subject of recession and unemployment and the impact of unemployment on inequality. Recessions and the resulting unemployment are the most important factors leading to underemployment in both labor and capital, and to a decline in the gross social product and thus in the long-term growth rate, not to mention that they are the factors that widen the gap even more severely in the form of the fault line between the working and the unemployed. Of course, this may be the self-restraint of Piketty as a researcher, who was once a sharp theorist, but in order to deepen his research, he dared to shut down his theories and specialize in empirical research (he judged that accumulation of data would be more productive for the time being due to insufficient theoretical accumulation to rely on after changing direction).

Chapter 10: Just a few steps away from Piketty

What do Piketty's opponents think about inequality?

In this concluding chapter, I would like to turn the spotlight on Piketty's opponents and ask what they would like to see happen in the future. - or rather, what they don't want - and consider other directions, while sharing the same interest in inequality.

Let's start with a slightly twisted comparison between the Piketty of the 1990s and the Piketty of today, the majority of the "inequality renaissance" who emphasized human capital spillovers and the Piketty who turned to pessimism about the effect of human capital on reducing inequality.

Of course, Piketty's rivals, those who are committed to optimism about human capital, have not only been engaged in theory building. Some of the criticisms of *Capital in the Twenty-First Century* have been based on empirical studies of human capital from this very perspective, such as David Weil's report to a session on Piketty at the American Economic Association conference in early 2015. Weil is a well-known growth theorist, whose textbook *Economic Growth* has been translated into Japanese, and who, as a collaborator of Galor, has been involved in the development of a "unified theory" and, with Greg Mankiw, the chair of the session in question, has produced a famous empirical study on the contribution of human capital to global inequality and its convergence. It seems that the battle between Piketty and human capital optimism will continue even at the empirical level.

Since I cannot yet make a final judgment on the outcome of this battle, I will conclude this book with a comparison of the two from a slightly different angle. This is, so to speak, the difference in moral philosophical stance between the two camps of (rather) egalitarian economists.

For those who argue that "reducing inequality and correcting inequality are positive for growth" by expecting the externalities of human capital, such as Benabou and Galor, it would be better to comment on the moral philosophical character of their stance. Their egalitarianism is, at first glance, utilitarianism and welfare in the broadest sense, or a position similar to it.

Let's take a rough look at what "utilitarianism" or "welfarism" means in terms of ethics, moral philosophy, public philosophy and justice theory. The moral rightness of a person's actions is measured by the extent to which they contribute to the increase of the interests of more people, including oneself (or, in the style of Jeremy Bentham, "the greatest happiness of the greatest number"). The criteria for evaluating the merits of social institutions and policies are basically the same.

The original utilitarianism holds that people's interests and happiness can be evaluated and compared using the same universal scale, and that it is possible to derive the "interests and happiness of society as a whole" by aggregating them. In contrast, many neoclassical economists are skeptical about the possibility of comparing and aggregating happiness among individuals, so they are not utilitarians in the narrow sense. However, they often believe that the evaluation of economic actions is basically measured by the benefits they produce, the happiness (utility) of people. This position is called "Welfarism".

The position represented by Benabou and Galor can be simply argued that "equalization of income and wealth distribution is fundamentally good and deserves to be pursued because it contributes to more efficient use of resources, maximization of gross social product, and higher growth". In other words, it is a position that treats equality not necessarily as a goal in itself, but as a means to other goals - in this case, production maximization (which for utilitarianism would be the maximization of happiness and utility for society as a whole). At first glance, this may seem like a position that does not respect equality very much, but in fact it is the opposite - or rather, a little different. It is fine to regard equality as a worthy goal in its own right. At the same time, however, if the pursuit of equality also has the side effect of contributing to other values, such as production and the maximization of happiness, then it will be more persuasive when appealing to people about the significance of pursuing equality. I don't know if Benabou, Galor, Weil, and others are utilitarians or welfarists in the strict sense. But at least they are making arguments that do not antagonize utilitarians and welfarists. They are egalitarians in terms of policy, but in fact they do not explicitly discuss the value of equality itself, nor do they take a stance that does not require such discussion. In this sense, we can even say that they are actually closer to Smith in the composition of the confrontation since "Rousseau versus Smith.

Is Piketty's stance on "equality" wavering?

I am not sure how seriously Piketty takes this, but at the end of *Capital in the 21st Century*, he mentions the names of John Rawls, famous for his critique of utilitarianism, and Amartya Sen, Nobel laureate in economics, who succeeded Rawls in a more profound way.

Rawls not only criticizes the optimism in utilitarianism regarding the comparison and aggregation of utility between individuals, but also distances himself somewhat from utilitarianism and welfarism in his moral evaluation of individual actions and social institutions and policies. For him, influenced by Kant, the basis of moral evaluation is "whether or not it violates people's basic rights and their personal dignity as subjects of rights," before asking "to what extent the action or policy has contributed to people's happiness as a result". Therefore, while utilitarianism focuses on individual actions and policies and their effects, Rawls (and Kantianism as Rawls understands it) focuses on the more general principles that govern all actions, the laws and political principles as general rules that govern individual actions and policies. Rawls argues that in order for everyone to live a life of dignity as a person – as a subject of rights, we need not only a legal system, but also an economic system that guarantees a minimum standard of living (and more, if possible). Although Sen is also an economist, he tries to distance himself somewhat from the welfarist direction, so to speak, and to embody Rawls' concept of rights in a more institutional and policy way.

From the way he writes in *Capital in the 21st Century*, it seems that Piketty is more sympathetic to the rights-oriented position of Rawls and Sen than to utilitarianism and welfarism. However, Piketty's pessimism about the effect of human capital on reducing inequality seems to prevent him from asking, for example, "Why should we pursue equality?" or "What are the benefits of pursuing equality and what are the harms of neglecting it?" Even if he relies entirely on Rawls and Sen, the narrative of *Capital in the 21st Century* unfortunately does not sufficiently reveal how he reads and understands Rawls and Sen.

In the world of philosophy after Rawls, criticism of egalitarianism has been actively conducted in recent years, not in order to deny egalitarianism, but from the awareness of the problem of confirming the true intention and purpose of egalitarianism. For example, the "leveling-down objection", which I mentioned in Chapter 0, may have been implied in Smith's critique of Rousseau, but it is a sharp attack on the perversion that can be brought about by making equality a self-purpose. Derek Parfit, one of today's most influential and radical utilitarian philosophers, poses the problem that "equality is not an end worth pursuing in itself, but is worth pursuing because the pursuit of equality almost inevitably leads to the salvation of the weak, and the priority of the weak is more important than the pursuit of equality among people". His stance is called "prioritarianism". In addition, Harry Frankfurter and others have argued that "the Rawlsian guarantee of rights is not egalitarianism, but rather a guarantee of a minimum standard, and that is sufficient." Harry Frankfurter and others have proposed the theory of "sufficientism," which says, "If equality is still worth pursuing, it is only because it is useful for guaranteeing minimum standards."

From Rawls to Sen, the philosophical debate on egalitarianism has been centered on "what equality is about" and "what should be guaranteed equally". This is becoming a hot issue. In light of such controversies, it seems to me that the position of Galor and Benabou is surprisingly deep and robust, while that of Piketty may be a bit wobbly.

If I may be so bold as to imagine, Piketty may be trying to stand on the side of Rousseau in the "Rousseau vs. Smith" scenario. In other words, what is damaged by inequality, more than the private well-being of the poor, is the opportunity for public political participation. In order to guarantee this, it is more important to guarantee rights than to simply raise the standard of living at the bottom. However, Piketty is a bit shaky in this area. Sometimes he says, "The problem is not the reduction of inequality per se, but the universal guarantee of rights," and sometimes he says, "Then the problem should not be the concentration of wealth at the top per se, but the universal guarantee of rights". But, if so, isn't it okay to take the financial resources from the middle class rather than the top class?

In the long run, there is no need to sacrifice the wealth of the bottom class in order to enrich the top class, because the market economy that we are considering is not a zero-sum situation. However, when considering domestic politics, the opportunities for public political participation and the degree to which influence is exercised may in some cases be zero-sum, with the expansion of influence at the top eclipsing that at the bottom. (Of course, one could disagree.) If we go further into this narrow realm of politics, we might be able to make a strong case for egalitarianism against the ideas of utilitarianism, prioritarianism, and sufficientism, but Piketty doesn't go into that yet.

Inequality issues not discussed by Piketty

I would like to add one more thing about global inequality as a problem of inequality, which is basically "out of the scope" of Piketty. Basically, domestic inequality in developed countries, countries that have fully developed market economies and achieved industrialization, is the subject of *Capital in the 21st Century* and Piketty's theoretical work in the 1990s, but it is difficult to say that global inequality among nations has been analyzed in depth. If I may say so, as I suggested earlier, we can talk about catching up with rapid growth and convergence of disparities.

However, it is necessary to consider what is to be explained in terms of global disparity, that is, the disparity between developed and developing countries. In other words, which is the main question, "it is natural for growth to occur, so if it does not, what is preventing it?" or "it is natural that growth does not occur, so if it does, ask what does cause it?" In economics, it is usually assumed that humans have a selfish and rational subjectivity, which is also called "homo economicus," so we tend to think that "growth is the norm. However, since Theodore Schultz's study of smallholders in developing countries, there has been an understanding that "even profit-oriented rational economists may become conservative and not aim for investments that lead to growth or expansion of management, or they may aim to take from each other rather than produce for themselves". If you think about it, this is the classic understanding since Thomas Hobbes' Leviathan.

The question is, what is this "surrounding environment"? In Galor's "Unified Theory of Growth," which I mentioned earlier, the argument is that it is mainly technological factors that cause the shift from high fertility and high death rates, which were trapped in the "Malthusian Trap," to low fertility and low death rates after the population shift and the industrial revolution, and that the situation changes drastically depending on whether productivity exceeds a certain critical level or not.

In a similar vein to the "dual structure" argument, Indian economists Mukesh Eswaran and Ashok Kotwal's analysis of the mechanism by which rural poverty drags down the entire developing economy, including the advanced industrial sector, is also interesting. According to their analysis, the wages of workers, whether rural or urban, agricultural or industrial, will be pulled down to the lowest level if the market is efficient enough. All the profits made by advanced industries will be siphoned off to the capitalists, and inequality will not improve. In order to improve the standard of living of the entire national economy, it is first necessary to increase the absolute productivity of the rural areas at the bottom, and only when the income and purchasing power of the rural poor increase and they demand not only food but also industrial products, can the productivity increase in the industrial sector lead to an improvement in the standard of living of the rural areas and workers. Note that this is not only an analysis of the disparity between developing and developed countries, but also an analysis of the disparity within developing countries (including Japan in the past), which is a different type of disparity from that analyzed by Piketty.

There are also many commentators who focus on "institutions" rather than production technology as a barrier to growth. The term "institutions" here refers to the "rule of law," which mainly focuses on property rights - a situation in which there is a system of laws concerning property rights and a governance structure that ensures that people properly observe them. Of course, this type of argument that focuses on "institutions" has its origin in Hobbes, and Rousseau's *The Origin of Human Inequality* and *The Social Contract* are also in the same lineage.

A representative of this approach in modern political economy and mathematical politics is Daron Acemoglu, a theorist of "skill-biased technological change" and a scholar of the external economy of human capital. Their book, *Why Nations Fail*, co-authored with James Robinson for the general public, has been translated into Japanese. They argue that the failure of many developing countries to take off toward growth is mainly due to the lack of "institutions" in the sense of the rule of law.

When there is either no unified governing power (a Hobbesian "state of nature"), or when there is a governing body but the rule by arbitrary tyranny for personal gain, that is, the "rule of law" is lacking, people cannot work and invest with confidence, because of fear of being robbed by outlaws or dictators.

It is also true that establishing the "rule of law" is not necessarily a disadvantage for dictators. If the source of the dictator's revenue is mainly tax revenue from the people under his control, it is not only possible to raise tax revenue by tightening and raising tax rates, but also to raise the absolute amount of tax revenue without raising tax rates by loosening the tax rates appropriately, releasing the people's will to work, and making them increase their absolute output.

Therefore, Acemoglu and Robinson use game theory to determine when a dictator/ruling class is willing to exploit and when it is willing to adopt an open-minded strategy. Very simply put, dictators tend to avoid open-minded policies even if it increases tax revenues, as long as it increases the income of the people, raises their standard of living, and increases the risk that the people will become powerful enough to launch a revolution. In addition, when the source of income of the dictator/ruler is not only the tax revenue from the ruled, but also the revenue from his own property, there is an increased possibility that he will use heavy taxes to disarm the people rather than to increase the tax revenue.

In order to break out of this equilibrium of mutual distrust between the rulers and the people, and to achieve sustainable growth under the "rule of law," it is ultimately necessary for the governing power to become a neutral "institution" rather than the private property of a particular person - that is, the rulers become "institutions" that hold positions in the power structure rather than owners of power. In other words, it is desirable to achieve a so-called "civil revolution," but this is quite difficult to achieve in practice. Even if the uprising by the rulers succeeds in overthrowing and ousting the dictator, it will be the same thing in the end if the newcomers to power privatize it.

Acemoglu & Robinson's work is not only an analysis of economic wealth, but more than that, an analysis of the disparity of political power, not the inequality of distribution in a market economy after it has been established, but the disparity in a situation where neither a market economy nor even a private property system as a precondition for a market economy has been established. In this sense, it does not conflict with Piketty's work per se - it is a separate work. They can either be dismissed as unrelated, or they can be thought of as complementing each other's work, when necessary, to gain a greater vision. However, Acemoglu & Robinson criticize Piketty for saying that "r > g is not a general law" (Piketty persistently says that this is not a claim of theoretical law, and has withdrawn from such quality work in the first place), which Piketty never said. I don't really understand what he means.

Piketty does comment on their work in *Capital in the 21st Century*. It is, however, a different work than the one in question, *Why Nations fail*. The article by Acemoglu et al. that was discussed and critically reviewed there argues that Anglo-Saxon capitalism has a higher capacity for innovation than continental European or Nordic capitalism. If it was a response to this, it would be foolish.

In general, at the end of the 20th century, from the 80s to the 90s, the dominant view of poverty and inequality was that it was a problem of the global North-South divide, and that poverty and inequality within developed countries, if it existed at all, was a problem of the few (and therefore serious). To put it in a negative light, Acemoglu-Robinson's discussion was also overshadowed by this tendency. Piketty's argument, on the other hand, was very powerful in shedding new and intense light on mass poverty and inequality in an advanced country, in a world of modern law and rule, and in an established market economy, but it was also driven by the winds of the times.

Epilogue

So, let's wrap up by looking back from Piketty.
The main points to be noted in Thomas Piketty's *Capital in the 21st Century* are, first of all, the following in the context of the flow of economics on inequality

1. He focuses on economic inequality in developed countries where free market economies are prevalent and sustained technological innovation and economic growth are the norm.

2. While most economists who focused on inequality within developed countries focused on labor income rather than capital income, and on human capital rather than physical capital, he focused more on physical capital. (Not only that, he also questioned the usefulness of the concept of human capital itself.

This can be seen as a return from neoclassical thinking to classical Marxism.
At the same time, however, inequality theorists who emphasize human capital, a trend that Piketty once embraced but now distances himself from, are in a sense breaking with the neoclassical norm and returning to classical thinking. This is because, based on the framework of the endogenous growth theory, they seek the driving force in human capital (knowledge and skills) with spillover effects, and argue that the distribution of this human capital affects production and growth. If the separation of the distributional and production (and growth) problems in the economy was the prevailing view in the neoclassical tradition, this is a departure from it. Of course, the former classicals, while also deeply concerned with the relationship between distribution and production/growth, tended to believe that production would increase and grow more if there was inequality (more wealth concentrated in the hands of capitalists as investors). On the other hand, the theorists of the "Inequality renaissance" at the end of the 20th century argued the opposite: "At least in an economy dominated by human capital, not only human capital but also physical capital, if the spillover effect of investment is high, equality will lead to higher production and higher growth."
Against this backdrop, it is extremely interesting to note that Piketty, who wrote theoretical papers on the external effects of human capital in the 1990s, eventually turned to statistical evidence and came to emphasize the importance of physical capital in inequality. If Benabou, Galor, and the other human capital optimists were on the same axis as classical economics, but in the opposite direction, Piketty is more in the same direction as the classical, which makes him even closer to the classics and especially to Marx. However, unlike Marx, who rejected the capitalist market economy as a system that could not grow without inequality, Piketty, as a modern economist after the collapse of socialism, can criticize but not reject the capitalist market economy.

The human capital optimist seeks to address the spillover effect of human capital as a "market failure" by redistributive policies rather than by internalizing it into the market, thereby reducing inequality, which is considered a chronic disease of capitalism, and further growth through this very means. (To use an analogy from environmental policy, rather than "emissions trading," in which the government puts a price on the right to pollute and makes people buy and sell it, we should have a "carbon tax," in which the government taxes polluters directly and makes them pay for their pollution.) But Piketty is not optimistic about such a prospect.

In *Capital in the 21st Century*, Piketty does not abandon the theory of endogenous growth, nor does he underestimate the effect of knowledge, skills, and education on it. The enhancement of public education is also emphasized as a mission of the welfare state (social state). However, while Piketty emphasizes the reduction of inequality through redistributive policies as a goal in itself, he does not, as in the case of human capital optimism, have an optimistic vision of "growth only through equalization"!

It is for this reason that he advocates strong taxation of asset income under international cooperation to prevent capital flight, based on the perspective that redistribution of physical capital, more than human capital, is the key to equalization. He does not believe that the correction of inequality and equalization can be done without undermining growth. However, he is not inclined to take the easy way out and say, "Equalization and growth can go hand in hand". And perhaps they also believe that "equalization, to some extent, is a goal that should be pursued even at the expense of growth." But nevertheless, we are not yet prepared to argue it convincingly.

In any case, in order to understand the significance of Piketty's work, it is surely desirable to properly determine its limitations and keep in mind the various research trends that complement it. In this book, I have reviewed the analysis of inequality, starting from pre-economic political thought and ending with contemporary economics, focusing mainly on its theoretical aspects. Piketty, too, should be seen in the context of such historical trends. Just as Piketty himself took a long historical perspective and looked at economic inequality from that perspective, we readers should read his work by placing it in the context of history. I hope this book will help you to do so.

Mathematical Appendix

1. Zero-growth steady state

First, let us introduce the following Cobb-Douglas production function as a concrete example of a production function with constant returns to scale and diminishing returns to capital and labor, where K is capital, L is labor, and Y is the product, $Y = AK^\alpha L^{1-\alpha}$, where A is total factor productivity, and is constant in (1) through (4). In other words, there is no technological change. $\alpha$ denotes the substitutability of capital and labor, $0 < \alpha < 1$. To make it easier to visualize, here is a graph with the parameters A=1 and $\alpha$ =0.5.

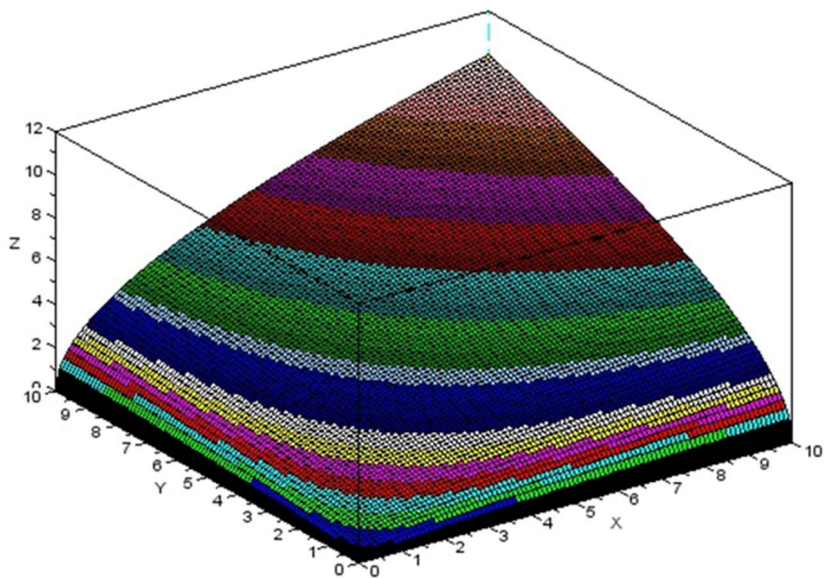

Fig.A 1 ：$Y = K^{0.5}L^{0.5} = \sqrt{KL}$ （A=1、 $\alpha$ =0.5）

The X-axis of the horizontal plane corresponds to capital K, the Y-axis to labor L, and the vertical Z-axis to output Y. Here, for example, if we fix the amount of capital as K=1, we can draw a 2D graph ($Y = L^{0.5} = \sqrt{L}$) ), which is a cross section of this 3D graph cut at K=1. (The same applies to labor L.) The shape of the graph itself is exactly the same as in Figure 2, which we will look at later, so I'll skip it. The gradual decrease in slope shows the diminishing returns of labor and capital, respectively. If you make a cross section at K=L, you will get an image of Y=K=L, or "constant harvest with respect to scale," so try it.

To simplify the discussion, we will assume that L is constant and invariant, but $L \geq 1$. Also, let us assume that each individual is allocated the same amount and quality of labor by 1. In this case, L is also the working population. (The total amount of labor L is a real number, but this is a natural number. Note that the differentiation of L is only done for the former. If we assume that average labor productivity y=Y/L and average capital equipment rate k=K/L, then $y = Ak^\alpha$.

If the capital equipment rate of individual $i$ is $k_i$ and production is $y_i$, then $y_i = Ak_i^\alpha$, and $\sum_{i=1}^{L} k_i = K$.

$Y = \sum_{i=1}^{L} y_i = \sum_{i=1}^{L} Ak_i^\alpha = A \sum_{i=1}^{L} k_i^\alpha$, $y = \frac{Y}{L} = \sum_{i=1}^{L} y_i$, $k = \frac{K}{L} = \frac{1}{L}\sum_{i=1}^{L} k_i$.

When the total amount of social capital, K, is constant, how can Y be maximized if capital is allocated to each individual? Solving the constrained maximization problem, it is easy to see that $k_i$ should be equal for all individuals i, i.e., $k_i = k$.

$$\max_{k_i} A \sum_{i=1}^{L} k_i^\alpha \quad \text{s.t.} \sum_{i=1}^{L} k_i = K.$$

Let the Lagrangian of this problem be L. Then
L= $A \sum_{i=1}^{L} k_i^\alpha + \lambda(\sum_{i=1}^{L} k_i - K)$.
For all $i$, partial differentiation over $k_i$ yields

$\frac{\partial}{\partial k_i}$L= $\alpha A k_i^{-(1-\alpha)} + \lambda = 0$

where the maximization of $Y = \sum_{i=1}^{L} Ak_i^\alpha$ is achieved. This means that $k_i = k(= \left(\frac{\alpha A}{-\lambda}\right)^{\frac{1}{1-\alpha}})$ for all i, that is, capital should be equally distributed.

Here is a graph to help you visualize this.

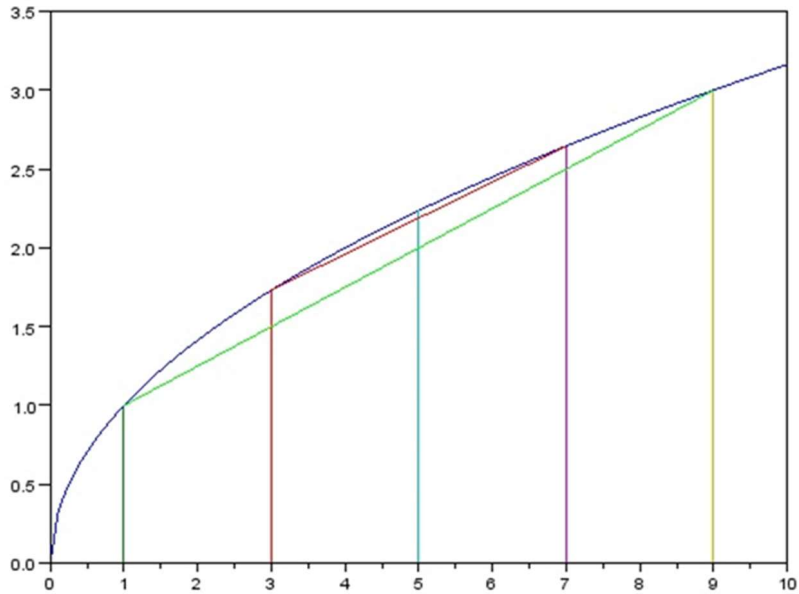

Fig.A2：production per capita, or labor productivity, $y = k^{0.5} = \sqrt{k}$

where the curve $y = \sqrt{k}$ represents the production function in terms of labor productivity per worker. The horizontal axis is the capital equipment rate per capita, $k = \frac{K}{L}$, and the vertical axis is labor productivity, $y = \frac{Y}{L}$. Dividing both sides of $Y = \sqrt{KL}$ by L, we get $\frac{Y}{L} = y, \frac{\sqrt{KL}}{L} = \frac{\sqrt{K}}{\sqrt{L}} = \sqrt{\frac{K}{L}} = \sqrt{k}$, so $y = \sqrt{k}$. The "diminishing returns on capital" is expressed here in the form of a slowing down of the growth of labor productivity as the rate of capital equipment increases.

Now let's look at the effect of equality and inequality in the distribution of capital on the average labor productivity of society as a whole. To make it easier to see, we have drawn a line segment at $k = 1, 3, 5, 7, 9$, respectively, which extends vertically along the vertical axis and reaches the curve representing the production function. At these points, we have $y = 1, \sqrt{3}, \sqrt{5}, \sqrt{7}, 3$, respectively. Furthermore, $(k, y) = (1, 1)$ and $(k, y) = (9, 3)$, $(k, y) = (3, \sqrt{3})$ and $(k, y) = (7, \sqrt{7})$ and (a) and (b), respectively. Then, let's consider the intersection of these line segments with the line segment representing k=5.

(a) $(k, y) = (1, 1)$ and $(k, y) = (9, 3)$, (b) $(k, y) = (3, \sqrt{3})$ and $(k, y) = (7, \sqrt{7})$, what we want to show is the two different patterns of the combination of capital allocation (capital equipment ratio) and corresponding labor productivity in an economy with a population of two and total capital of 10. It is clear that the distribution is more unequal in (a) than in (b). Let's also include (c), a perfectly equal combination where ¥$(k, y) = (5, \sqrt{5})$ is two people. If we were to represent (c) on this diagram, it would be the intersection of the $y = \sqrt{k}$ curve representing the production function and the line segment representing k=5.

The average labor productivity in (a), (b), and (c) can be found by looking at the line segment representing k=5. The average labor productivity in (a) can be seen by looking at the intersection of the line segment in (a) and the line segment representing k=5, and the average labor productivity in (b) can be seen by looking at the intersection of the line segment in (b) and the line segment representing k=5, both of which are slightly lower than the average labor productivity in the case of equal distribution in (c). And in case (a), where the inequality in the distribution of capital is greater, the average labor productivity is even lower than in case (b).

The generation overlap model of impure altruism

Finally, we will move on to a holistic model that considers the utility function of consumers in addition to the production function described above. Let's start with (1) and (2) using the overlapping generations model.

The lifetime utility of an individual belonging to each generation is $U = \ln c + \frac{1}{1+\theta} \ln b$.

where c is consumption, b is the legacy to the next generation, and $\theta$ is the subjective discount rate. Each person is interested in increasing the utility of themselves and their immediate children; if $\theta$ is positive, then they are giving a little priority to their own utility over the next generation.

If we introduce the time (generation) dimension explicitly, the lifetime utility of an individual in the *t*th generation of "dynasty" (family tree) *i* is

$U_i(t) = \ln c_i(t) + \frac{1}{1+\theta} \ln b_i(t)$

which is

Case 1)

The budget constraint is

$c_i(t) + b_i(t) = f(k_i(t)) - r(t)(k_i(t) - b_i(t-1))$, $y_i(t) = f(k_i(t)) = A k_i(t)^\alpha$

So

$\max_{c_i(t), b_i(t), k_i(t)} \ln c_i(t) + \frac{1}{1+\theta} \ln b_i(t)$ s.t. $c_i(t) + b_i(t) = f(k_i(t)) - r(t)(k_i(t) - b_i(t-1))$

Let $Li(t)$ be the Lagrangian for this problem (but for each subject, not for the economy as a whole), then

$Li(t) = \ln c_i(t) + \frac{1}{1+\theta} \ln b_i(t) + \lambda_i(t)(c_i(t) + b_i(t) - f(k_i(t)) + r(t)(k_i(t) - b_i(t-1)))$

This is then subjected to partial differentiation for $c_i(t), b_i(t), k_i(t)$ and

$$\frac{\partial}{\partial c_i(t)} Li(t) = \frac{1}{c_i(t)} + \lambda_i(t) = 0$$

$$\frac{\partial}{\partial b_i(t)} Li(t) = \frac{1}{(1+\theta)b_i(t)} + \lambda_i(t) = 0$$

$$\frac{\partial}{\partial k_i(t)} Li(t) = \lambda_i(t)(-f'(k_i(t)) + r(t)) = 0.$$

Thus, the condition for utility maximization is $c_i(t) = (1+\theta)b_i(t)(= -\frac{1}{\lambda_i(t)})$, and

$f'(k_i(t)) = \alpha A k_i(t)^{-(1-\alpha)} = r(t) = f'(k(t)) = \alpha A k(t)^{-(1-\alpha)}$ or $\lambda_i(t) = 0$.

However, in order for $\lambda_i(t) = 0$, $c_i(t) = (1+\theta)b_i(t) = -\infty$, so we can exclude that possibility.

In other words, the income of each individual in the current period is determined by the individual inheritance from the previous generation and the production in the current period, which is socially determined by the market interest rate. However, since $k(t) = \frac{1}{L}\sum_{i=1}^{L} b_i(t-1) = b(t-1)$, the total (average) amount of social capital that determines the current period's production is equal to the total amount of the previous period's inheritance. In other words, each individual's income in the current period is determined by both the individual and total legacies.

$$b_i(t) = \frac{1}{2+\theta}\{f(b(t-1)) - f'(b(t-1))(b(t-1) - b_i(t-1))\}$$

$$= \frac{1}{2+\theta}\{A(b(t-1))^\alpha - \alpha A(b(t-1))^{-(1-\alpha)}(b(t-1) - b_i(t-1))\}$$

$$= \frac{A(b(t-1))^{-(1-\alpha)}}{2+\theta}\{(1-\alpha)(b(t-1)) + \alpha b_i(t-1)\}$$

Averaging this out, we get $b(t) = \frac{A(b(t-1))^\alpha}{2+\theta}$.

Here we see that $\frac{1}{2+\theta}$ is the savings rate for the estate. If we solve this as $b(t) = b(t-1) = b^*$, we get the steady-state social average savings, but since we don't know the relationship between this and individual savings, we will consider that below.

If we assume that the average social saving in the steady state is $b^*$ and that of each individual is $b_i^*$, then

$$b_i^* = \frac{1}{2+\theta}\{f(b^*) - f'(b^*)(b^* - b_i^*)\}$$

from

$$b_i^* = \frac{f(b^*) - b^*f'(b^*)}{2+\theta - f'(b^*)}$$

which is equal for everyone, so

$$b_i^* = \frac{f(b^*) - b^* f'(b^*)}{2 + \theta - f'(b^*)} = b^*$$

and everyone's savings ≈ inheritance are equal, which means that consumption and income are also equal. Solving this again, the steady state is

$$b^* = \left(\frac{A}{2+\theta}\right)^{\frac{1}{1-\alpha}}$$

The steady state becomes Incidentally, the relationship between the discount rate and the marginal productivity of capital is

$\alpha(2 + \theta) = f'(b^*) = \alpha A b^{*-(1-\alpha)}$.

The growth rate to the steady state is

$$1 + g(t) = \frac{b(t)}{b(t-1)} = \frac{A(b(t-1))^{-(1-\alpha)}}{2+\theta}$$

$$1 + g_i(t) = \frac{b_i(t)}{b_i(t-1)} = \frac{A(b(t-1))^{-(1-\alpha)}}{2+\theta}\{(1-\alpha)\left(\frac{b(t-1)}{b_i(t-1)}\right) + \alpha\}$$

(1).

Case 2.

In the absence of a capital market, it is simpler, and each person's budget constraint is

$$c_i(t) + b_i(t) = f(k_i(t)) = f(b_i(t-1))$$

The Lagrangian is Let the Lagrangian be $Li(t)$.

$Li(t) = \ln c_i(t) + \frac{1}{1+\theta} \ln b_i(t) + \lambda_i(t)(c_i(t) + b_i(t) - f(b_i(t-1)))$

$\frac{\partial}{\partial c_i(t)} Li(t) = \frac{1}{c_i(t)} + \lambda_i(t) = 0$

$\frac{\partial}{\partial b_i(t)} Li(t) = \frac{1}{(1+\theta) b_i(t)} + \lambda_i(t) = 0$

The condition for utility maximization is $c_i(t) = (1+\theta)b_i(t)$, therefore

$$b_i(t) = \frac{1}{2+\theta} f(b_i(t-1)) = \frac{A b_i(t-1)^\alpha}{2+\theta}$$

The steady state is also

$$b^* = \left(\frac{A}{2+\theta}\right)^{\frac{1}{1-\alpha}}$$

The growth rate is

$$1 + g_i(t) = \frac{b_i(t)}{b_i(t-1)} = \frac{A b_i(t-1)^{-(1-\alpha)}}{2+\theta}$$

$$1 + g(t) = \frac{b(t)}{b(t-1)} = \frac{Ab(t-1)^{-(1-\alpha)}}{2+\theta}$$

The result is as follows.

Below is a graph that intuitively illustrates that this generation overlap model converges to a steady-state path no matter what initial state we start from.

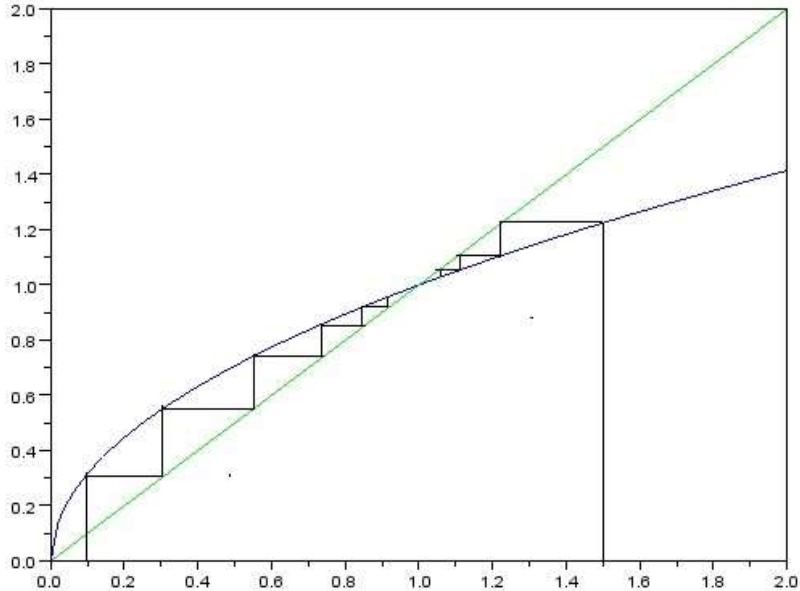

Fig.A 3：An image of convergence to a steady state in the canonical OLG model.
（horizontal axis: $b(t)$, vertical axis: $b(t+1)$）

The Ramsey Model

For reference, I will also present a model for (3) and (4), which will play a supporting role in this book. The subject here is an individual who lives forever. As for time, as in the overlapping generations model, we can use discrete time represented by natural numbers, but here we use continuous time represented by real numbers (no gaps). For the discrete time case, please refer to Acemoglu and Hayashi (2012).

The instantaneous utility of individual i at time t is $Ui(t) = \frac{c_i(t)^{1-\gamma}}{1-\gamma}$

The present value of individual i's lifetime utility at time 0 is $\int_0^\infty \frac{c_i(t)^{1-\gamma}}{1-\gamma} e^{-\theta t}\, dt$.

The budget constraint at each point in time in the case ③ is the asset $a_i$ at i.
$\dot{a_i}(t) = f(k_i(t)) - r(t)(k_i(t) - a_i(t)) - c_i(t)$ (dot over variables) (☐) denotes the time

derivative.)

Maximization of the lifetime utility of i under this constraint is obtained by solving the Hamiltonian.

Let each Hamiltonian be $Hi(t)$

$$Hi(t) = \frac{c_i(t)^{1-\gamma}}{1-\gamma} e^{-\theta t} + \lambda_i(t)\big(f(k_i(t)) - r(t)(k_i(t) - a_i(t)) - c_i(t)\big).$$

The condition for $\int_0^\infty \frac{c_i(t)^{1-\gamma}}{1-\gamma} e^{-\theta t}\, dt$ to be maximized is

$$\frac{\partial}{\partial c_i(t)} Hi(t) = c_i(t)^{-\gamma} e^{-\theta t} - \lambda_i(t) = 0,$$

$$\frac{\partial}{\partial k_i(t)} Hi(t) = \lambda_i(t)\left(f'(k_i(t)) - r(t)\right) = 0,$$

$$\frac{\partial}{\partial a_i(t)} Hi(t) = \lambda_i(t)\, r(t) = -\dot{\lambda}(t).$$

As per From there

$$\frac{-\dot{\lambda}_i(t)}{\lambda_i(t)} = \frac{-\left(c_i(t)^{-\gamma} e^{-\theta}\right)^{\cdot}}{c_i(t)^{-\gamma} e^{-\theta}} = \frac{-(c_i(t)^{-\gamma})\dot{}\, e^{-\theta t} + \theta c_i(t)^{-\gamma} e^{-\theta t}}{c_i(t)^{-\gamma} e^{-\theta t}} = \frac{-(c_i(t)^{-\gamma})\dot{}}{c_i(t)^{-\gamma}} + \theta = \gamma c_i(t)^{-\gamma-1} \frac{(\dot{c_i(t)})}{c_i(t)^{-\gamma}} + \theta =$$

$$\gamma \frac{(\dot{c_i(t)})}{c_i(t)} + \theta = r(t) = f'(k_i(t)),$$

$$g_i(t) = \frac{\dot{c_i(t)}}{c_i(t)} = \frac{f'(k_i(t)) - \theta}{\gamma} = \frac{\alpha A k_i(t)^{-(1-\alpha)} - \theta}{\gamma}.$$

In steady state, g=0, so for all individuals i

$$f'(k_i(t)) = \alpha A k_i(t)^{-(1-\alpha)} = \theta.$$

Somewhat surprisingly, this is independent of each individual's asset ownership $a_i(t)$ and is the same for all individuals. This means that in the steady state, all individuals use the same amount of capital.

However, this growth process is made possible by each individual borrowing scarce capital from others/loaning excess capital to others at the interest rate $r(t)$ through the market. Therefore, the coincidence of capital equipment rates in the steady state only means the coincidence of capital to be put into operation, not the coincidence of capital ownership to be owned and earn income from it. Individuals whose asset holdings happen to match the average capital equipment ratio at any given time neither lend nor borrow, but each person lends and borrows according to the deviation of his or her capital holdings from this average. In other words, after the end of lending and borrowing, everyone's capital equipment ratio and the marginal productivity of capital should be the same. Therefore, before the steady state, each person's output at each point in time and the growth rate of consumption will already be the same. However, since the asset gap is maintained, the gap in the absolute amount of consumption is also maintained. The interest rate at each point in time up to the steady state is $r(t) = f'(k(t)) = \alpha A k(t)^{-(1-\alpha)}$, where the average capital equipment rate at each point in time is $k(t) = \frac{1}{L}\sum_{i=1}^{L} k_i(t)$. Thus, the marginal productivity of each individual's capital in the steady state is also equal to the rate of interest, $r(t) = \theta$.

Thus

$$g_i(t) = \frac{\dot{c_i}(t)}{c_i(t)} = \frac{f'(k(t)) - \theta}{\gamma} = \frac{\alpha A k(t)^{-(1-\alpha)} - \theta}{\gamma} = \frac{r(t) - \theta}{\gamma} = g(t).$$

Note that the "golden rule of transformation" (usually translated as "golden rule of correction") r = θ + γg, which Piketty refers to in Capital in the 21st Century, holds here. The budget constraint is also

$$\dot{a_i}(t) = f(k_i(t)) - r(t)(k_i(t) - a_i(t)) - c_i(t) = f(k(t)) - f'(k(t))(k(t) - a_i(t)) - c_i(t) = Ak(t)^\alpha - \alpha A k(t)^{-(1-\alpha)}(k(t) - a_i(t)) - c_i(t) = (1-\alpha)Ak(t)^\alpha + \alpha A k(t)^{-(1-\alpha)} a_i(t) - c_i(t).$$

As mentioned earlier, in the steady state, the interest rate finally matches the subjective discount rate. Comparing consumption at this time (which is also income since savings ≈ investment is zero), we find

$$c_i(t) = f(k(t)) - r(t)(k(t) - a_i(t)) = Ak(t)^\alpha - \alpha A k(t)^{-(1-\alpha)}(k(t) - a_i(t)) = (1-\alpha)Ak(t)^\alpha + \alpha A k(t)^{-(1-\alpha)} a_i(t),$$

and we see that the consumption gap is equal to the asset gap.

In case 4), the budget constraint at each point in time is $\dot{k_i}(t) = f(k_i(t)) - c_i(t) = A k_i(t)^\alpha - c_i(t)$.

Maximizing the lifetime utility of i under this constraint, we have the Hamiltonian as H$i(t)$

$$\mathrm{H}i(t) = \frac{c_i(t)^{1-\gamma}}{1-\gamma} e^{-\theta t} + \lambda_i(t)\left(f(k_i(t)) - c_i(t)\right),$$

$$\frac{\partial}{\partial c_i(t)} \mathrm{H}i(t) = c_i(t)^{-\gamma} e^{-\theta t} - \lambda_i(t) = 0,$$

$$\frac{\partial}{\partial k_i(t)} \mathrm{H}i(t) = \lambda_i(t) f'(k_i(t)) = -\dot{\lambda_i}(t),$$

$$\frac{-\dot{\lambda_i}(t)}{\lambda_i(t)} = \gamma \frac{(\dot{c_i}(t))}{c_i(t)} + \theta = f'(k_i(t)),$$

$$g_i(t) = \frac{\dot{c_i}(t)}{c_i(t)} = \frac{f'(k_i(t)) - \theta}{\gamma} = \frac{\alpha A k_i(t)^{-(1-\alpha)} - \theta}{\gamma}.$$

However, since $a_i(t) = k_i(t)$ here, asset ownership is equal in the steady state among all individuals. Therefore, there is no consumption gap. However, the adjustment to that point will take time.

## 2. Endogenous growth

In this section, we will consider the simplest model, the so-called AK model. Specifically, we have

$Y = AK^\alpha L^{1-\alpha}, A = \bar{A}K^{1-\alpha}$, thus $Y = \bar{A}KL^{1-\alpha}$

in the previous section. In the previous chapter, we assumed that total factor productivity A was a constant, but here it is a function of capital K. In other words, as capital accumulation progresses, total factor productivity increases. To put it in more words, the characteristic of this production function is that the harvest is constant with respect to capital, and the harvest is increasing with respect to scale. This assumption seems somewhat arbitrary, but it is characterized by the fact that a steady-state growth path is reached, as we will see below.

The 3-dimensional graph corresponding to Supplementary Figure 1 ($Y = KL^{0.5} = K\sqrt{L}$) is shown in Supplementary Figure 4. It would be interesting to create various 2D graphs of cross sections for this as well (if we only present the equations, $Y = \sqrt{L}$ for K=1 cross section, Y=K for L=1 cross section, and $Y = K\sqrt{K} = L\sqrt{L}$ for K=L cross section). Please try to draw it yourself.

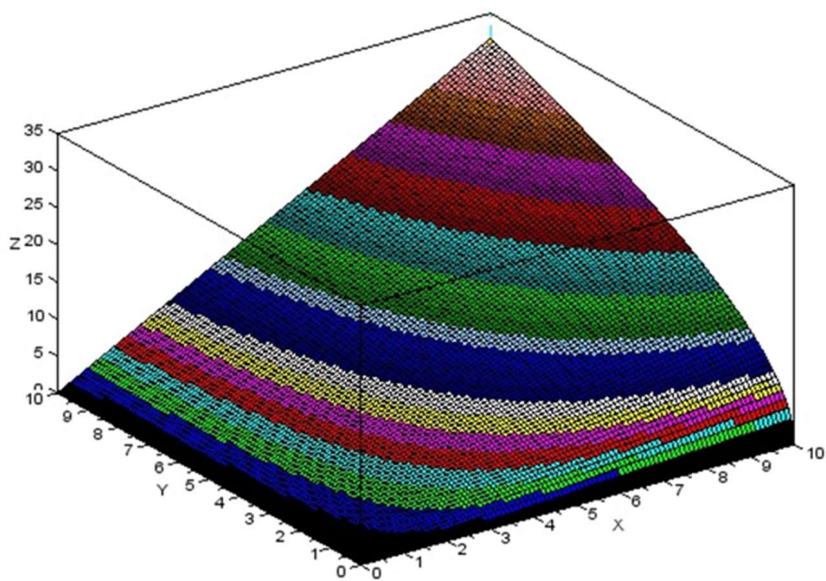

Fig.A 4 ：$Y = KL^{0.5} = K\sqrt{L}$ （A=1、 $\alpha$ =0.5）

In the following, as in the previous chapter, let us consider output per worker = labor productivity and capital equipment ratio. At first glance, if we average this out

$$= \frac{Y}{L} = F\left(\frac{K}{L}, 1\right) = f\left(\frac{K}{L}\right) = f(k) = \bar{A}K^{1-\alpha}k^\alpha = \bar{A}L^{1-\alpha}k$$

and looking at each one of them, we get

$$y_i = f(k_i) = \bar{A}K^{1-\alpha}k_i^\alpha = \bar{A}L^{1-\alpha}k^{1-\alpha}k_i^\alpha = \bar{A}L^{1-\alpha}\left(\frac{k}{k_i}\right)^{1-\alpha}k_i$$

which becomes.

Here, the marginal productivity of social capital can be obtained by differentiating Y by K, where $\bar{A}L^{1-\alpha}$, but since there is no central planning authority, investment is made at the level of individuals and individual households, according to their interests. In this case (1)', unlike the case of (1), where the harvest is constant with respect to scale, there is a Marshallese external economy in the return on investment, and the marginal productivity of capital for the individual, the marginal return on investment on an individual basis, is not the same as the social one. To put it more concretely, it is only $k_i$ that can be directly manipulated at the individual level, and each individual subject has to treat k and K as given, and in individual-level decision making, the production function is not as if $Y = \bar{A}KL^{1-\alpha}$, but as in (1), $AK^\alpha L^{1-\alpha}$, or $y_i = Ak_i^\alpha$ as in (1), and the marginal productivity of capital at the individual level is not $\bar{A}L^{1-\alpha}$ but $\alpha\bar{A}\left(\frac{K}{k_i}\right)^{1-\alpha} = \alpha\bar{A}L^{1-\alpha}\left(\frac{k}{k_i}\right)^{1-\alpha}$, whose average is $\alpha\bar{A}L^{1-\alpha}$.

Based on the above, considering the utility maximization of each individual in (1)' in the same way as in (1), the budget constraint is

$$c_i(t) + b_i(t) = f(k_i(t)) - r(t)(k_i(t) - b_i(t-1)),$$

$$y_i(t) = f(k_i(t)) = \bar{A}K(t)^{1-\alpha}k_i(t)^\alpha = \bar{A}L^{1-\alpha}k(t)^{1-\alpha}k_i(t)^\alpha,$$

in

$$\max_{c_i(t), b_i(t), k_i(t)} \ln c_i(t) + \frac{1}{1+\theta}\ln b_i(t) \quad \text{s.t.} \quad c_i(t) + b_i(t) = f(k_i(t)) - r(t)(k_i(t) - b_i(t-1)).$$

Let $Li(t)$ be the Lagrangian for this problem (but for each subject, not for the economy as a whole), then

$$Li(t) = \ln c_i(t) + \frac{1}{1+\theta}\ln b_i(t) + \lambda_i(t)(c_i(t) + b_i(t) - f(k_i(t)) + r(t)(k_i(t) - b_i(t-1)))$$

This is then subjected to partial differentiation for $c_i(t), b_i(t), k_i(t)$ and

$$\frac{\partial}{\partial c_i(t)}Li(t) = \frac{1}{c_i(t)} + \lambda_i(t) = 0,$$

$$\frac{\partial}{\partial b_i(t)}Li(t) = \frac{1}{(1+\theta)b_i(t)} + \lambda_i(t) = 0,$$

$$\frac{\partial}{\partial k_i(t)}Li(t) = \lambda_i(t)(-f'(k_i(t)) + r(t)) = 0.$$

Thus, the condition for utility maximization is $c_i(t) = (1+\theta)b_i(t)(= -\frac{1}{\lambda_i(t)})$, and

$$f'(k_i(t)) = \alpha \bar{A} L^{1-\alpha} \left(\frac{k(t)}{k_i(t)}\right)^{1-\alpha} = r(t) = f'(k(t)) = \alpha \bar{A} L^{1-\alpha} \text{ or } \lambda_i(t) = 0,$$

but again we can ignore the possibility that ¥lambda_i(t)=0. Then, formulate the same way as before to get

$$b_i(t) = \frac{1}{2+\theta}\{f(b(t-1)) - f'(b(t-1))(b(t-1) - b_i(t-1))\} = \frac{1}{2+\theta}\{\bar{A}L^{1-\alpha}b(t-1) - \alpha\bar{A}L^{1-\alpha}(b(t-1) - b_i(t-1))\} = \frac{\bar{A}L^{1-\alpha}}{2+\theta}\{(1-\alpha)(b(t-1)) + \alpha b_i(t-1)\}.$$

In other words, the steady state here is a state of sustained positive growth, not a halt or zero growth as in (1). In other words

$$\frac{b(t)}{b(t-1)} = \frac{\bar{A}L^{1-\alpha}}{2+\theta}.$$

(assuming that $\bar{A}L^{1-\alpha} \geq \bar{A} > 2+\theta$).

For each individual

$$\frac{b_i(t)}{b_i(t-1)} = \frac{\bar{A}L^{1-\alpha}}{2+\theta}\{(1-\alpha)\left(\frac{b(t-1)}{b_i(t-1)}\right) + \alpha\}.$$

So, if $b_i$ is above (below) b, its growth rate is below (above) the growth rate of b (average and steady-state growth rate). Therefore, the growth rate of b_i tends to converge to the growth rate of b. To go a little further, we have

$$\frac{b_i(t)}{b(t)} = \frac{\bar{A}L^{1-\alpha}}{2+\theta}\left\{\frac{(1-\alpha)(b(t-1))}{b(t)}\right\} + \alpha\frac{b(t-1)b_i(t-1)}{b(t)b(t-1)} = (1-\alpha) + \alpha\frac{b_i(t-1)}{b(t-1)}$$

$$= 1 + \alpha(\frac{b_i(t-1)}{b(t-1)} - 1)$$

Thus

$$\lim_{t\to\infty} \frac{b_i(t)}{b(t)} = 1.$$

Thus, there is convergence not only in the growth rate but also in the level of income and wealth.

Similarly for (2)', referring to (2), we can formulate

$$b_i(t) = \frac{1}{2+\theta}f(b_i(t-1)) = \frac{\bar{A}L^{1-\alpha}}{2+\theta}\left(\frac{b(t-1)}{b_i(t-1)}\right)^{1-\alpha} b_i(t-1)$$

$$= \frac{\bar{A}L^{1-\alpha}}{2+\theta}(b(t-1))^{1-\alpha}(b_i(t-1))^{\alpha},$$

$$\frac{b_i(t)}{b(t)} = \frac{\bar{A}L^{1-\alpha}}{2+\theta} \frac{(b(t-1))^{1-\alpha}(b_i(t-1))^{\alpha}}{b(t)} = \frac{(b_i(t-1))^{\alpha}}{(b(t-1))^{\alpha}} = \left(\frac{b_i(t-1)}{b(t-1)}\right)^{\alpha}$$

Here again

$$\lim_{t \to \infty} \frac{b_i(t)}{b(t)} = 1$$

and convergence of not only the growth rate but also the level occurs in all families.

The steady-state growth rate is also

$$\frac{b(t)}{b(t-1)} = \frac{\bar{A}L^{1-\alpha}}{2+\theta}$$

The growth rate for each family line is

$$\frac{b_i(t)}{b_i(t-1)} = \frac{\bar{A}L^{1-\alpha}}{2+\theta} \left(\frac{b(t-1)}{b_i(t-1)}\right)^{1-\alpha}$$

So the trend is the same.

So, there is not much difference between (1)' and (2)', but that is not the case. First, for (1)', let's look again at

$$b_i(t) = \frac{\bar{A}L^{1-\alpha}}{2+\theta}\{(1-\alpha)(b(t-1)) + \alpha b_i(t-1)\}$$

Let's consider this from

$$\frac{1}{L}\sum_{i=1}^{L} b_i(t) = b(t)$$

So

$$b(t) = \frac{1}{L}\sum_{i=1}^{L} b_i(t) = \frac{\bar{A}L^{1-\alpha}}{2+\theta}(1-\alpha)(b(t-1)) + \frac{\bar{A}L^{1-\alpha}}{2+\theta}\left\{\alpha \frac{1}{L}\sum_{i=1}^{L} b_i(t-1)\right\}$$

$$= \frac{\bar{A}L^{1-\alpha}}{2+\theta}\{(1-\alpha)(b(t-1)) + \alpha b(t-1)\} = \frac{\bar{A}L^{1-\alpha}}{2+\theta} b(t-1)$$

That is

$$\frac{b(t)}{b(t-1)} = \frac{\bar{A}L^{1-\alpha}}{2+\theta}$$

is actually always true even before convergence to the steady state.

However, in (2)', we have

$$b_i(t) = \frac{\bar{A}L^{1-\alpha}}{2+\theta}(b(t-1))^{1-\alpha}(b_i(t-1))^{\alpha}$$

Therefore

$$b(t) = \frac{1}{L}\sum_{i=1}^{L} b_i(t) = \frac{\bar{A}L^{1-\alpha}}{2+\theta}(b(t-1))^{1-\alpha} \frac{1}{L}\sum_{i=1}^{L}(b_i(t-1))^{\alpha}$$

In this case, $0 < \alpha < 1$, so from Herder's inequality

$$\frac{1}{L}\sum_{i=1}^{L}(b_i(t))^\alpha \leq \left(\frac{1}{L}\sum_{i=1}^{L}b_i(t)\right)^\alpha = b(t)^\alpha$$

(The equality sign only holds if b_i¥left(t¥right)=¥ b¥left(t¥right) for all i). Thus

$$b(t) = \frac{1}{L}\sum_{i=1}^{L}b_i(t) = \frac{\bar{A}L^{1-\alpha}}{2+\theta}(b(t-1))^{1-\alpha}\frac{1}{L}\sum_{i=1}^{L}(b_i(t-1))^\alpha \leq \frac{\bar{A}L^{1-\alpha}}{2+\theta}b(t-1),$$

$\frac{b(t)}{b(t-1)} = \frac{\bar{A}L^{1-\alpha}}{2+\theta}(b(t-1))^{-\alpha}\frac{1}{L}\sum_{i=1}^{L}(b_i(t-1))^\alpha = \left(\frac{\bar{A}L^{1-\alpha}}{2+\theta}\right)\frac{\frac{1}{L}\sum_{i=1}^{L}(b_i(t-1))^\alpha}{\left(\frac{1}{L}\sum_{i=1}^{L}b_i(t-1)\right)^\alpha} = \left(\frac{\bar{A}}{2+\theta}\right)\frac{\sum_{i=1}^{L}(b_i(t-1))^\alpha}{\left(\sum_{i=1}^{L}b_i(t-1)\right)^\alpha} \leq$

$\frac{\bar{A}L^{1-\alpha}}{2+\theta}$.

That is, in (2)', until convergence to the steady state, the average growth rate will be consistently lower than in (1)', other things being equal, including the initial state and the distribution of wealth at the starting point. After convergence to the steady state, the average growth rate is equal thereafter, so the average income and wealth levels in 2´ remain consistently lower than in 1´. The following is a graph of this in the setting shown in Figure 4. This is Figure 1 in Chapter 7 of this book.

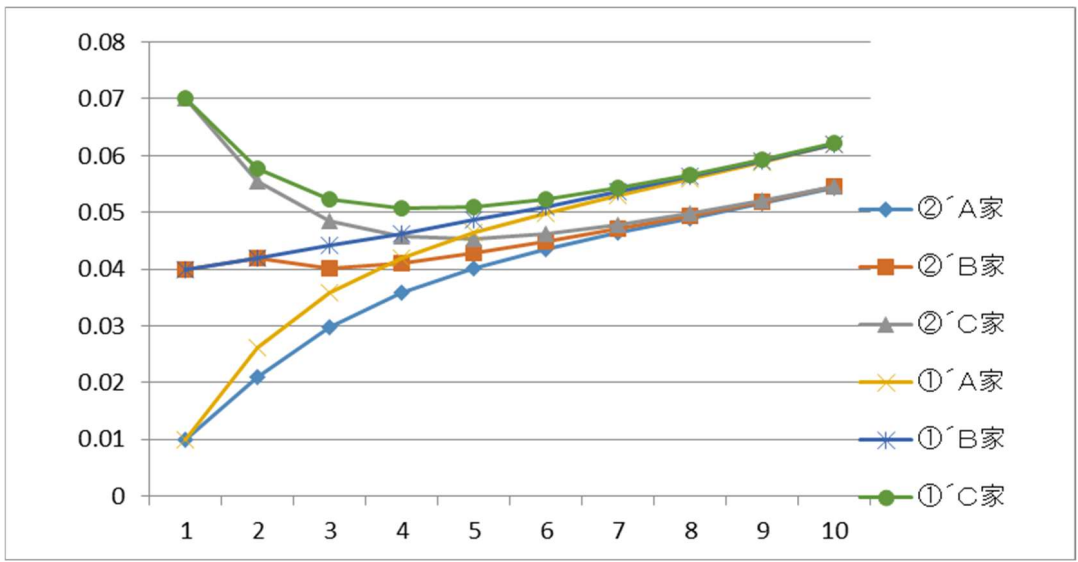

Fig. 1 in chapter 1

Also, the average level of income and wealth in (2)' will be lower the greater the wealth inequality at the starting point. For example, if we consider the most extreme inequality, specifically a situation where all capital is concentrated in one person, the growth rate at that point will be $\frac{\bar{A}}{2+\theta}$, while if there is perfect equality at the starting point, the individual growth rate ≈ average growth rate from the beginning will be $\frac{\bar{A}L^{1-\alpha}}{2+\theta}$.

$$1 \leq \frac{\sum_{i=1}^{L}(b_i(t-1))^{\alpha}}{(\sum_{i=1}^{L}b_i(t-1))^{\alpha}} \leq L^{1-\alpha}$$

$$\left(\sum_{i=1}^{L}b_i(t-1)\right)^{\alpha} \leq \sum_{i=1}^{L}(b_i(t-1))^{\alpha} \leq L^{1-\alpha}\left(\sum_{i=1}^{L}b_i(t-1)\right)^{\alpha}$$

In general, if X and Y are random variables and Y is a mean-preserving expansion of X-- E(X) = E(Y), σ²(X) < σ²(Y)--then the function f( ) for X and Y is a concave function Then, Ef(X) ≤Ef(Y). (cf. Aghion, Caroli, and Garcia-Peñalosa (1999)).

From there, we can consider $b(t)$ as a function of $\{b_1(t-1),..,b_L(t-1)\},¥$ . If we apply a mean-preserving expansion to $\{b_1(t-1),..,b_L(t-1)\}$, the value of $b(t)$ will be lower. This is illustrated in the following figure. This is Figure 2 in Chapter 7.

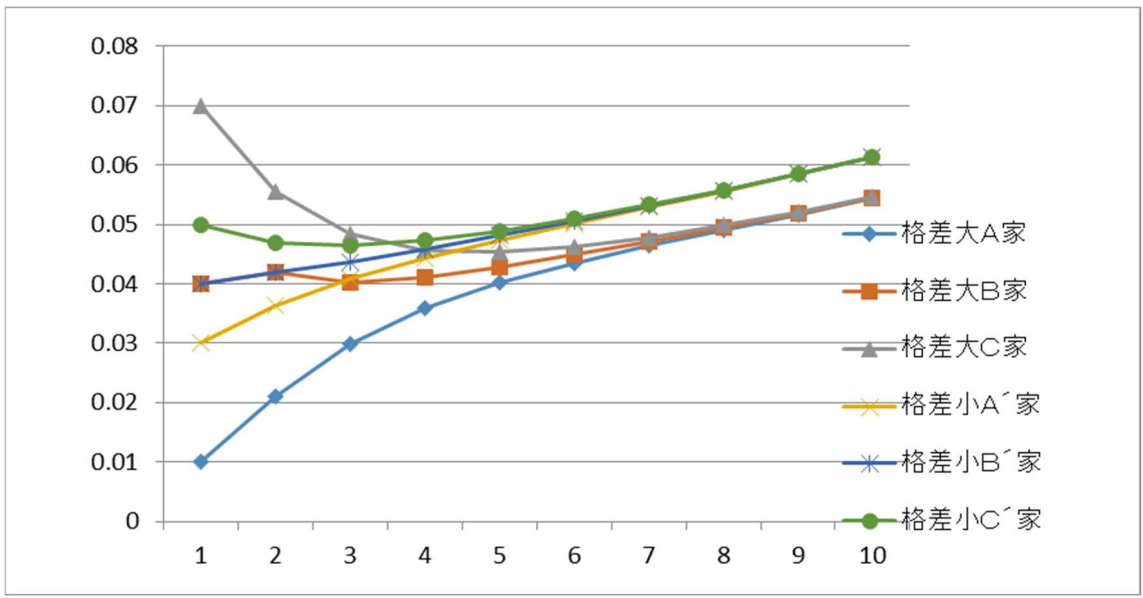

Fig.2 in chapter 7

As for (3)', if we identify the production function with the AK model as before, the growth rate of each individual consumption is

$$g_i(t) = \frac{\dot{c}_i(t)}{c_i(t)} = \frac{f'(k_i(t))-\theta}{\gamma} = \frac{\alpha\bar{A}L^{1-\alpha}\left(\frac{k(t)}{k_i(t)}\right)^{1-\alpha}-\theta}{\gamma}.$$

However, if we take into account the existence of capital markets, this is not the case in practice. As before, the capital market will work to adjust for the difference between the capital owned by each individual and the optimal capital equipment ratio, which is not shown here, so that everyone's capital equipment ratio is equal and

$$g(t) = \frac{\dot{c}(t)}{c(t)} = \frac{f'(k(t)) - \theta}{\gamma} = \frac{\alpha \bar{A} L^{1-\alpha} - \theta}{\gamma} = g$$

This results in a steady-state growth of The growth rate of each individual's personal consumption is constant, and thus asset, income, and consumption inequality will continue to hold in proportion to initial assets.

However, in 4′ , the growth rate for each individual is realized.

In addition, the budget constraint for each individual is

$$\dot{k_i}(t) = \bar{A} L^{1-\alpha} \left(\frac{k(t)}{k_i(t)}\right)^{1-\alpha} k_i(t) - c_i(t) = \bar{A} L^{1-\alpha} k(t)^{1-\alpha} k_i(t)^\alpha - c_i(t)$$
$$= \bar{A} K(t)^{1-\alpha} k_i(t)^\alpha - c_i(t)$$

If the capital owned by each individual is above the social average, the marginal productivity of that individual's capital and the growth rate will fall, and if it is below, it will rise, resulting in convergence to steady-state growth, which involves equalization of capital ownership.

Now let's examine the circumstances leading to this convergence. If we add the above budget constraint equation for each individual to the economy as a whole and move the consumption term to the left side, we get the equality of national income [left side] and gross national product [right side]. Then, again from Herder's inequality

$$\sum_{i=1}^{L} [c_i(t) + \dot{k_i}(t)] = \bar{A} K(t)^{1-\alpha} \sum_{i=1}^{L} k_i(t)^\alpha \leq \bar{A} K(t)^{1-\alpha} \left[\sum_{i=1}^{L} k_i(t)\right]^\alpha L^{1-\alpha} = \bar{A} K(t) L^{1-\alpha}$$

(The equality sign is only valid when $k_i(t)$ is equal for all i). Without going into details, in (3)', since everyone's capital-labor ratio is aligned through capital lending and borrowing, we have

$$\sum_{i=1}^{L} [c_i(t) + \dot{k_i}(t)] = \bar{A} L^{1-\alpha} \sum_{i=1}^{L} k_i(t) = \bar{A} K(t) L^{1-\alpha}$$

always holds. Hence, similar to the relationship between (1)' and (2)', the average growth rate in (4)' will be consistently lower than in (3)' until it converges to the steady state, other things being equal, including the initial state and the distribution of wealth at the starting point. Also, as in the case of (1)' and (2)', we now assume that gross national product is equal to the capital distribution $\{k_1(t), \ldots, k_L(t)\}$, which is a concave function, so that $\{k_1(t), \ldots, k_L(t)\}$ with a mean-preserving expansion, the value of gross national product will be lower.

The graphs were drawn using Microsoft Excel for two-dimensional graphs and Scilab (http://www.scilab.org/) for three-dimensional graphs.

Chapters 7 and 8.

David N. Weil (2008) *Economic Growth, 2nd Edition*, Pearson..

This is a textbook for beginning and intermediate undergraduate students. There is an explanation of the Galois-Zyra model in the chapter on "Inequality.

Elhanan Helpman (2004) *The Mystery of Economic Growth,* Belknap Press.

A general introduction to the study of economic growth since the theory of inclusive growth. There is also a chapter on inequality, which is informative.

Giuseppe Bertola, Reto Foellmi & Josef Zweimüller (2005) *Income Distribution in Macroeconomic Models,* Princeton University Press.

Philippe Aghion & Peter Howitt（2008）*The Economics of Growth*, The MIT Press.

Daron Acemoglu (2009) *Introduction to Modern Economic Growth,* Princeton University Press.

The above three are generally graduate level textbooks. There is no need for amateurs to read them.

Oded Galor & Joseph Zeira (1993) "Income Distribution and Macroeconomics," *The Review of Economic Studies* 60(1), 35-52.

Alberto Alesina and Dani Rodrik (1994), "Distributive Politics and Economic Growth," *The Quarterly Journal of Economics* 109(2), 465-490.

Roland Bénabou (1996), "Inequality and Growth," in B. S. Bernanke and J. J. Rotemberg (eds.), *NBER Macroeconomics Annual 1996*, 11-74.

Thomas Piketty (1997), "The Dynamics of the Wealth Distribution and the Interest Rate with Credit Rationing," *The Review of Economic Studies* 64(2), 173-174. Economic Studies 64(2), 173-189.

Chapter 9

Thomas Piketty (2014) *Capital in the Twenty-First Century,* Harvard University Press.

Oded Galor (2011) *Unified Growth Theory,* Princeton University Press.

Oded Galor & Omer Moav (2006) "Das Human-Kapital: A Theory of the Demise of the Class Structure," *The Review of Economic Studies*, 73(1), 85-117.